\documentclass[numberedappendix]{emulateapj}
\usepackage[colorlinks,citecolor=cyan]{hyperref}

\usepackage{graphicx}
\usepackage{amsmath}
\usepackage{psfig}
\usepackage{apjfonts}

\def\kg{\textrm{kg}}
\def\sec{\textrm{s}}
\def\Joule{\textrm{J}}
\def\Watt{\textrm{W}}
\def\Newton{\textrm{N}}
\def\Amp{\textrm{A}}
\def\Coulomb{\textrm{C}}
\def\Farad{\textrm{F}}

\def\nm{\textrm{nm}}
\def\um{\mu\textrm{m}}
\def\mm{\textrm{mm}}
\def\cm{\textrm{cm}}
\def\meter{\textrm{m}}
\def\km{\textrm{km}}

\def\pc{\textrm{pc}}
\def\kpc{\textrm{kpc}}
\def\Mpc{\textrm{Mpc}}
\def\Gpc{\textrm{Gpc}}

\def\yr{\textrm{yr}}
\def\kyr{\textrm{kyr}}
\def\Myr{\textrm{Myr}}
\def\Gyr{\textrm{Gyr}}

\def\Hz{\textrm{Hz}}

\def\GHz{\textrm{GHz}}
\def\THz{\textrm{THz}}

\def\mJy{\textrm{mJy}}
\def\Jy{\textrm{Jy}}

\def\uKelv{\mu\textrm{K}}

\def\Kelv{\textrm{K}}

\def\arcmin{\textrm{arcmin}}

\def\kgm2{\textrm{kg\ m}^{-2}}
\def\mgm2{\textrm{mg\ m}^{-2}}
\def\ugm2{\mu\textrm{g\ m}^{-2}}
\def\rhoUnits{\textrm{g\ cm}^{-3}}
\def\kms{\textrm{km\ s}^{-1}}

\def\Msun{\textrm{M}_{\odot}}
\def\Lsun{\textrm{L}_{\odot}}
\def\MEarth{\textrm{M}_{\oplus}}

\def\MPa{\textrm{MPa}}
\def\bit{\textrm{bit}}
\def\Gbit{\textrm{Gbit}}
\def\Ohm{\Omega}
\def\kOhm{\textrm{k}\Omega}
\def\amu{\textrm{amu}}
\def\ton{\textrm{ton}}
\def\Mton{\textrm{Mton}}
\def\AU{\textrm{AU}}
\def\dayUnit{\textrm{day}}

\def\RDipole{{\cal R}_{\lambda/2}}
\def\RRad{{\cal R}_{\rm rad}}
\def\ZImp{{\cal Z}}
\def\XReact{{\cal X}}
\def\RResist{{\cal R}}
\def\EDotHarvest{\dot{E}_{\rm harvest}}
\def\etaAbs{\overline{\eta_{\rm abs}}}
\def\etaWork{\overline{\eta_{\rm work}}}
\def\etaRec{\overline{\eta_{\rm rec}}}
\def\PBarSpect{\overline{P(\nu)}}
\def\PSendSpect{P_{\rm send}(\nu)}
\def\PRecSpect{P_{\rm rec}(\nu)}
\def\PRecBol{\overline{P_{\rm rec}}}
\def\dBar{\overline{d}}
\def\Gain{{\cal G}}

\def\GHAT{\^G}
\def\icarus{Icarus}

\def\ga{\gtrsim}
\def\la{\lesssim}

\newcommand{\mean}[1]{\ensuremath{\langle #1 \rangle}}

\shorttitle{(No) Type III Blackboxes}
\shortauthors{Lacki}

\begin{document}

\title{Type III Societies (Apparently) Do Not Exist}
\author{Brian C. Lacki$^\varnothing$}
\email{astrobrianlacki@gmail.com}
\noaffiliation

\begin{abstract}
Whether technological societies remain small and planet-bound like our own, or ultimately span across galaxies is an open question in the Search for Extraterrestrial Intelligence.  Societies that engineer on a galactic scale are classified as Type III on Kardashev's scale.  I argue that Type III societies can take the form of blackboxes, entire galaxies veiled in an opaque screen.  A blackbox has a temperature that is just above that of the cosmic microwave background, for the maximum possible thermodynamic efficiency.  The screen can be made from artificial dust pervading the galaxy, establishing feedback on the diffuse interstellar medium itself.  I show that there is enough material in galaxies to build blackboxes if the dust is fashioned into dipole antennas, especially if they are made of carbon nanotubes.

The thermal emission of a blackbox makes it a bright microwave source.  I examine the \emph{Planck} Catalog of Compact Sources (PCCS2) to constrain the abundance of blackboxes.  None of the 100 GHz sources has the spectrum expected of a blackbox.  The null result rules out shrouded galaxy clusters out to $z \approx 1$ and shrouded Milky Ways out to (comoving) 700 Mpc.  The reach of the results includes 3 million galaxies containing an estimated 300 quadrillion terrestrial planets, as well as tens of thousands of galaxy clusters.  A more detailed search of the PCCS2 can find or rule out blackboxes in 30 million galaxies, and the South Pole Telescope Sunyaev-Zeldovich survey can search another 20 million galaxies.

Combined with the null results from other searches for Type III societies, I conclude that they are so rare that they basically do not exist within the observable Universe.  A hypothesis of ``Cosmic Pessimism'' is discussed, in which we are alone, our long-term chances for survival are slim, and if we do survive, our future history will be checkered.  Our loneliness is suggested by the lack of Type III societies.  I discuss the remaining forms of Type III societies not yet well constrained by observation.  I argue that the ease of building blackboxes on planetary and Solar System scales may lead, within a few centuries, to environmental catastrophes vastly more devastating than anything we are doing now, boding ill for us.
\end{abstract}


\section{Introduction}

What are the limits to technological achievement?  That question has always been implicit in discussions for the Search for Extraterrestrial Intelligence (SETI).  As far as we can tell, the Universe as a whole is pristine and natural.  If it's easy for technological societies to engineer whole sections of the cosmos, then the simplest explanation seems to be that we are alone.  If such ``cosmic engineering'' is difficult or impossible, then there may be a menagerie of societies out there that are too subtle to detect yet.  Interstellar travel, which has not yet been achieved by humans, is one possible limit to technology.  The apparent lack of starfarers in the local Universe leads to Fermi's Paradox, which is the observation that the Galaxy is old enough for even fairly slow starfarers, traveling at a few percent of $c$, to spread across entirely (\citealt{Brin83} and \citealt{Cirkovic09-Fermi} review possible solutions and implications).  While some societies may decide against a large-scale expansion program, it seems unbelievable that any unified ban could be maintained across multitudes of completely alien societies and across billions of years \citep{Hart75}.

Technological limits can be classified according to how much energy a society harvests.  \citet{Kardashev64} introduced a widely-used scale that roughly spans the imaginable possibilities.  A Type I society harnesses the power its planet receives from its sun.  A Type II society catches all of the power radiated by its home sun.  And a Type III society harvests all the energy radiated by all of the stars in its home galaxy \citep[see also][]{Cirkovic15}.  If technology is limited, then SETI should search mainly for Type I (or smaller) societies, which may be nearby.  If, on the other hand, societies can and do achieve starfaring, then the Fermi-Hart argument suggests we should look for distant but easily observable Type III societies \citep{Wright14-SF}. 

Most SETI efforts so far have concentrated on small and nearby societies \citep{Tarter01}, but as galaxy surveys have become commonplace, there has been more discussion of how to look for Type III societies.  Surveys for Type III societies can follow at least two basic strategies.  They can look for galaxies with strange morphologies, to see if aliens have physically rearranged or hidden stars (\citealt{Carrigan12,Voros14}; \citealt{Badescu06} describes how stars could be pushed around).  Or, they look for signs that the radiated starlight of the entire galaxy is collected.  Then the galaxy will look abnormally dark in the optical \citep{Annis99,Zackrisson15}, but will emit waste power as thermal infrared \citep{Wright14-SF,Garrett15}.  Either the optical faintness or the thermal emission are signs. 

In order to isolate excess thermal emission, the temperature of the artificial structures must be assumed.  This is especially important because the dust mixed into galaxies' gas catches about half of the ultraviolet/optical starlight and thus also glows thermally, with typical temperatures of order 20 K in $z = 0$ normal star-forming galaxies \citep[e.g.,][]{Draine11}.  The typical assumption is that the thermal emission comes from billions of Dyson spheres, each a shell of habitats surrounding a star at $\sim$AU distances and kept at habitable temperatures.  This maximizes the habitable volume around each individual star.  Then the excess thermal emission comes out in the mid-infrared (MIR; \citealt{Dyson60,Sagan66,Slysh85}).  The most comprehensive survey yet for engineering of this kind, Glimpsing Heat from Alien Technology (\GHAT), searched for excess MIR thermal emission from $10^5$ galaxies in the Wide-field Infrared Survey Explorer (WISE) All-Sky source catalog, and found no evidence for galaxies with their stellar populations surrounded by Dyson spheres (\citealt{Wright14-SF,Wright14-Results,Griffith15}; \citealt{Garrett15} re-examined this sample).

This is not the only possible way to harvest all of the starlight from a galaxy, though.  I will motivate the idea that Type III societies can plausibly take the form of \emph{blackboxes}.  A blackbox is, simply, an entire galaxy (or even cluster of galaxies) shrouded within an opaque screen.  The activities of the societies within it are thus completely unknown to us, and largely irrelevant as far as detecting them goes.  A blackbox is fundamentally a galactic-scale entity, not just a collection of stellar Type II societies.  Since the emitting area covers the entire galaxy, and not just AU-sized bubbles around individual stars, its effective temperature is no more than about a K (Kelvin) over that of the cosmic microwave background (CMB).  Although a structure that big may seem far-fetched, especially since Dyson spheres already defy the modern limits of engineering, I show that it could be possible with the use of artificial dust.  

A serendipitous survey for blackboxes already exists: the \emph{Planck} maps of the microwave sky.  After all, \emph{Planck} is designed specifically to look for temperature fluctuations in the CMB, and a blackbox appears on the sky as a fairly large region of the sky that is slightly hotter than the CMB.  Sunyaev-Zeldovich surveys also are good for detecting blackboxes for similar reasons.  

Yet these experiments have found no blackboxes.  As I will show, this means that \emph{there are essentially no blackboxes anywhere in the observable Universe}.  Combined with other null results, it appears that Type III societies do not exist within the observable Universe, not one in millions of galaxies.  I am left to conclude, from the Fermi-Hart argument, that we are in fact alone in the observable Universe, or we are doomed, and very plausibly both.

\begin{deluxetable*}{lll}
\tablecaption{Constants in SI units}
\tablehead{\colhead{Name} & \colhead{Value} & \colhead{Explanation}}
\startdata
$c$                & $2.998 \times 10^8\ \meter\ \sec^{-1}$                 & Speed of light in vacuum\\
$\mu_0$            & $4 \pi \times 10^{-7}\ \Newton\ \Amp^{-1}$             & Magnetic permeability of vacuum\\
$\epsilon_0$       & $8.854 \times 10^{-12}\ \Farad\ \meter^{-1}$           & Electric permittivity of vacuum \\
$h$                & $6.626 \times 10^{-34}\ \Joule\ \sec$                  & Planck's constant\\
$G$                & $6.674 \times 10^{-11}\ \Newton\ \meter^2\ \kg^{-2}$   & Newton's gravitational constant\\
$q_e$              & $1.602 \times 10^{-19}\ \Coulomb$                      & Electric charge of electron\\
$m_e$              & $9.109 \times 10^{-31}\ \kg$                           & Mass of electron\\
$\amu$             & $1.661 \times 10^{-27}\ \kg$                           & Atomic mass unit\\
$k_B$              & $1.381 \times 10^{-23}\ \Joule\ \Kelv^{-1}$            & Boltzmann's constant\\
$\sigma_{\rm SB}$  & $5.671 \times 10^{-8}\ \Watt\ \meter^{-2}\ \Kelv^{-4}$ & Stefan-Boltzmann constant\\
$\ZImp_0$          & $376.7\ \Ohm$                                          & Impedance of free space ($\mu_0 c$)\\
$\RDipole$         & $73.2\ \Ohm$                                           & Radiation resistance of thin half-wavelength dipole\\
$\Jy$              & $10^{-26}\ \Watt\ \meter^{-2}\ \Hz^{-1}$               & Jansky unit of flux density\\
$\MPa$             & $10^6\ \Newton\ \meter^{-2}$                           & MegaPascal, unit of pressure\\
$\ton$             & $1000\ \kg$                                            & Metric ton\\
$\yr$              & $3.1557 \times 10^7\ \sec$                             & Earth year\\
$\AU$              & $1.496 \times 10^{11}\ \meter$                         & Astronomical unit\\
$\MEarth$          & $5.973 \times 10^{24}\ \kg$                            & Mass of the Earth\\
$\Lsun$            & $3.828 \times 10^{26}\ \Watt$                          & Bolometric luminosity of the Sun\\
$\Msun$            & $1.989 \times 10^{30}\ \kg$                            & Mass of the Sun\\
$\kpc$             & $3.0857 \times 10^{19}\ \meter$                        & Kiloparsec\\
$L_{\rm MW}$       & $6.5 \times 10^{10}\ \Lsun$                            & Bolometric luminosity of the Milky Way\\
$u_{\rm EBL} (0)$  & $1.589 \times 10^{-15}\ \Joule\ \meter^{-3}$           & Present energy density of extragalactic background light\\
$T_{\rm CMB} (0)$  & $2.725\ \Kelv$                                         & Present temperature of the CMB\\
$H_0$              & $67.7\ \km\ \sec^{-1}\ \Mpc^{-1}$                      & Hubble's constant\\
$\Omega_m$         & 0.309                                                  & Cosmological (baryonic+dark) matter density\\
$\Omega_{\Lambda}$ & 0.691                                                  & Cosmological constant energy density
\enddata  
\tablecomments{I find $\RDipole$ from \citet{Heald12}, equation 9.73.  The value for $L_{\rm MW}$ comes from \citet{Strong10}.  \citet{Finke10} provides the current extragalactic background light energy density.  Cosmological values are from \emph{Planck} \citep{Ade15-Params}, except for $T_{\rm CMB} (0)$, which is taken from \citet{Fixsen09}.  The other values are taken from \citet{Olive15}.}
\label{table:Constants}
\end{deluxetable*}

As this paper includes calculations of the electrical properties of bulk materials, particularly the resistances of metal wires, I use the SI unit system.  The values of physical and astronomical constants in these units is given in Table~\ref{table:Constants}.  I also use a flat $\Lambda$CDM cosmology when calculating distances.  The given distances in this paper are comoving unless otherwise stated.

\section{What Form Would a Type III Society Take?}

When Kardashev elaborated on the idea of a Type III society, he described, as an example, a single structure that actually was galactic in size \citep{Kardashev85}.  The entire edifice would be a habitat for life and a home for sentient beings.  While Kardashev focuses on a structure with an Earthly surface temperature of 300 K, he does mention that temperatures as low as 3 K are possible. 

Current ideas of Type III societies are more conservative: they are now imagined mainly as ``just'' a galaxy-sized collection of Type II societies in the form of Dyson spheres.  Although \citet{Slysh85} considers cold, diffuse Dyson spheres with temperatures down to 50 K, and \citet{Timofeev00} note the possibility of 3 K Dyson spheres, most Dyson sphere searches assume that the spheres are compact, AU-scale structures with characteristic temperatures of a few hundred Kelvin (going down to 100 K in \citealt{Carrigan09} and \GHAT).  But I will show that there is a problem with this concept that has not been discussed in the literature: it would take too much matter to surround a red giant with a shell of habitats.  The solution -- capturing light with dipole antennas -- also makes building true galactic-size structures easy.  This motivates a search for truly big structures, like those envisioned by \citet{Kardashev85}, with temperatures much colder than 300 K. 

\subsection{Chilly Ways: Why it is hard to catch a galaxy's starlight with Dyson spheres}
\label{sec:ChillyWays}
Imagine every star in the Milky Way, besides the Sun, was surrounded by a Dyson sphere.  This is the kind of galaxy \GHAT~is looking for.  Everything outside of the Dyson Spheres could be unchanged.  If we looked into this altered Milky Way at night, it would be invisible to our eyes, in optical light.  But in MIR light, it would basically look the same as it does now in near-infrared and optical (ignoring interstellar dust).  We would still see a disk and a bulge; they'd just appear to be made of big, cold stars.  No convenient name has been given to these converted galaxies in the literature, as far as I am aware.  In this paper, I will call them \emph{Chilly Ways} -- the diffuse, white band of starlight that we know as the Milky Way is transformed into a diffuse band of mid-infrared light from chilly Dyson spheres.  

Dyson spheres were originally conceived as something living beings would build around their home stars, which are presumed to be similar to the Sun.  The spheres are rigid structures, or swarms of rigid structures, that they would physically live on; they maintain a habitable temperature of a few hundred K, the same as the home planet of the builders \citep{Dyson60}.  But not all stars have the same luminosity, and therein lies the problem with extrapolating one of these Type II societies to an entire galaxy.  The surface area of a Dyson sphere must scale directly with the luminosity of the star in order to maintain a habitable temperature.  A 1 AU wide Dyson sphere around Vega (luminosity of $40\ \Lsun$), for example, would have a surface temperature of $\ga 700\ \Kelv$, comparable to Venus' surface temperature.

The problem with Chilly Ways is that the majority of starlight in galaxies comes from bright blue stars or red giants.\footnote{Using the local bolometric luminosity function of stars from \citet{Gilmore83} as a rough guide, about 97\% of the luminosity comes from stars brighter than $1\ \Lsun$, $\sim$85\% from stars brighter than $10\ \Lsun$, 75\% from stars brighter than $50\ \Lsun$, half from stars brighter than $800\ \Lsun$, and 15\% from stars brighter than $30000\ \Lsun$.  According to the V-band luminosity function from \citet{Bahcall80}, 90\% of the V-band light of the Galaxy comes from stars brighter than the Sun, three quarters from stars $3$ times brighter than the Sun, half from stars $\sim 30$ times brighter than the Sun, one quarter from stars $\sim 140$ times brighter than the Sun, and 16\% from stars $\sim 260$ times brighter than the Sun.  \GHAT~rules out Chilly Ways that convert 85\% of their starlight into MIR light in their sample and allow 15\% to escape, with weaker limits on partial Chilly Ways that capture half and one quarter of their starlight \citep{Griffith15}.}  Red giants, in particular, have typical luminosities that are several hundred times that of the Sun.  So, aspiring Chilly Way builders would need to build structures with several hundred times the mass of a Solar Dyson Sphere.  And the mass of a traditional Solar Dyson Sphere is already something not to be undertaken lightly:
\begin{equation}
\label{eqn:DysonMass}
M_{\rm Dyson} = 60\ \MEarth\ \left(\frac{L}{\Lsun}\right)\ \left(\frac{\Sigma_{\rm Dyson}}{1000~\kgm2}\right).
\end{equation}
The column density of the Dyson sphere, $\Sigma_{\rm Dyson}$, is presumed to be comparable to a building wall, since these are supposed to be habitats \citep{Dyson60}.  

Where could Chilly Way builders get the materials to build Dyson Spheres around brighter stars?  A habitable Dyson sphere around a Solar mass red giant would have a mass of up to a thousand Jupiters -- or about the entire mass of the host star itself.  In principle, aliens could mine the material in stars with a starlifting process, in which light is reflected back onto a star, causing its outer layers to heat and escape \citep{Criswell85}.  The early phases of a Chilly Way might then look like a galaxy filled with planetary nebulas, an interesting signature in its own right.  

Aside from the mass of the sphere, there's the problem of its composition.  The majority of the mass in gas giant planets and host stars is in hydrogen and helium, but habitat structures would probably be mostly made of heavier elements.  The Chilly Way builders would need to build fusion reactors, each big enough to be a small star in its own right and having a lifetime shorter than the Universe's age, otherwise the Dyson spheres couldn't be built in time for us to see them.  Like rampant starlifting, these artificial stars might themselves be more spectacular signatures of cosmic engineering than the Dyson spheres.

A Chilly Way is a far more daunting engineering project than usually considered for these reasons.  These problems all arise from the assumption that the Dyson sphere is a collection of habitats.  If the builders are simply out to collect all the energy of a star, the column density of a Dyson sphere can be much lower, since the swarm elements may be more like efficient solid-state solar panels, as discussed by \citet{Carrigan09}.  Depending on the details of the structures, they might be a lot thinner than $1000~\kgm2$, easily allowing much bigger Dyson swarms to be built.  

But then, there's no reason to assume that the surface temperature of a Dyson sphere is habitable.  In fact, a colder sphere might be preferred for thermodynamic reasons \citep[c.f.,][]{Bradbury00}.  Since the solar panels might be a lot thinner than $1000~\kgm2$, there could easily be enough material to build a very large Dyson sphere.  We therefore have reason to consider looking for structures that are far more diffuse and far colder than the habitable Dyson spheres of Chilly Ways.

\subsection{A blackbox as a \emph{genius loci}}
\label{sec:GeniusLoci}
Once we go beyond the idea that Type III societies will simply maximize habitable surface area, then there are many more things that can be done with a galaxy.  

A galaxy is not empty space sprinkled with some stars.  A Type III society may wish to engineer the interstellar medium (ISM) between the stars.  This would grant them the power to adjust the rate of star formation and to harvest the metals produced by dying stars.  They may turn on and off the supermassive black hole at the center of their galaxy, to power more radical projects.  Or, they may wish to blast the gases out into intergalactic space.

Dust grains are one way to act on the ISM.  Since dust grains contain heavy elements, they could be engineered to house microscopic computers, sensors, and broadcasting equipment.  This kind of dust is called \emph{smart dust}.  Smart dust is the ultimate endpoint of the current trend towards miniaturizing electronic devices \citep[e.g.,][]{Kahn99,Warneke01}.  It has been proposed that satellites and space probes can be replaced by swarms of spaceborne smart dust \citep[e.g.,][]{Teasdale01,Barnhart09,Atchison10}.  But a Type III society may take this further; rather than just sending a few discrete swarms to individual targets, they could fill an entire galaxy with smart dust.  When the smart dust is numerous enough, the galaxy becomes opaque to its own thermal emission and it becomes a kind of blackbox.

While interstellar smart dust cannot house biological beings, there are many things that it can do.  The smart dust can be used to form a galactic-scale information network.  A Type III society needs some way of communicating with itself across interstellar distances; the interstellar smart dust could function in this capacity much like our own Internet does for us.  Any artificial intelligences that the society may develop might dwell in the smart dust, since they are not limited by the cold temperatures of interstellar space.  A blackbox, then, is a possible end point of the machine intelligence societies described in \citet{Scheffer94}, even those that did not have cosmic engineering as a goal per se.  In addition, non-intelligent computation and automation could be distributed among the vast number of smart dust grains, as a kind of literal cloud computing.  Furthermore, the dust can mechanically couple with the gases in the ISM.  If the dust grains can propel themselves, perhaps using radiative thrusters or by deflecting themselves on galactic magnetic fields, they can shape the ISM.  Finally, dust grains are relatively efficient at capturing the starlight of a galaxy.  I provide quantitative estimates of these capabilities in the Appendix.

Once the opacity of the dust becomes high enough, it is possible to create a galactic-scale greenhouse effect using artificial dust.  Natural dust in starbursting regions can raise the temperature to $\sim 100\ \Kelv$ on scales of a few hundred parsecs \citep[e.g.,][]{Thompson05,Sakamoto08}.  In the most extreme natural star clusters, the temperatures might reach habitable levels \citep{Thompson13}.  Artificial dust, in principle, could have much higher opacities than natural dust.  Then, artificial dust could be used to establish literal galactic habitable zones extending over kiloparsecs.  For comparison, the total habitable volume of a Chilly Way with luminosity $L$ is $(L/\Lsun) \times \AU^3 \approx 10^{-5}\ (L / 10^{11}\ \Lsun) \pc^{-3}$.  Actually achieving this on 10 kpc scales requires an optical depth of order $\sim (300\ \Kelv/3\ \Kelv)^4 \approx 10^8$, which is probably too high to actually achieve.  In addition, convection could set in and result in faster heat transport out of the blackbox.  But smaller habitable zones are possible, as are large but colder zones that might be comfortable for something native to a world like Titan \citep[as in ][]{McKay05} or the Kuiper Belt \citep[as in ][]{Dyson03}.  The ultimate result could be to create an actual ecosystem where quasi-biological entities subsist off the ISM and starlight.  The dust grains themselves could be self-replicating, and there might be larger lifeforms that feed on them and reprocess them, analogous to ecological and geological cycles.  Different optical depths would have different ``climates'', hosting different populations of beings.\footnote{Once established, the blackbox may be self-perpetuating even if the original builders retreat or go extinct, so there may be no intelligence \emph{per se} in the blackbox.  On the other hand, it is imaginable, if not likely, that the builders have access to computational powers that are qualitatively more powerful than any kind of intelligence, like an NP-oracle.  Between these possibilities, I am reluctant to use the term \emph{intelligence} for what this paper is looking for.  Hence my use of the generic ``aliens'' instead of the usual ``extraterrestrial intelligence'' of the literature.}

A truly \emph{galactic} entity exploits the spaces and matter between the stars.  The inhabitants of a Chilly Way are restricted to AU-scale bubbles surrounding stars, leaving most of the galaxy's volume untouched.  But by inhabiting the ISM itself, as might be done with smart dust, the Type III society is truly pervasive -- less a collection of individual worlds and more a \emph{genius loci} that reaches everywhere in the galaxy.  

\emph{The case for blackboxes} -- From a thermodynamic view, a Chilly Way is sub-optimal at harnessing the power of starlight.  The maximum efficiency of a heat engine working between a hot heat reservoir with temperature $T_{\rm hot}$ and a cold heat reservoir with temperature $T_{\rm cold}$ is
\begin{equation}
\eta_{\rm Carnot} = 1 - \frac{T_{\rm cold}}{T_{\rm hot}}.
\end{equation}
On a galactic scale and with conventional physics, the temperature of the cold reservoir is at least the temperature of the CMB, $T_{\rm CMB} (z) = 2.725\ (1 + z)\ \Kelv$ at redshift $z$ \citep{Fixsen09}.  In one sense, a Dyson sphere can be very efficient, of course, since it harnesses optical light emitted by the surfaces of stars with temperatures of $T_{\rm hot} \approx 5000\ \Kelv$ and thermalizes it to a temperature $T_{\rm cold} \approx 300\ \Kelv$, releasing it as waste heat.  But there's no reason why a second Dyson swarm couldn't surround the first, using the inner swarm as its hot reservoir and stepping it down to a colder temperature still (perhaps $100\ \Kelv$).  The ``waste heat'' of the inner swarm does not have to be waste at all, but can still be put to useful work.  The second swarm can then be surrounded by a third swarm, and so on, until the outermost swarm has a temperature approaching that of the CMB.  The idea of Dyson swarms within swarms, called a Matrioshka brain, was proposed as a way for maximizing the computation that a Type II society could perform \citep{Bradbury00}.

Landauer's principle provides a fundamental reason why a society seeking to maximize computational power would tend to build a galaxy-wide blackbox instead of a Chilly Way.  According to Landauer's principle, each erasure of a bit requires the use of $(\ln 2) k_B T$ energy \citep{Landauer61}.\footnote{The actual constraints from Landauer's Principle are a bit subtle, since any computation can be instantiated in a fully reversible manner that does not cost work \citep{Bennett73}.  However, the energy cost does apply if the computer transfers the entropy into the outside world, as would happen when correcting errors \citep[see the discussion in][]{Lloyd00}.}  Colder temperatures allow for more energy-efficient computation.  But put another way, Landauer's principle states that each bit erasure produces one unit of entropy.  Aside from black hole formation, the greatest source of entropy production in the present Universe is the absorption and re-emission of starlight by dust in the ISM \citep{Bousso07,Egan10}.  And indeed, this is because ISM dust is both effective at capturing starlight and very cold, typically just around 20 K.  A society that attempts to maximize entropy production -- whether to convert more of the starlight into usable work, or to maximize computation -- will probably use something like dust to harness the starlight, if they are using conventional physics at all.  Along these lines, \citet{Cirkovic06} proposed that Type II--III societies would build structures in interstellar space far from their home galaxy's core to take advantage of the frigid conditions, and even suggested they might use something like smart dust, although they didn't quite extend the idea to capturing all of a galaxy's light.

\subsection{Blackboxes as dipole clouds}
\label{sec:DipoleClouds}
\emph{The uses of antennas} -- But is it really possible to render a galaxy totally opaque to its own thermal radiation, on kiloparsec scales?  The typical Milky Way-like spiral galaxy absorbs about half of its ultraviolet radiation \citep{Xu95,Bell03}, but dust grains are much less effective at absorbing light at longer wavelengths.  The most intense starburst regions of star-forming galaxies do appear to be opaque to their own thermal radiation \citep[e.g.,][]{Sakamoto08}.  Yet these regions are generally only about a few hundred parsecs across.  The large optical depths arise from the large column densities of gas; hiding an entire Milky Way-sized galaxy requires either too much gas, or too high a metal abundance.  Even at high redshift, where actively star-forming galaxies can contain large amounts of gas, the biggest starbursts are about a kiloparsec across \citep[e.g.,][]{Walter09}.  Natural blackboxes are therefore unlikely to be more than a kiloparsec across, too small to hide a fully engineered galaxy.

The problem with natural dust is that it is three-dimensional.  A lot of the mass is hidden inside the center of a dust grain, where it does not absorb infrared light.  Small dust grains are very poor at coupling with far infrared light, but large grains about one wavelength across require disproportionately more mass because of the cube law.

Wire antennas have a far more effective geometry for absorbing and emitting long wavelength light.  The simplest antenna geometry is the dipole antenna floating in free space.  It is a straight wire of length $\ell$, typically cut into two segments that are wired into a feed point that is connected to whatever devices are powered by the antenna.  The system can be modeled as an RLC circuit.  The antenna itself has an associated impedance $\ZImp_A (\nu)$ at each frequency $\nu$; the real part is called the radiation resistance ($\RRad$), and the imaginary part is the reactance $\XReact_A$ \citep{Best09}.

A half-wavelength dipole is a special case that interacts well with light of wavelength $\lambda \approx 2 \ell$.  At this frequency, the antenna reactance is zero, and light can effectively be absorbed or scattered by the dipole.  For this reason, this frequency is said to be the resonant frequency of the antenna.  When the antenna has no thickness, the resonance is exactly at $\lambda = 2 \ell$, at which $\RRad = \RDipole \equiv 73.2\ \Ohm$.  The resonance occurs at slightly greater wavelengths for thick antennas \citep{Laport52,Carr12}.  Away from resonance, the reactance diverges from 0 at a rate $d\XReact_A/d\nu > 0$ that depends on the thickness of the antenna.  Thick antennas have a relatively small $d\XReact_A/d\nu$ and therefore have a big bandwidth, while thin antennas are narrowband instead.  However, the bandwidth depends only weakly on the length-to-width ratio of the dipole once the ratio is $\gg 1$ \citep{Laport52}, so a collection of narrow antennas may be more effective than a thick antenna in terms of spectrum coverage per material used.  There are also ways to emulate a thick antenna with relatively little material, by using a hollow wire or connecting a few parallel thin wires, for example \citep[e.g.,][]{Carr12}.

We are interested in how much power the antenna can deliver into the rest of the circuit, even if it is not in resonance.  For simple circuits, the rest of the circuit can be characterized by an impedance of $\ZImp_L$ which can be regarded as in series with $\ZImp_A$ for dipole antennas near resonance (\citealt{Best09} discusses some of the caveats for circuit models of antennas).  $\ZImp_L$ includes the effects of both dissipative Ohmic losses in the antenna and the circuitry attached to the feed point.  Then the efficiency of power delivery to the rest of the circuit,
\begin{equation}
\eta_L (\nu) = 1 - |\Gamma_L (\nu)|^2,
\end{equation}
follows from a calculation of the reflection constant
\begin{equation}
\Gamma_L (\nu) = \frac{\ZImp_A - \ZImp_L}{\ZImp_A + \ZImp_L}.
\end{equation}
The efficiency reaches 1 when the impedances are matched ($\ZImp_A = \ZImp_L$; \citealt{Best09,Carr12}).  Note that the fraction of power delivered to a circuit element attached to the feed point is less than $\eta_L$ because of Ohmic losses in the antenna conductor; another reflection constant must be considered to calculate this fraction \citep{Best09}.  However, I am mainly interested in calculating whether the antenna is an effective absorber, so I include Ohmic losses because they convert the incident light into heat.  

As an example, consider what would happen if the rest of the circuit had a purely resistive nature, with a purely real $\ZImp_L$ equal to $\RResist_L$:
\begin{equation}
\label{eqn:etaLResistiveLoad}
\eta_L (\nu) = \frac{4 \RResist_L \RRad}{(\RRad + \RResist_L)^2 + \XReact_A^2}
\end{equation}
At resonance ($\XReact_A = 0$), $\eta_L$ starts off small for low resistance systems ($4 \RResist_L/\RRad$), peaks at 1 when $\RResist_L = \RRad$, and then falls off for high resistance systems ($4 \RRad/\RResist_L$) \citep[c.f.,][]{Best09}.  According to equation~\ref{eqn:etaLResistiveLoad}, the reactance only matters when $\XReact_A^2 \ga (\RRad + \RResist_L)^2$.  Antennas attached to a highly resistive circuit ($\RResist_L \ga \XReact_A, \RRad$) are actually wideband, in that they maintain a constant $\eta_L \approx 4 \RRad / \RResist_L$ over a broad span of frequencies.  This suggests one strategy at implementing an economical blackbox: very thin wires have high Ohmic resistance, but they use very little mass so that a lot of them can be built, each effective at absorbing light of a variety of frequencies \citep[c.f.,][]{Wright82}.  

The great advantage of using antennas is clear when considering their effective aperture, which is their absorption cross section.  When averaged over all angles and all polarizations, as appropriate for an antenna floating in a thermal radiation background, it is \citep{Condon10,Heald12,Zavrel16}
\begin{equation}
\mean{A_{\rm abs} (\lambda)} = f_{\rm pol} \eta_L (\lambda) \frac{\lambda^2}{4 \pi}.
\end{equation}
The $f_{\rm pol} = 1/2$ factor accounts for the fact that a dipole antenna has a single polarization, and radiation interacts with it only if its polarization is the same.  A dipole antenna can easily be as effective at absorbing light as a standard three dimensional dust grain, using just a thousandth of the mass.  This makes them well-suited for building blackboxes, where the limiting factor is the amount of materials available (equation~\ref{eqn:DysonMass}).\footnote{From a historical perspective, this effectiveness was exploited by quasi-steady state cosmologies, which explained the CMB as starlight thermalized by intergalactic conducting ``needles'' \citep{Wickramasinghe75,Hoyle94,Li03}.  These needles needed to fill the entire Universe, most of which is relatively empty, so the amounts required approached the metal budget.  A similar mechanism generates the CMB in a cold Big Bang scenario \citep{Wright82,Aguirre00}.  In later sections, I point out that absorbers in intergalactic space should at least absorb the EBL from the current stellar population and heat up several mK above the CMB.  Significant opacities during the epoch of EBL generation would distort the CMB spectrum.  Furthermore, the distortions would be distributed anisotropically on the sky if the absorption is inhomogeneous, which would be expected if they formed in or around galaxies that are the sources of the Universe's metals.  Whether any needle-like or antenna-like dust ever forms in the Universe is unknown \citep{Dwek04,Gomez05}, but whatever the case, there's no reason they could not be artificially constructed for the express purpose of absorbing light.}

Antennas can also scatter or reflect light.  The effective cross section of the antenna, which can be written as
\begin{equation}
\mean{A_{\rm s} (\lambda)} = f_{\rm pol} \eta_s (\lambda) \frac{\lambda^2}{\pi},
\end{equation}
depends on a scattering efficiency $\eta_s (\lambda)$.  For a resonant dipole antenna, the scattering opacity is largest when dissipative losses in the antenna (represented by $\RResist_L$) are small.  When the load and radiation impedances are matched in the antenna circuit, $A_s \approx A_{\rm abs}$, and in resistive antennas, $\eta_s$ drops quickly as $\RResist_L$ increases and reflection is negligible \citep{Best09,Zavrel16}.

\emph{Estimating the number of motes in a blackbox} -- The problem of building a blackbox amounts to tiling its entire surface area $A_{\rm box}$ with a number of absorbers, each of cross section $A_{\rm abs}$.  These two quantities determine the amount of material needed to construct the blackbox.  

In what follows, I imagine the blackbox to be made of smart dust elements that are called \emph{motes}.  Each mote has a dipole antenna of length $\ell$.  These lengths vary, but the total population of motes has a length distribution function of $dN_{\rm motes}/d\ell$ that is nonzero for $\ell_{\rm min} \le \ell \le \ell_{\rm max}$.  I shall also assume the antennas are half-wavelength dipoles with a narrowband resonance of logarithmic bandwidth $\beta_a$:
\begin{equation}
\label{eqn:etaAbs}
\eta_{\rm abs} (\lambda, \ell) = \left\{ \begin{array}{ll} 
                                 \overline{\eta_{\rm abs}} & (\lambda_a e^{-\beta_a/2} \le \lambda \le \lambda_a e^{\beta_a/2}) \\
																 0                    & (\lambda < \lambda_a e^{-\beta_a/2}~{\rm or}~\lambda > \lambda_a e^{\beta_a/2})
																 \end{array} \right.
\end{equation}
The central wavelength $\lambda_a$ of the resonance is related to the length as $\lambda_a = 2 f_{\ell} \ell$.  

In order for a blackbox to be black, its absorption optical depth $\tau_{\rm abs}$ must be at least $1/(1 + \tau_s$) for all frequencies, where $\tau_s$ is the optical depth to scattering \citep{Rybicki79}.  I take this quantity,
\begin{align}
\nonumber \tau_{\rm abs} (\lambda) & = \int_{\ell_{\rm min}}^{\ell_{\rm max}} \frac{dN_{\rm motes}}{d\ell} \frac{\mean{A_{\rm eff}}}{A_{\rm box}} d\ell \\
                                   & = \frac{f_{\rm pol} \overline{\eta_{\rm abs}} \lambda^2}{4 \pi A_{\rm box}} \int_{\lambda e^{-\beta_a/2}/(2 f_{\ell})}^{\lambda e^{\beta_a/2}/(2 f_{\ell})} \frac{dN_{\rm motes}}{d\ell} d\ell.
\end{align}
to be a constant over a wavelength range $\lambda_{\rm min} \le \lambda \le \lambda_{\rm max}$.  The wavelength range must include most of the thermal emission from the blackbox surface.  

As long as $\beta_a \la 1$, I can estimate the mote length distribution function as
\begin{equation}
\label{eqn:dNdlnl}
\frac{dN_{\rm motes}}{d\ln \ell} \Big|_{\ell = \lambda/(2 f_{\ell})} = \frac{4 \pi \tau_{\rm abs} A_{\rm box}}{f_{\rm pol} \overline{\eta_{\rm abs}} \beta_a \lambda^2}.
\end{equation}
From this, it is an easy step to integrate up the total number of motes in the blackbox:
\begin{align}
\label{eqn:NMotes}
N_{\rm motes} & = \int_{\ln\ \ell_{\rm min}}^{\ln\ \ell_{\rm max}} \frac{dN_{\rm motes}}{d\ln \ell} d\ln \ell = \frac{2 \pi \tau_{\rm abs} A_{\rm box}}{f_{\rm pol} \overline{\eta_{\rm abs}} \beta_a} \left[\frac{1}{\lambda_{\rm min}^2} - \frac{1}{\lambda_{\rm max}^2}\right].
\end{align}

The total mass of the dipole cloud $M_{\rm box}$ determines whether the blackbox is feasible with the materials in a galaxy, and it depends on the mass of each individual dipole.  To estimate it, I imagine each dipole antenna to have a density $\rho$ and a cross-sectional area $A_{\rm dipole} (\ell)$.  Note that the mote cross-section may be hollow or disconnected due to skin effects, as I discuss in Appendix~\ref{sec:SkinEffects}.  Since I am setting a lower limit on the cloud mass, I also assume that any other parts of the mote have negligible mass.  Then the mass of each antenna is 
\begin{equation}
m_{\rm dipole} (\ell) = \rho \ell A_{\rm dipole}(\ell),
\end{equation}
for a total dipole cloud mass of 
\begin{equation}
\label{eqn:DipoleCloudMass}
M_{\rm box} = \frac{2 \pi \tau_{\rm abs} \rho A_{\rm box}}{f_{\ell} f_{\rm pol} \overline{\eta_{\rm abs}} \beta_a} \int_{\lambda_{\rm min}}^{\lambda_{\rm max}} \frac{A_{\rm dipole} (\lambda/(2 f_{\ell}))}{\lambda^2} d\lambda.
\end{equation}

Each dipole has a minimum cross-sectional area set by three constraints, if it is to be an effective antenna.  First, there are Ohmic losses in the antenna conductor caused by non-zero resistivity $\sigma^{-1}$.  These losses introduce a resistance of $\RResist_{\Omega} = \sigma^{-1} \ell / A_{\rm dipole}$.  If $\overline{\eta_{\rm abs}} \approx 1$, then $\RResist_{\Omega} \la \RRad$.  This condition sets a minimum cross section $A_{\rm dipole}^{\Omega} \ga \sigma^{-1} \ell / \RRad$.  If $A_{\rm dipole}$ is smaller than this value, $\overline{\eta_{\rm abs}}$ falls proportionately, and $M_{\rm box}$ remains approximately the same.  As long as the conductors are wide enough, the resistivity approaches its bulk value of $\sigma_{\rm bulk}^{-1}$.  This sets a dipole cloud mass of
\begin{equation}
\label{eqn:MinMassConductivity}
M_{\rm box}^{\rm \Omega-bulk} \ga \frac{\pi \tau_{\rm abs} A_{\rm box} \rho \sigma_{\rm bulk}^{-1} {\cal B}}{f_{\ell}^2 f_{\rm pol} \overline{\eta_{\rm abs}} \RRad},
\end{equation}
where ${\cal B} = \ln (\lambda_{\rm max} / \lambda_{\rm min}) / \beta_a$.  Note that this is actually wavelength independent.  This allows for relatively low mass blackboxes that trap long-wavelength thermal infrared and even radio emission.  The minimum mass can also be written as a minimum column density, independent of the surface area of the blackbox:
\begin{equation}
\label{eqn:SigmaOhmic}
\Sigma_{\rm \Omega-bulk} = \frac{\pi \tau_{\rm abs} \rho \sigma_{\rm bulk}^{-1} {\cal B}}{f_{\ell}^2 f_{\rm pol} \overline{\eta_{\rm abs}} \RRad}.
\end{equation}

Second, since the dipoles are made of atoms, the antennas must be at least a few atoms across.  This sets a minimum cross section $A_{\rm atomic}$ which is at least about a square nanometer.  Then the dipole cloud mass must be
\begin{equation}
\label{eqn:MinMassAtoms}
M_{\rm box}^{\rm atomic} \ga \frac{2 \pi \tau_{\rm abs} A_{\rm box} A_{\rm atomic} \rho}{f_{\ell} f_{\rm pol} \overline{\eta_{\rm abs}} \beta_a \lambda_{\rm min}}.
\end{equation}
The minimum cross-section limit becomes negligible for long wavelength antennas, in which case the Ohmic losses are the determining factor.  This again supports the idea that long-wavelength blackboxes are feasible.  Once again, there is a characteristic column density required by a blackbox:
\begin{equation}
\label{eqn:SigmaAtomic}
\Sigma_{\rm atomic} = \frac{2 \pi \tau_{\rm abs} A_{\rm atomic} \rho}{f_{\ell} f_{\rm pol} \overline{\eta_{\rm abs}} \beta_a \lambda_{\rm min}}.
\end{equation}

A third constraint arises from the physics of conductivity, which is limited by the rate at which electrons collide with defects or other structures in metals.  According to Drude's electron model of conducting metals,
\begin{equation}
\label{eqn:sigmaDrude}
\sigma_{\rm Drude}(\nu) = \frac{n_e q_e^2 \tau}{m_e (1 - 2 \pi i \nu t_e)} \approx \frac{n_e q_e^2 t_e}{m_e}
\end{equation}
at low frequencies ($\nu \la 1/(2 \pi t_e)$) in a metal with free electron density $n_e$ and a characteristic electron collision time $t_e$, related to the electron mean free path \citep{Jackson99,Heald12}.  Collisions on the wall of a wire limit the mean free path to at most $2 r$, for a cylindrical wire with radius $r$.  When the wire is small enough, it becomes the limiting factor for the electrons' mean free path, and the conductivity drops proportionally with radius \citep[e.g.,][]{Li03}.  In this limit, $t_e = 2 r / v_F$, where $v_F$ is the Fermi energy of the conducting electrons in the metal \citep{Heald12}.  The increasing resistivity of wires with radii $\la 100\ \nm$ was confirmed by \citet{Bid06}.

The corrected value of the resistivity can be calculated using $n_e$ in a metal with atomic mass $m_{\rm atom}$ and $N_{\rm free}$ free electrons per atom, $n_e = \rho N_{\rm free} / m_{\rm atom}$.  I found $N_{\rm free}$ from the number of electrons in the outermost shell, using the electron configurations given in \citet{Olive15}.\footnote{The $n_e$ values I calculated are close to those listed at http://hyperphysics.phy-astr.gsu.edu/hbase/tables/fermi.html at the HyperPhysics web site, which concisely describes how to calculate $n_e$.}  Cold metals have degenerate electrons that fill up energy levels to the Fermi energy (\citealt{Pauling70}, equation 17-2; \citealt{Harrison89}, equation 15-4),
\begin{equation}
E_F = \frac{h^2}{8 \pi^2 m_e} (3 \pi^2 n_e)^{2/3} = \frac{h^2}{8 \pi^2 m_e} \left(\frac{3 \pi^2 \rho N_{\rm free}}{m_{\rm atom}}\right)^{2/3},
\end{equation}
which sets the Fermi velocity $v_F = \sqrt{2 E_F / m_e}$.  Combining these relations gives the maximum conductivity of this wire:
\begin{equation}
\label{eqn:sigmaDrudeThin}
\sigma_{\rm Drude}^{\rm narrow} = \frac{4 q_e^2 r}{3 \pi h} \left(\frac{3 \pi^2 \rho N_{\rm free}}{m_{\rm atom}}\right)^{2/3}.
\end{equation}
Then, in order to avoid Ohmic losses in the antenna, the wire must have a minimum radius
\begin{align}
\nonumber r_{\rm Drude} & = \left(\frac{3 h \ell}{4 \RRad q_e^2}\right)^{1/3} \left(\frac{m_{\rm atom}}{3 \pi^2 \rho N_{\rm free}}\right)^{2/9}\\
\nonumber               & = 104\ \nm\ N_{\rm free}^{-2/9} \left(\frac{\ell}{\mm}\right)^{1/3} \left(\frac{\RRad}{\RDipole}\right)^{-1/3} \left(\frac{m_{\rm atom}}{50\ \amu}\right)^{2/9}\\
                        & \times  \left(\frac{\rho}{10\ \rhoUnits}\right)^{-2/9}.
\end{align}
Since the Fermi velocity is usually $\sim 10^6\ \meter\ \sec^{-1}$, the low frequency approximation to the conductivity is valid up to THz frequencies.  Above $\nu_{\sigma} = 1/(2 \pi t_e)$, the conductivity turns imaginary.  For thin antennas (equation~\ref{eqn:sigmaDrudeThin}),
\begin{equation}
\nu_{\sigma} = \frac{1}{8 \pi^2 m_e} \left(\frac{4 \RRad q_e^2 h^2}{3 \ell}\right)^{1/3} \left(\frac{3 \pi^2 \rho N_{\rm free}}{m_{\rm atom}}\right)^{5/9}.
\end{equation}
When this happens, light waves attempting to propagate into the metal become evanescent and are attenuated on length scales of roughly $\omega_p / c$, where $\omega_p = \sqrt{n_e q_e^2 / (\epsilon_0 m_e)}$ is the plasma frequency of the metal \citep{Jackson99,Heald12}.  As long as the metal is thicker than this, the light is reflected.  For the metals I considered, the attenuation length is about 10 nm.  Thus, the dipoles could still contribute to the optical depth by scattering radiation, but the effective cross section of the antenna is the geometrical cross section, which can be thousands of times smaller than $\lambda^2/(4\pi)$ for thin antennas.  

When the temperature is just a few K, it is this condition that sets the minimum mass for blackboxes, since the wire thicknesses must be of $\sim 100\ \nm$ instead of the few nm predicted using the large-scale conductivity.  Using a dipole wire cross section of $A_{\rm dipole} = \pi r_{\rm Drude}^2$, the minimum mass for a blackbox is
\begin{multline}
M_{\rm box}^{\rm Drude} = \frac{3 \pi^2 \tau_{\rm abs} A_{\rm box} \rho}{8 f_{\ell} f_{\rm pol} \overline{\eta_{\rm abs}} \beta_a} \left(\frac{24 h}{f_{\ell} \RRad q_e^2}\right)^{2/3} \left(\frac{3 \pi^2 \rho N_{\rm free}}{m_{\rm atom}}\right)^{-4/9} \\
\times \left(\frac{1}{\lambda_{\rm min}^{1/3}} - \frac{1}{\lambda_{\rm max}^{1/3}}\right),
\end{multline}
which slowly decreases as $\lambda_{\rm min}$ increases.  The minimum blackbox column density is, of course,
\begin{multline}
\Sigma_{\rm Drude} = \frac{3 \pi^2 \tau_{\rm abs} \rho}{8 f_{\ell} f_{\rm pol} \overline{\eta_{\rm abs}} \beta_a} \left(\frac{24 h}{f_{\ell} \RRad q_e^2}\right)^{2/3} \left(\frac{3 \pi^2 \rho N_{\rm free}}{m_{\rm atom}}\right)^{-4/9} \\
\times \left(\frac{1}{\lambda_{\rm min}^{1/3}} - \frac{1}{\lambda_{\rm max}^{1/3}}\right).
\end{multline}
In order for the blackbox to be opaque, the column density of its component dipole antennas must be at least $\Sigma_{\rm blackbox} = \max(\Sigma_{\rm \Omega-bulk}, \Sigma_{\rm atomic}, \Sigma_{\rm Drude}$).  

Numerical evaluation of the minimum column densities depends on factors that require detailed engineering work, or failing that, subjective judgments on the details of the antennas.  For starters, a minimal blackbox is just massive enough to be opaque, so I take $\tau_{\rm abs} = 1$, ignoring the possible contribution of scattering.  Dipole antennas only absorb one polarization of light effectively, so $f_{\rm pol} = 1/2$.  The bandwidth of the antenna depends on the resistance of the antenna and its geometry.  I assume that the antennas are resonant half-wavelength dipoles with $f_{\ell} = 1$, but that they are fashioned to be fairly broadband.  This might be done by emulating a thick antenna, with a cage dipole for example \citep{Carr12}, so I nominally choose $\beta_a = 1/2$.  I also assume that the thinnest possible antenna has a disk cross section with radius of 1 nanometer: $A_{\rm atomic} = \pi (\nm)^2$.  Finally, I choose a wavelength range based on a presumed 3 K temperature: the assumed range $\lambda_{\rm min} = 0.5\ \mm$ to $\lambda_{\rm max} = 5\ \cm$ (6 to 600 GHz) covers about 98.6\% of the blackbody radiation.  

\subsubsection{Metal wire dipoles}
\label{sec:MetalDipoles}
The dipole antennas must be made of conductors in order to be efficient, and solid metals are an obvious choice.  I list the properties of some relatively abundant metallic elements in Table~\ref{table:ColdMetalResistivities}.  The cold temperatures ($\sim 3\ \Kelv$) of galaxy-sized blackboxes lower the (bulk) resistivities of these metals and $\Sigma_{\rm \Omega-bulk}$.  For pure metals, the dependence can steepen to $\sigma_{\rm bulk}^{-1} \propto T^5$ once the temperature is about an order of magnitude below its Debye temperatures \citep{Chi79-Alkali,Chi79-Alkaline}.  Most metals have impurities, though, so measurements of the intrinsic resistivity becomes imprecise with actual metals in the lab \citep{Chi79-Alkali,Chi79-Alkaline,Matula79,Desai84}.  Instead of trying to extrapolate values down to 3 K, I use the values in the references at the lowest provided temperature.

The maximum frequency before the conductivity turns imaginary is also listed in the table, derived using the Drude conductivity formula.  The bulk resistivities listed in the Table no longer apply above $\nu_{\rm max}^{\rm bulk}$, with typical values of tens of GHz.  This is right in the middle of the thermal emission range, so thick antenna conductors made of pure metals would fail.  I find that the antenna conductors  must usually be thin enough that the actual conductivities are smaller.  Therefore, the conductivities should be valid up to $\nu_{\rm max}^{\rm Drude}$, around 1 THz, near the edge of my assumed wavelength range.

\begin{deluxetable*}{lcccccccccccc}
\tablecaption{Properties of cold metallic conductors}
\tablehead{\colhead{Material} & \colhead{$m_A$} & \colhead{$N_{\rm free}$} & \colhead{$\rho$} & \colhead{$T_{\rm ref}$} & \colhead{$\sigma_{\rm bulk}^{-1} (T_{\rm ref})$} & \colhead{$\nu_{\sigma}^{\rm bulk}$} & \colhead{$\nu_{\sigma}^{\rm Drude}$} & \colhead{$\Sigma_{\rm \Omega-bulk}$} & \colhead{$\Sigma_{\rm atomic}$} & \colhead{$\Sigma_{\rm Drude}$} & \colhead{$\Sigma_{\rm blackbox}$} & \colhead{Notes} \\ & \colhead{$(\amu)$} & & \colhead{$(\rhoUnits)$} & \colhead{$(\Kelv)$} & \colhead{$(\Ohm\ \meter)$} & \colhead{$(\GHz)$} & \colhead{$(\THz)$} & \colhead{$(\kgm2)$} & \colhead{$(\kgm2)$} & \colhead{$(\kgm2)$} & \colhead{$(\Msun\ \kpc^{-2})$}}
\startdata
Lithium    & 6.9  & 1 & 0.53 & 20 & $5 \times 10^{-11}$   & 10  & 1.3 & $2 \times 10^{-8}$   & $8 \times 10^{-11}$   & $1.3 \times 10^{-6}$ & 600 & (a)\\
Beryllium  & 9.0  & 2 & 1.9  & 40 & $4 \times 10^{-11}$   & 50  & 3   & $6 \times 10^{-8}$   & $3 \times 10^{-10}$   & $2 \times 10^{-6}$ & 1100 & (b)\\
Sodium     & 23.0 & 1 & 0.97 & 20 & $1.5 \times 10^{-10}$ & 17  & 0.9 & $1.2 \times 10^{-7}$ & $1.5 \times 10^{-10}$ & $3 \times 10^{-6}$ & 1400 & (a)\\
Magnesium  & 24.3 & 2 & 1.7  & 30 & $2 \times 10^{-10}$   & 80  & 1.8 & $3 \times 10^{-7}$   & $3 \times 10^{-10}$   & $3 \times 10^{-6}$ & 1500 & (b)\\
Aluminum   & 27.0 & 3 & 2.7  & 20 & $7 \times 10^{-12}$   & 6   & 3   & $1.5 \times 10^{-8}$ & $4 \times 10^{-10}$   & $4 \times 10^{-6}$ & 1700 & (c)\\
Potassium  & 39.1 & 1 & 0.86 & 20 & $1.2 \times 10^{-8}$  & 700 & 0.6 & $8 \times 10^{-6}$   & $1.3 \times 10^{-10}$ & $4 \times 10^{-6}$ & 4000 & (a)\\
Calcium    & 40.1 & 2 & 1.6  & 40 & $1.3 \times 10^{-9}$  & 300 & 1.3 & $1.7 \times 10^{-6}$ & $3 \times 10^{-10}$   & $4 \times 10^{-6}$ & 1900 & (b)\\
Iron       & 55.8 & 2 & 7.9  & 20 & $1.0 \times 10^{-10}$ & 80  & 3   & $6 \times 10^{-7}$   & $1.2 \times 10^{-9}$  & $1.1 \times 10^{-5}$ & 5000 & (d)\\
Nickel     & 58.7 & 2 & 8.9  & 20 & $9.5 \times 10^{-11}$ & 80  & 3   & $7 \times 10^{-7}$   & $1.4 \times 10^{-9}$  & $1.2 \times 10^{-5}$ & 6000 & (d)\\
Copper     & 63.5 & 1 & 8.9  & 20 & $8 \times 10^{-12}$   & 3   & 3   & $6 \times 10^{-8}$   & $1.4 \times 10^{-9}$  & $1.7 \times 10^{-5}$ & 8000 & (e)
\enddata  
\tablecomments{The density values apply at 300 K, and are slightly higher near absolute zero.  The bulk, or large scale, resistivity $\sigma_{\rm bulk}^{-1}$ for each element applies at the temperature $T_{\rm ref}$.  At lower temperatures, the bulk resistivity decreases as quickly as $T^5$ in very pure samples \citep{Chi79-Alkali,Chi79-Alkaline,Desai84}, but $\Sigma_{\rm \Omega-bulk}$ is never the limiting factor at millimeter wavelengths and $\sigma_{\rm bulk}$ becomes complex at MHz frequencies (equation~\ref{eqn:sigmaDrude}).  The listed $\sigma_{\rm bulk}$ are valid in thick wires for frequencies smaller than $\nu_{\rm max}^{\rm bulk}$, but $\Sigma_{\rm Drude}$ should be valid up to $\nu_{\rm max}^{\rm Drude}$.  The values listed assume $\tau_{\rm abs} = f_{\ell} = \overline{\eta_{\rm abs}} = 1$, $f_{\rm pol} = 1/2$, $\beta_a = 1/2$, $A_{\rm atomic} = \pi (\nm)^2$, $\RRad = \RDipole$, $\lambda_{\rm min} = 0.5\ \mm$, and $\lambda_{\rm max} = 5\ \cm$.}
\tablenotetext{a}{The alkali metal properties are given in \citet{Chi79-Alkali}.}
\tablenotetext{b}{The alkaline metal properties are given in \citet{Chi79-Alkaline}.}
\tablenotetext{c}{The conductivity of aluminum is from \citet{Desai84}, while the density is from \citet{Olive15}.}
\tablenotetext{d}{The conductivities of iron and nickel are from \citet{Kemp56}, while the densities are from \citet{Olive15}.}
\tablenotetext{e}{The conductivity and density of copper is given in \citet{Matula79}.}
\label{table:ColdMetalResistivities}
\end{deluxetable*}

Different metals have distinct advantages and drawbacks for blackbox builders.  The minimum mass of a dipole cloud depends on the product of the density and resistivity, if Ohmic losses set the cross-sectional area of the antenna, or just the density, if the cross-sectional area reaches the atomic limit.  Copper is quite conductive, but it is dense and very rare.  Iron and nickel are relatively poor conductors and dense, but iron in particular is very abundant in the ISM, so it might be easier to mine the necessary amount.  The alkaline metals calcium and magnesium are also relatively poor conductors, but they are lightweight, and about as abundant as nickel.  Pure aluminum seems to be a very good material, being extremely conductive and lightweight.  I also included lithium (poor conductor and very lightweight) and beryllium (poor conductor and lightweight), but their abundances in the ISM are negligible.

With these uncertainties in mind, I find that the minimum column densities for blackboxes made of these metals are actually very small -- of order a few $\mgm2$.  In astronomical units, the column densities are about a few thousand $\Msun\ \kpc^{-2}$.  In all cases, the width of each dipole conductor is set by electrons colliding with the edges of the wires, so that $\Sigma_{\rm Drude}$ is the correct value to use.

\begin{deluxetable*}{lccccc}
\tablecaption{Minimum masses of dipole cloud blackboxes}
\tablehead{\colhead{Material} & \colhead{$Z$} & \multicolumn{4}{c}{Dipole cloud minimum masses} \\ & & \multicolumn{2}{c}{10 kpc} & \multicolumn{2}{c}{3 Mpc} \\ & & \colhead{$M_{\rm box}$} & \colhead{$M_{\rm mine}$} & \colhead{$M_{\rm box}$} & \colhead{$M_{\rm mine}$} \\ & & \colhead{$(10^6 \Msun)$} & \colhead{$(10^9 \Msun)$} & \colhead{$(10^9 \Msun)$} & \colhead{$(10^{15} \Msun)$} }
\startdata
Lithium   & $5.7 \times 10^{-11}$ & 0.8 & $1.4 \times 10^7$ & 70  & $1.2 \times 10^6$\\
Beryllium & $1.6 \times 10^{-10}$ & 1.3 & $8 \times 10^6$   & 120 & $7 \times 10^5$\\
Sodium    & $2.5 \times 10^{-5}$  & 1.9 & 70                & 170 & 7\\
Magnesium & 0.00061               & 1.9 & 3                 & 170 & 0.3\\
Aluminum  & $4.7 \times 10^{-5}$  & 2   & 50                & 200 & 4\\
Potassium & $3.5 \times 10^{-6}$  & 5   & 1400              & 400 & 130\\
Calcium   & $6.1 \times 10^{-5}$  & 2   & 40                & 200 & 3\\
Iron      & 0.0012                & 7   & 6                 & 600 & 0.5\\
Nickel    & $7.4 \times 10^{-5}$  & 7   & 97                & 600 & 9\\
Copper    & $7.6 \times 10^{-7}$  & 10  & 13000             & 900 & 1200\\
\hline
Carbon (nanotubes) & 0.0022  & 0.07   & 0.03   & 6   & 0.003
\enddata  
\tablecomments{I list the mass fraction $Z$ of each element in Solar abundance gas as derived from \citet{Asplund05}.  I assume a density of $2\ \rhoUnits$ for carbon nanotubes.}
\label{table:BlackboxMasses}
\end{deluxetable*}

I convert the column densities into minimum masses and list them in Table~\ref{table:BlackboxMasses}.  Two cases are listed -- one that is about the size of the Milky Way ($R_{\rm box} = 10\ \kpc$), and one that is about the size of a galaxy cluster ($R_{\rm box} = 3\ \Mpc$; roughly based on their virial radii in \citealt{Reiprich02}).  The mass of the blackbox itself is given as $M_{\rm box}$.  The mass of a blackbox the size of the Milky Way is around a few million $\Msun$, a small fraction of the Galactic baryon budget.  A cluster-sized blackbox needs a mass of a few hundreds of billion $\Msun$, also a small minority of a cluster's baryon budget.  Lithium has the best performance, whereas the high density of transition metals implies relatively large masses.

The catch is that most of the baryons in a galaxy or a cluster are not in these metals, but locked in hydrogen and helium.  This is the same problem that comes up when building Dyson spheres around red giants and blue stars in the Chilly Way scenario.  Maybe blackbox builders can convert stars into metal factories somehow, but it's not clear that manipulating nucleosynthesis so brazenly is even possible under the laws of physics.  A far more conservative possibility is that blackbox builders mine the ISM for metals.  They must sift through and refine
\begin{equation}
M_{\rm mine} = M_{\rm box} / Z
\end{equation}
in order to accumulate enough mass to build a blackbox, where $Z$ is the mass abundance of the metal.

These masses, too, are listed in Table~\ref{table:BlackboxMasses}, and are much less favorable.  For comparison, the Milky Way's ISM mass is about $7 \times 10^9\ \Msun$ within 20 kpc of the Galactic Center \citep{Draine11}.  The two most favorable elements are magnesium and iron.  Neither is peculiar in terms of its density or its conductivity, but they are very abundant.  There is enough magnesium in the ISM to build a blackbox around the Milky Way.  Adding iron dipoles might help, but iron and other ferromagnetic materials have pronounced skin effects (Appendix~\ref{sec:SkinEffects}), which might require more wasteful antenna geometries.

The masses are inversely proportional to the metallicity of the ISM; there may not be enough metals in a dwarf galaxy for a blackbox, for example.  High redshift star-forming galaxies have bigger gas fractions, meaning more ISM to mine, although it is unclear whether technological societies appear quickly enough to be around during this era.  There is about ten times as much mass in the Galaxy's stars as in the ISM, and over time, this mass is cycled through the ISM as stars die.  Furthermore, mass is accreted onto the Galaxy over Gyr timescales, providing more material to be mined.  Still, the feasibility of blackboxes made of metals is unclear, and may require some cleverness on the part of its builders beyond the naive estimates here.

There are similar, but worse problems for cluster blackboxes.  Galaxy clusters have baryonic masses of order a few $10^{13}\ \Msun$ in their inner regions (where the density is $\ge 500$ times the critical density), which have radii of $\sim 1\ \Mpc$ \citep{Gonzalez13}.  The amount of iron or magnesium falls short by a factor of $\sim 10$, if the properties listed in Table~\ref{table:ColdMetalResistivities} are correct.  Including all baryons within the cluster virial radius would only help by a factor $\sim 2$ \citep{Reiprich02}.  The requirements are loosened if only the inner Mpc is shrouded; then, there should be enough metals in the ICM to build the blackbox.

\subsubsection{Carbon nanotubes as dipoles}
\label{sec:CarbonNanotubes}

Carbon atoms can be arranged in thin cylinders called nanotubes with desirable properties.  In addition to having enormous tensile strength, a nanotube can be engineered to be just about one nanometer across and to have high electrical conductivity \citep{Bernholc02}.  For our purposes, the advantages of carbon are that it is lightweight (with typical densities of $\sim 2\ \rhoUnits$) and very abundant in the Universe, so that the mass of carbon required to form a blackbox dipole cloud is relatively small.

Electrons in carbon nanotubes can achieve ballistic transport, in which the electrons move unimpeded from one end of the tube to the other.  When this happens, each nanotube has a resistance $2 \RResist_{\rm quantum}$, where $\RResist_{\rm quantum} = h/(2 q_e^2) = 12\ \kOhm$ is given by the fundamental conductance quantum \citep{Bernholc02,Bandaru07}.  The resistance appears to remain roughly constant for frequencies up to at least $\sim 25\ \THz$ \citep{Bernholc02}.  

I estimate the mass by assuming that a carbon nanotube antenna consists of $N_{\rm tubes} = 2 \RResist_{\rm quantum} / \RRad$ nanotubes working in parallel.  If each nanotube has a minimum area of $A_{\rm NT}$, then the total minimum cross-sectional area of the nanotube antenna is
\begin{equation}
\label{eqn:CNTArea}
A_{\rm dipole} = 2 \left(\frac{{\cal R}_{\rm quantum}}{\RRad}\right) A_{\rm NT}.
\end{equation}
As seen in Table~\ref{table:BlackboxMasses}, idealized carbon nanotubes easily outperform standard metal wires.  There appears to be enough carbon in the Galaxy's ISM alone to build a blackbox with $\tau_{\rm abs} = 50$.  Optical depths of around ten appear to be possible in a cluster-scale blackbox using carbon harvested from ICM.  It may be possible to do even better, in fact, if individual carbon nanotubes are used as resistive dipole antennas.  These would have high efficiency per mass than resonant dipoles, and could cover about a decade in frequency.

\subsubsection{Summary}
Dipole clouds are excellent at absorbing radiation for a minimum mass.  The main requirement is finding a conductor that has low resistivity, low density, and high abundance.  A Milky Way-scale blackbox requires a few million $\Msun$ of conductors, while a galaxy cluster blackbox needs about a few tens of billion $\Msun$ of conductors.  The actual masses are small compared to the baryonic masses of galaxies, but the low abundance of heavy metals is a challenge for blackbox builders.  Magnesium and perhaps iron appear to be abundant enough to build minimal blackboxes, although the uncertainties are very large.  Alternatively, carbon nanotubes might be fashioned into antennas; if they perform ideally, there is easily enough carbon to make a blackbox.  

I have tried to be conservative in imagining how a blackbox might be built.  I consider some other engineering issues for a dipole cloud blackbox in the Appendix.  But this is not to say that actual blackbox builders \emph{will} have these limits, or use dipole clouds.  For all we know, physics unknown to us could be harnessed to make a blackbox far more efficiently out of something we cannot even imagine yet.  My point is that the laws of physics and the materials available in galaxies allow serious contemplation of blackboxes as a project.  No miraculous advances are required, as far as I can tell, only an enormous scaling up of our technology.

\section{What Blackboxes Look Like}
\subsection{Thermal emission}
Blackboxes glow with thermal radiation at the frequencies they are opaque.  Like Dyson spheres, they should therefore be detectable if they emit any waste heat from internal processes.  In addition, a blackbox might absorb extragalactic light and convert it into waste heat, even if there's nothing inside.  The heating effect may be subtle for a very large blackbox, which can have an effective temperature just above that of the CMB as the internal heat is diluted over a wide surface area.  Even then, though, the temperature excess still leads to the sky appearing slightly brighter in microwaves on sightlines to the blackbox than elsewhere.  The blackbox appears as a hot spot on the CMB if it is resolved, or as a point source if unresolved. 

\subsubsection{Temperature of the blackbox}

From energy conservation, the thermal luminosity of a blackbox is just the sum of the internal waste heat $L_{\rm int}$ and the total power absorbed from external radiation $L_{\rm ext}$: 
\begin{equation}
\label{eqn:LBlackbox}
L_{\rm blackbox} = \varepsilon_{\rm waste} A_{\rm box} \sigma_{\rm SB} T_{\rm eff}^4 = L_{\rm int} + L_{\rm ext}.
\end{equation}
A blackbox may not be perfectly black, hence the $\varepsilon_{\rm waste}$ factor, which describes its albedo and optical depth at the frequencies it radiates away its waste heat.

The internal heating is the luminosity of the stars and anything else inside the blackbox, $L_{\star}$, times some dissipation efficiency factor $\varepsilon_{\star}$.  In principle, since the stars are very hot, most of $L_{\star}$ could be harnessed as work, producing very little waste heat.  In order for $\varepsilon_{\star} \to 0$ to be maintained for a significant time, some kind of energy storage must be found.  Chemical energy storage like batteries are limited to about an eV per atom and would overflow very rapidly.  But perhaps the energy could be stored in the gravitational potential energy of matter within the blackbox, maybe by moving stars.  In practice, though, most activities (including computation and biology) dissipate energy, leading to $\varepsilon_{\star} \approx 1$.

The external radiation includes the CMB and the extragalactic background light (EBL).  These too would have their own efficiency factors $\varepsilon_{\rm CMB}$ and $\varepsilon_{\rm EBL}$.  The coldest a blackbox can be is the temperature of the CMB; such a blackbox is necessarily good at absorbing the thermal CMB radiation if it can emit its own thermal radiation effectively ($\varepsilon_{\rm CMB} \approx \varepsilon_{\rm waste}$).  A very hot blackbox can be transparent at microwave frequencies and still be an effective radiator as long as it is opaque at infrared frequencies, but then the CMB makes a negligible contribution to its heating anyway.  

The blackbody's opacity to the optical and infrared EBL is less constrained.  A cold blackbox does not need to be opaque to optical light at its surface.  Instead, the aliens could capture it all close to the stars (in Dyson spheres, for example); the blackbox proper then only needs to be opaque to low energy radiation.  Then the effective cross sectional area to the optical EBL is negligible.  Similar considerations apply to the infrared EBL.  Still, since some of the EBL is emitted by cold dust, a small fraction of it is in the form of microwave radiation and would be absorbed by a cold blackbox, so $\varepsilon_{\rm EBL}$ is at least $0.05 \varepsilon_{\rm CMB}$ \citep{Finke10}.  A hot blackbox would absorb the EBL effectively, but the energy density of the EBL is so low compared to the energy density of the blackbox's own radiation that it hardly matters then.  If individual motes are small enough, they can be heated strongly by each EBL photon and have no single temperature \citep{Purcell76}, but I will ignore this complication and it should not be a problem for the microwave EBL.

From these considerations, the absorbed power can be conveniently written in the form of the energy densities of the CMB and the EBL:
\begin{equation}
L_{\rm ext} = A_{\rm box} c (\varepsilon_{\rm CMB} u_{\rm CMB} + \varepsilon_{\rm EBL} u_{\rm EBL}) / 4,
\end{equation}
for a spherical blackbox.  Now we can find the temperature of the blackbox using equation~\ref{eqn:LBlackbox}.
\begin{equation}
\label{eqn:TBlackbox}
T_{\rm eff} = \left[\frac{\varepsilon_{\star}}{\varepsilon_{\rm waste}} \frac{L_{\star}}{A_{\rm box} \sigma_{\rm SB}} + \frac{\varepsilon_{\rm CMB}}{\varepsilon_{\rm waste}} T_{\rm CMB}^4 + \frac{\varepsilon_{\rm EBL}}{\varepsilon_{\rm waste}} \frac{c u_{\rm EBL}}{4 \sigma_{\rm SB}}\right]^{1/4}. 
\end{equation}

For simplicity, one can define an excess luminosity for the blackbox, that beyond the amount absorbed and re-emitted from the CMB:
\begin{equation}
L_{\rm excess} \equiv \varepsilon_{\star} L_{\star} + \varepsilon_{\rm EBL} A_{\rm box} c u_{\rm EBL} / 4,
\end{equation}
and compare it to the power absorbed from the CMB, $L_{\rm CMB} \equiv \varepsilon_{\rm CMB} A_{\rm box} c u_{\rm CMB} / 4$.  These two powers add up to the total blackbox luminosity: $L_{\rm blackbox} = L_{\rm CMB} + L_{\rm excess}$.  Then the effective temperature of the blackbox is
\begin{equation}
T_{\rm eff} = T_{\rm CMB} \left[\frac{\varepsilon_{\rm CMB}}{\varepsilon_{\rm waste}} \left(1 + \frac{L_{\rm excess}}{\varepsilon_{\rm CMB} L_{\rm CMB}}\right)\right]^{1/4}.
\end{equation}

Two different limits to the temperature can be derived: one for the case when the excess heating exceeds the CMB heating, and one for when the CMB heating is more important.  When $L_{\rm excess} \gg L_{\rm CMB}$, a Taylor series approximation gives us
\begin{equation}
T_{\rm eff} \approx \left(\frac{L_{\rm excess}}{\varepsilon_{\rm waste} A_{\rm box} \sigma_{\rm SB}}\right)^{1/4}.
\end{equation}
If the blackbox's heating is mostly from the CMB ($L_{\rm excess} \ll L_{\rm CMB}$), we have
\begin{equation}
T_{\rm eff} \approx T_{\rm CMB} \left(\frac{\varepsilon_{\rm CMB}}{\varepsilon_{\rm waste}}\right)^{1/4} \left(1 + \frac{L_{\rm excess}}{4 L_{\rm CMB}}\right).
\end{equation}
Since the thermal emission likely comes out at the same millimeter wavelengths that CMB photons have, the temperature excess over the CMB $\Delta T_{\rm eff} \equiv T_{\rm eff} - T_{\rm CMB}$ is
\begin{equation}
\Delta T_{\rm eff} \approx T_{\rm CMB} \frac{L_{\rm excess}}{4 L_{\rm CMB}}.
\end{equation}

\begin{figure}
\centerline{\includegraphics[width=9cm]{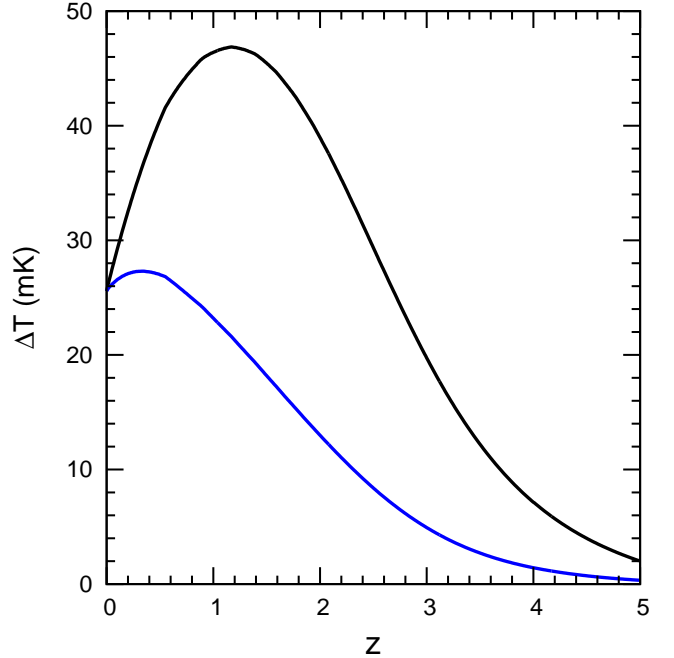}}
\figcaption{The temperature excess on the blackbox's surface caused by the thermalization of the EBL if $\varepsilon_{\rm CMB} = \varepsilon_{\rm EBL} = 1$ .  These temperature excesses were calculated from the EBL spectra of \citet{Finke10}.  The black line is the physical temperature excess in the blackbox frame ($\Delta T_{\rm eff}$), while the blue line is the observed temperature excess after redshifting ($\Delta T_{\rm obs}$).  The actual excess scales as $\varepsilon_{\rm EBL}$, which is unlikely to be less than $0.05\ \varepsilon_{\rm CMB}$.\label{fig:EBLTempExcess}}
\end{figure}

The heating from the EBL means that the blackbox still has an excess temperature even if there are absolutely no internal heat sources inside.  For a blackbox that is opaque at all frequencies (so all $\varepsilon$ are 1) and with $L_{\star} = 0$, the temperature excess is 
\begin{equation}
\frac{\Delta T_{\rm eff}}{T_{\rm CMB}} \approx \frac{1}{4} \frac{u_{\rm EBL}}{u_{\rm CMB}}.
\end{equation}
Using the EBL spectra of \citet{Finke10}, the excess is 25 mK at $z = 0$.  The ratio of $u_{\rm EBL}/u_{\rm CMB}$ was actually higher up to redshifts of about 1, which should help detection (Figure~\ref{fig:EBLTempExcess}).  Even if $\varepsilon_{\rm EBL} = 0.05$, the observed excess should be at least a mK for $z \la 1$.  A resolved cold blackbox is therefore an easily visible hotspot in today's CMB instruments.

\subsubsection{Observed flux from the blackbox}
A blackbox at redshift $z$ appears to have a temperature of $T_{\rm obs} = T_{\rm eff} / (1 + z)$:
\begin{equation}
T_{\rm obs} = T_{\rm CMB} (0) \left[\frac{\varepsilon_{\rm CMB}}{\varepsilon_{\rm waste}} \left(1 + \frac{L_{\rm excess}}{\varepsilon_{\rm CMB} L_{\rm CMB}}\right)\right]^{1/4}.
\end{equation}
Likewise the temperature excess has a scaling of $\Delta T_{\rm obs} \equiv T_{\rm obs} - T_{\rm CMB} (0) = \Delta T_{\rm eff} / (1 + z)$.

Assuming that $\varepsilon_{\rm waste}$ is constant over the frequencies the blackbox emits, the specific surface brightness of the blackbox on the sky is given by the Planck law,
\begin{equation}
I^{\rm obs}_{\nu} = \varepsilon_{\rm waste} B_{\nu} (T_{\rm obs}) = \frac{2 \varepsilon_{\rm waste} h \nu^3}{c^2} \frac{1}{e^{x_{\rm obs}} - 1},
\end{equation}
where $x_{\rm obs} = h \nu / (k T_{\rm obs})$, and $B_{\nu} (T_{\rm obs})$ is the Planck emission function.  Of course, the CMB contributes a similar thermal background everywhere on the sky.  What's actually measured is the excess surface brightness that remains once the CMB is subtracted off,
\begin{equation}
I^{\rm excess}_{\nu} = \frac{2 \varepsilon_{\rm waste} h \nu^3}{c^2} \left[\frac{1}{e^{x_{\rm obs}} - 1} - \frac{1}{e^{x_{\rm CMB}} - 1}\right],
\end{equation}
with $x_{\rm CMB} = h \nu / [k_B T_{\rm CMB} (0)]$.  

We are now in a position to calculate the excess flux from the blackbox.  The blackbox covers a solid angle $\Omega_{\rm blackbox}$ on the sky; a spherical blackbox with radius $R_{\rm box}$ appears as a disk with $\Omega_{\rm blackbox} = \pi [R_{\rm box} / D_A (z)]^2$, where $D_A (z)$ is its angular diameter distance \citep[see][]{Hogg99}.  The excess specific flux is then simply $\Omega_{\rm blackbox} I^{\rm excess}_{\nu}$:
\begin{equation}
\label{eqn:FExcessFull}
F^{\rm excess}_{\nu} = \frac{2 \pi \varepsilon_{\rm waste} R_{\rm box}^2 h \nu^3}{D_A(z)^2 c^2} \left[\frac{1}{e^{x_{\rm obs}} - 1} - \frac{1}{e^{x_{\rm CMB}} - 1}\right].
\end{equation}

In the low frequency Rayleigh-Jeans tail ($h \nu \ll k_B T_{\rm CMB} (0)$), this simplifies to
\begin{equation}
\label{eqn:FExcessRJ}
F^{\rm excess}_{\nu} \approx F^{\rm RJ}_{\nu} \equiv \frac{2 \pi \varepsilon_{\rm waste} R_{\rm box}^2 \nu^2 k_B}{D_A(z)^2 c^2} \left[T_{\rm obs} - T_{\rm CMB}(0)\right]
\end{equation}
Even for conservative values, the excess flux easily reaches mJy levels:
\begin{multline}
\label{eqn:FiducialExcessFlux}
F^{\rm excess}_{\nu} \approx 9.7\ \mJy\ \varepsilon_{\rm waste} \left(\frac{\nu}{100\ \GHz}\right)^2  \left(\frac{\Delta T_{\rm obs}}{0.1\ \Kelv}\right) \\
\times \left(\frac{R_{\rm box}}{10\ \kpc}\right)^2 \left(\frac{D_A}{\Gpc}\right)^{-2}.
\end{multline}
It's this sheer brightness at the thermal peak that allows powerful constraints to be set on blackboxes across the Universe.

\begin{figure}
\centerline{\includegraphics[width=9cm]{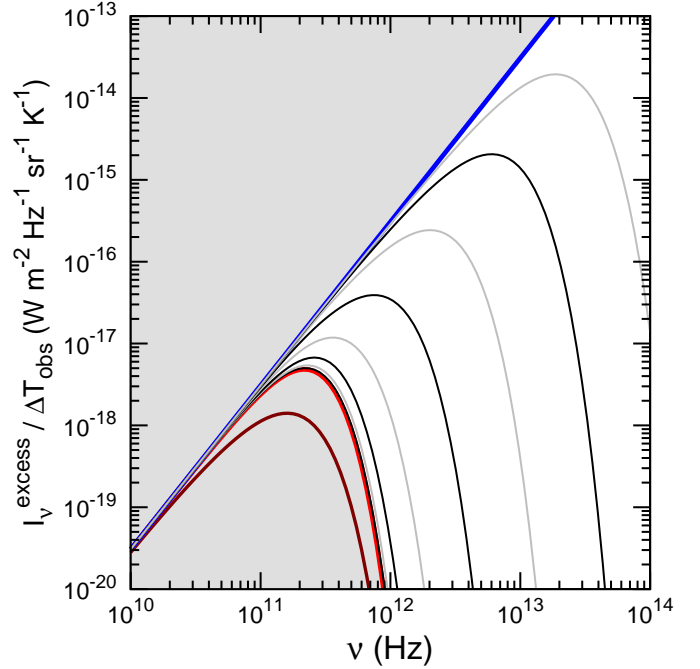}}
\figcaption{The excess intensity from a blackbox with $\varepsilon_{\rm waste} = 1$, after subtracting off the CMB and dividing out the temperature excess, has a characteristic shape.  The alternating black and grey lines are the spectra for $\log_{10} (\Delta T_{\rm obs} / \Kelv)$ of $-1.0$, $-0.5$, $0.0$, $0.5$, $1.0$, $1.5$, $2.0$, and $2.5$.  The spectrum of a cold blackbox follows the red line, while hot blackboxes have a Rayleigh-Jeans spectrum (blue line).  The excess spectrum never lies in the shaded regions. For comparison, the spectral shape of the CMB $[B_{\nu} (T_{\rm CMB}(0))/T_{\rm CMB}(0)]$ is plotted as the dark red line in the lower shaded region.\label{fig:FluxExcessShapes}}
\end{figure}

As the temperature excess $\Delta T_{\rm obs}$ vanishes, the excess flux spectrum approaches a limiting form of
\begin{equation}
F^{\rm excess}_{\nu} \to F^{\rm cold}_{\nu} \equiv \frac{dB_{\nu}}{dT}\Big|_{T_{\rm CMB} (0)} \Delta T_{\rm obs} \varepsilon_{\rm waste} \Omega_{\rm blackbox}.
\end{equation}
The shape is bluer than the CMB spectrum, with more flux at the exponential high-frequency tail of the spectrum:
\begin{align}
\label{eqn:FCold}
\nonumber F^{\rm cold}_{\nu} & = \frac{2 h \nu^3}{c^2} \left(\frac{\Delta T_{\rm obs}}{T_{\rm CMB} (0)}\right) \frac{x_{\rm CMB} e^{x_{\rm CMB}}}{[e^{x_{\rm CMB}} - 1]^2} \varepsilon_{\rm waste} \Omega_{\rm blackbox}\\
                       & = B_{\nu}(T_{\rm CMB} (0)) \left(\frac{\Delta T_{\rm obs}}{T_{\rm CMB} (0)}\right) \frac{x_{\rm CMB}}{1 - e^{-x_{\rm CMB}}} \varepsilon_{\rm waste} \Omega_{\rm blackbox}
\end{align}
The differences in shape between the Planck function and $F^{\rm cold}_{\nu}$ can be seen in Figure~\ref{fig:FluxExcessShapes}, which shows the normalized spectra of blackboxes with varying temperatures.  The CMB Planck function (dark red) starts falling below the excess flux spectra (black and grey) at frequencies of a few tens of GHz.  The Rayleigh-Jeans limit (blue; equation~\ref{eqn:FExcessRJ}), in fact, provides a better approximation to the cold blackbox spectrum shape (red), and is roughly accurate up to about 100 GHz.  

Figure~\ref{fig:FluxExcessShapes} shows the normalized spectra of excess flux from blackboxes of varying temperatures.  These lie in between the cold blackbox limit (red) and the hot Rayleigh-Jeans limit (blue).  All have a shape distinct from the CMB spectrum (dark red), which falls off more rapidly at higher temperatures.  All also show a Rayleigh-Jeans tail ($F_{\nu} \propto \nu^2$) at low frequencies.   

For the rest of this paper, I assume $\varepsilon_{\rm waste} = \varepsilon_{\rm CMB} = 1$, unless explicitly stated otherwise.

\subsubsection{Spectral indexes}
As seen in Figure~\ref{fig:FluxExcessShapes}, the excess spectra all have the distinctive Rayleigh-Jeans spectrum at low frequency.  At high frequencies, they are all distinctly bluer than the Planck spectrum, meaning that $dF^{\rm excess}_{\nu}/d\nu > dB_{\nu}(T_{\rm CMB})/d\nu$.  These qualities make the spectral index of the measured flux a useful signature of a blackbox.

The spectral index of a source between frequencies $\nu_1$ and $\nu_2$ is defined as
\begin{equation}
\alpha_{1 \to 2} \equiv \frac{\ln[F^{\rm excess}_{\nu}(\nu_2)/F^{\rm excess}_{\nu}(\nu_1)]}{\ln(\nu_2/\nu_1)}.
\end{equation}
For a blackbox's excess flux (equation~\ref{eqn:FExcessRJ}), it is equal to 
\begin{multline}
\alpha_{1 \to 2} = 3 + \ln \left(\frac{e^{x_{\rm CMB1}} - e^{x_1}}{e^{x_{\rm CMB2}} - e^{x_2}} \frac{e^{x_2} - 1}{e^{x_1} - 1} \frac{e^{x_{\rm CMB2}} - 1}{e^{x_{\rm CMB1}} - 1}\right) \Big/ \ln\left(\frac{\nu_2}{\nu_1}\right),
\end{multline}
after defining $x_{\rm CMB1} = h \nu_1 / (k_B T_{\rm CMB} (0))$, $x_{\rm CMB2} = h \nu_2 / (k_B T_{\rm CMB} (0))$, $x_1 = h \nu_1 / (k_B T_{\rm obs})$, and $x_2 = h \nu_2 / (k_B T_{\rm obs})$.  There is also an instantaneous spectral slope given by
\begin{align}
\nonumber \alpha & \equiv \frac{d \ln F_{\rm excess}}{d \ln \nu} \\
                 & = \displaystyle 3 + x \left[1 + \frac{(\Delta T_{\rm obs} / T_{\rm CMB} (0))}{1 - e^{x - x_{\rm CMB}}} - \frac{1}{1 - e^{-x}} - \frac{T_{\rm obs} / T_{\rm CMB} (0)}{1 - e^{-x_{\rm CMB}}}\right].
\end{align}

The spectral slope has a useful property: it is always larger for hotter blackboxes at any fixed frequency.  This follows from the fact that
\begin{equation}
\frac{\partial^2 F^{\rm excess}_{\nu}}{\partial \nu \partial T_{\rm obs}} > 0,
\end{equation}
which is always true for positive temperatures and frequencies, as long as $\varepsilon_{\rm waste}$ is constant over frequency.  Cold blackboxes have an instantaneous spectral slope that approaches
\begin{equation}
\alpha_{\rm cold} = 4 + x_{\rm CMB} - \frac{2 x_{\rm CMB}}{1 - e^{-x_{\rm CMB}}},
\end{equation} 
as found from the flux $F_{\rm cold}$ in equation~\ref{eqn:FCold}.  When measured between $\nu_1$ and $\nu_2$, the spectral slope of a cold blackbox is
\begin{equation}
\label{eqn:alphaColdRange}
\alpha_{1 \to 2}^{\rm cold} = 4 + \left[(x_{\rm CMB2} - x_{\rm CMB1}) + 2 \left(\frac{e^{x_{\rm CMB1}} - 1}{e^{x_{\rm CMB2}} - 1}\right)\right] \big/ \ln\left(\frac{\nu_2}{\nu_1}\right).
\end{equation}
Hot blackboxes have a spectral slope that approaches the Rayleigh-Jeans limit of $+2$.  

It is also the case that
\begin{equation}
\frac{d\alpha}{d\nu} < 0 
\end{equation}
for all positive frequencies and $T_{\rm obs} > T_{\rm CMB} (0)$ with constant $\varepsilon_{\rm waste}$.  Thus, only those sources with negative spectral curvature are viable blackbox candidates.  

\subsubsection{Temperature spreads and limb darkening}
A solid blackbox may have a single surface temperature, but the thermal radiation of a diffuse blackbox should have a spread in temperatures, because the temperature of an atmosphere increases with optical depth.  A line of sight into a blackbox's atmosphere passes through layers with different temperatures, although radiation from below the photosphere is mostly absorbed.   For these reasons, a diffuse blackbox should display limb darkening, as lines of sight passing obliquely through the edges of the blackbox go through more matter per unit depth.   

The temperature spread in the thermal emission alters the shape of the blackbox's spectrum, if the blackbox is significantly hotter than the CMB.  These differences show up in the peak of the excess spectrum and the Wien exponential tail, where the shape depends critically on temperature.  But when observing at low frequencies, the thermal emission from each layer of the blackbox's atmosphere is well-approximated by a Rayleigh-Jeans tail.  The observed emission spectrum, as a weighted sum of these tails, should also have the characteristic $F_{\nu} \propto \nu^2$ shape.  Only the amount and the angular distribution of the emission is altered because of limb darkening at low frequencies.

A cold blackbox ($\Delta T_{\rm obs} \ll T_{\rm CMB}$) also displays limb darkening of a sort, with the \emph{excess} temperatures increasing deeper in the atmosphere, but the total temperature remains close to $T_{\rm CMB}$ for all lines of sight.  This can be understood readily for a purely grey atmosphere in the blackbox.  Then, radiative transfer of energy through the blackbox is characterized by a diffusion equation of the form $\nabla^2 u = 0$, where $u$ is the bolometric radiation energy density.  This equation is linear in $u$, so the radiation field can be separated into a uniform background $u_{\rm CMB}$ from the CMB and the excess radiation $u_{\rm int}$ from the blackbox's internal luminosity.  The dependence of $u_{\rm int}$ on optical depth is the same for hot and cold blackboxes; thus the CMB has no effect on the angular distribution of bolometric excess flux for a grey atmosphere.  The shape of the excess flux spectrum from a cold blackbox is not affected by limb darkening, since it is the same for all $\Delta T_{\rm obs} \ll T_{\rm CMB} (0)$ (equation~\ref{eqn:FCold}), and thus for each layer above the photosphere of the blackbox.

\subsection{Transparency of blackboxes to non-thermal emission}
A blackbox is opaque to its internal thermal radiation, in microwaves and possibly infrared, optical, and ultraviolet light.  That still leaves low frequency radio, X-rays, and gamma rays as potential messengers from the interior of the blackbox, as well as particles like neutrinos.  If aliens generate these forms of radiation within the blackboxes, they may still attempt to capture and process them, just like starlight.  

Normal star-forming galaxies typically emit only a tiny fraction of their bolometric power through non-thermal mechanisms.  The GHz synchrotron luminosity of a star-forming galaxy is only about a millionth of the infrared luminosity, for example \citep{Yun01}.  It is probably more advantageous to extract energy from the waste heat by building another layer around the blackbox and harnessing the temperature difference between the new, colder surface and the interior.  Yet, harvesting non-thermal radiation may be easier since it does not require working with small temperature differences.  Or, the occupants of a blackbox may try to stop non-thermal radiation simply to hide their activities or some other reason.

Radio emission can be absorbed quite efficiently with dipole antennas, similar to those that might be used to absorb the thermal radiation of a blackbox.  I showed in Section~\ref{sec:DipoleClouds} that the mass of an opaque dipole cloud does not depend on the wavelength at low frequencies -- the effective area of the antenna increases as the square of its length, while the cross-section must also increase as the square of its length to avoid Ohmic losses.  The dipole mass cloud grows proportionally to the number of decades of photon frequency that are stopped.  Still, doubling the total mass of a dipole cloud should still be allowed in the mass budget of a galaxy if carbon nanotube antennas are used.  For MHz radio waves, half-wavelength dipole antennas need to be kilometers long.  Yet antennas that are hundreds of meters tall are used on the Earth for transmitting radio.  Furthermore, interstellar space is a much gentler environment than the Earth, with negligible gravity, pressure, and temperature; building big antennas and setting them free in the ISM may work fine.

There is no obvious clever solution like dipole antennas for absorbing harder X-rays and gamma rays.  The optical properties of a material depend on structures that are smaller than the wavelength of a photon; when the wavelength of a photon is smaller than an atom, only chemistry and nuclear physics matters.  Hard ultraviolet light and soft X-rays are efficiently absorbed by the photoelectric effect in ISM hydrogen.  But beyond a few keV, a blackbox can only be opaque if it is Compton thick, with a column density of order $\Sigma_C \approx 25\ \kgm2$ \citep[e.g.,][]{Olive15}.  The amount of mass required to make a 10 kpc wide blackbox Compton thick, $4 \pi R_{\rm box}^2 \Sigma_C$, exceeds the baryonic content of a galaxy:
\begin{equation}
M_{\rm box} \ga 1.5 \times 10^{13}\ \Msun \left(\frac{R_{\rm box}}{10\ \kpc}\right)^2 \left(\frac{\Sigma}{25\ \kgm2}\right).
\end{equation}
The situation is much worse still for high energy gamma rays, which are typically stopped by columns of $\sim 1000\ \kgm2$ \citep{Olive15}.

\subsection{Blackboxes colder than the CMB?}
\label{sec:2LoT}
\citet{Wright14-Results} brings up the possibility that cosmic engineers are not bound by the conservation of energy as we understand it.  If they can summon astronomical amounts of energy, that can increase the waste heat.  As \citet{Wright14-Results} points out in the context of Chilly Ways, the excess power conjured against the First Law of Thermodynamics cannot be vastly greater than the typical luminosity of a galaxy, or else it would have been seen with WISE.  Similar considerations apply for blackboxes; they would have been detected by \emph{Planck} and earlier surveys.  If the First Law of Thermodynamics does not hold, then perhaps aliens destroy astronomical amounts of energy.  Then waste heat is reduced or eliminated, and it is harder to see a blackbox or a Chilly Way.

Instead of energy conservation, what if aliens know a way around the Second Law of Thermodynamics, unlikely as that is?  In the worst case from an observer's point of view (but not the aliens'), energy is continuously recycled and never released as waste heat.  As far as searches in the microwave and infrared regions are concerned, it might as well be as if the aliens are not there.  But there is another possibility -- the blackbox (or Chilly Way) can be refrigerated until it is much \emph{colder} than the CMB.  Then the aliens can harness the power contained within the CMB itself.  For a blackbox the size of a galaxy, this power is almost of the same magnitude as the starlight itself.  

Even within the laws of thermodynamics, a refrigerated blackbox is possible if it is opaque to low energy neutrinos or gravitational waves.  The temperature of the cosmic neutrino background from the Big Bang is expected to be only 1.95 K \citep{Egan10}.  A blackbox could potentially use the CMB as a hot reservoir and the neutrino background as a cold reservoir and extract useful work.  Just as the EBL heats a blackbox slightly over the CMB, the diffuse supernova neutrino background may heat a neutrino-opaque blackbox.  The energy density in the supernova neutrino background is probably comparable to that of the EBL \citep{Beacom10}, so this heating is at most a few hundredths of a Kelvin, compared to a potential cooling of a few tenths of a Kelvin.  The relic gravitational wave background, if there is one, is thought to be even colder, at less than 1 K \citep{Egan10}, providing another possible cold reservoir.  Of course, a galaxy-sized shroud that can absorb neutrinos and especially gravitational waves defies our knowledge of material physics.

A very large black hole would also appear colder than the CMB, if it was large enough to cast a detectable shadow, and would still be allowed by thermodynamics.  Rich galaxy clusters contain about $10^{15}\ \Msun$ of matter.  If it was all fed into a single black hole, the hole would grow to have a Schwarzschild radius of about 100 pc.  The shadow would then have a flux deficit of $\sim 1\ \mJy$ at 150 GHz if the hole was 100 Mpc away.  This is too small to be detected by \emph{Planck}, although the South Pole Telescope might detect the deficit.  

A blackbox that is colder than the CMB should appear to have a negative excess flux in the microwaves.  In fact, this is exactly what some surveys are looking for -- galaxy clusters appear to have negative flux at low frequencies thanks to the Sunyaev-Zeldovich effect \citep{Sunyaev72}.  But the Compton effect responsible for the Sunyaev-Zeldovich phenomenon merely boosts the energy of photons to slightly higher energy; the missing flux appears at high frequencies.  Thus, these galaxy clusters appear hotter at high frequencies \citep{Sunyaev72}.  A refrigerated blackbox appears as a negative source at all frequencies.

The CMB does not have exactly the same temperature in every direction, but is slightly cooler in a few parts of the sky than others.  A refrigerated blackbox appears like an extreme temperature anomaly, with a magnitude of order a K instead of a few dozen $\uKelv$.  Clearly, if any such blackboxes existed and were resolved in the sky by \emph{Planck}, it would have been noticed by now in searches for non-Gaussian anomalies.  Even if it was just the size of the Milky Way, and unresolved, the negative flux from the blackbox would be of order $-1\ \Jy$, easily noticed by \emph{Planck} and Sunyaev-Zeldovich surveys (equation~\ref{eqn:FiducialExcessFlux}).  Thus, there are probably no large, refrigerated blackboxes.

\section{Blackboxes Don't Exist}

\subsection{Which surveys to use?}
The thermal emission of a blackbox comes out in radio, microwave, and infrared bands.  Infrared surveys can be useful for hot blackboxes, just like they are for Dyson spheres \citep{Griffith15}, but they are essentially useless for cold blackboxes with $T_{\rm obs} \approx T_{\rm CMB} (0)$.  All (conventional) blackboxes should emit microwave thermal emission, since they are at least as hot as the CMB, but all-sky microwave surveys tend to be relatively shallow.  Finally, although blackboxes are much fainter at GHz radio frequencies, a radio survey might pick up their emission anyway, much like how radio telescopes can study the extended photospheres of giant stars \citep{Reid97}.  Fortunately, there are now many all-sky and large field surveys for all three, which I list in Table~\ref{table:FiguresOfMerit}.  But which of these is most constraining?

This depends on whether one knows where to look.  A survey reaches some minimum flux $F^{\rm min}_{\nu}$ and covers some solid angle on the sky $\Omega$.  There are three possibilities to consider.  

First, what if one wanted to catalog all the blackboxes in some well-defined region of space, like a given galaxy cluster?  Then, it does not matter if the survey is wide-field or not, as long as it covers the target region.  The effectiveness of the survey depends only on the ratio of the expected flux to the minimum flux, which can be parameterized by
\begin{equation}
\Omega_{\rm char}^{-1} = \frac{I^{\rm excess}_{\nu}}{F^{\rm min}_{\nu}}.
\end{equation}
A blackbox that covers a solid angle $\Omega_{\rm blackbox}$ on the sky is detectable if $\Omega_{\rm blackbox} \Omega_{\rm char}^{-1} \ga 1$.  

What if one wanted to construct a luminosity-selected catalog of blackboxes, or to measure the fraction of galaxies sealed in blackboxes in the local Universe?  Blackboxes with the necessary luminosity are visible out to some comoving distance $R_{\rm max}$.  The survey effectiveness is measured by how much comoving volume it examines: ${\cal V} = (\Omega/3) D_{\rm max}^3$, where $\Omega$ is the solid angle covered by the survey.  If the blackboxes are relatively faint, then $D_{\rm max}$ is small compared to cosmological scales and has the form $D_{\rm max} = \overline{R} \sqrt{\pi / \Omega_{\rm char}}$, where $\overline{R}$ is a reference radius of a blackbox.  The sampled comoving volume is then proportional to 
\begin{equation}
{\cal V} = \frac{\Omega \overline{R}^3}{3} \left(\frac{\pi}{\Omega_{\rm char}}\right)^{3/2}.
\end{equation}

On the other hand, if the blackboxes are very bright, they can be visible anywhere in the Universe.  This is actually true for blackboxes that contain rich galaxy clusters.  Then, $D_{\rm max}$ has no dependence on the survey sensitivity.  The only quantity that matters is the survey area $\Omega$.

\begin{deluxetable*}{lccccccccc}
\tablecaption{Figures of merit}
\tablehead{\colhead{Survey} & \colhead{$\nu$} & \colhead{$\lambda$} & \colhead{$h \nu / k_B$} & \colhead{$F_{\nu}^{\rm min}$} & \colhead{$4 \pi \Omega$} & \multicolumn{2}{c}{Cold blackboxes} & \multicolumn{2}{c}{Hot blackboxes} \\ & & & & & & \colhead{$\Omega_{\rm char}^{-1}$} & \colhead{${\cal V}$} & \colhead{$\Omega_{\rm char}^{-1}$} & \colhead{${\cal V}$} \\ & \colhead{(GHz)} & \colhead{(mm)} & \colhead{(K)} & \colhead{(mJy)} & \colhead{(sky)} & \colhead{($\arcmin^{-2}$)} & \colhead{($\Gpc^3$)} & \colhead{($\arcmin^{-2}$)} & \colhead{($\Gpc^3$)}}
\startdata
NVSS                & 1.4   & 210    & 0.067 & 2.5   & 0.796  & 2.0     & 0.0022  & 2.0   & 0.0022\\
AT20G               & 20    & 15     & 0.96  & 40    & 0.487  & 26      & 0.060   & 26    & 0.061\\
\emph{Planck} CSSC2 & 30    & 10     & 1.4   & 427   & 1.000  & 5.4     & 0.012   & 5.5   & 0.012\\
                    & 44    & 6.8    & 2.1   & 692   & 1.000  & 6.9     & 0.017   & 7.3   & 0.019\\
									  & 70    & 4.3    & 3.4   & 501   & 1.000  & 23      & 0.10    & 25    & 0.12\\
									  & 100   & 3.0    & 4.8   & 269   & 0.850  & 75      & 0.53    & 97    & 0.77\\
									  & 143   & 2.1    & 6.9   & 177   & 0.850  & 180     & 2.0     & 300   & 4.2\\
										& 217   & 1.4    & 10    & 152   & 0.659  & 270     & 2.7     & 810   & 14\\
									  & 353   & 0.85   & 17    & 304   & 0.476  & 83      & 0.34    & 1100  & 16\\
									  & 545   & 0.55   & 26    & 555   & 0.470  & 9.1     & 0.012   & 1400  & 23\\
									  & 857   & 0.35   & 41    & 791   & 0.463  & 0.16    & $2.7 \times 10^{-5}$ & 2400 & 52\\ 
SPT-SZ              & 90    & 3.2    & 4.6   & 11    & 0.0187 & 1700    & 1.2     & 2100  & 1.8\\
                    & 150   & 2.0    & 7.2   & 4.4   & 0.0187 & 7700    & 12      & 13000 & 2.7\\
									  & 220   & 1.4    & 11    & 11    & 0.0187 & 3700    & 4.0     & 11000 & 2.2\\
AKARI               & 1870  & 0.16   & 90    & 6300  & 1.000  & \nodata & \nodata & 1500  & 52\\
                    & 2140  & 0.14   & 100   & 1400  & 1.000  & \nodata & \nodata & 8500  & 750\\
                    & 3330  & 0.090  & 160   & 550   & 1.000  & \nodata & \nodata & 53000 & 11000\\
                    & 4610  & 0.065  & 220   & 2400  & 1.000  & \nodata & \nodata & 23000 & 3300\\
WISE                & 13600 & 0.022  & 650   & 6.0   & 1.000  & \nodata & \nodata & $8.1 \times 10^7$    & $6.8 \times 10^8$\\
                    & 25000 & 0.012  & 1200  & 1.0   & 1.000  & \nodata & \nodata & $1.6 \times 10^9$    & $6.2 \times 10^{10}$\\
									  & 65200 & 0.0046 & 3100  & 0.11  & 1.000  & \nodata & \nodata & $1.0 \times 10^{11}$ & $3.0 \times 10^{13}$\\
                    & 88200 & 0.0034 & 4200  & 0.080 & 1.000  & \nodata & \nodata & $2.5 \times 10^{11}$ & $1.2 \times 10^{14}$
\enddata
\tablecomments{I used fiducial values of $\varepsilon_{\rm waste} = 1$, $\Delta T_{\rm obs} = 1\ \Kelv$, and $\overline{R} = 10\ \kpc$.  The hot blackbox values are valid only if $T_{\rm obs} \gg h \nu / k_B$ and $T_{\rm obs} \gg T_{\rm CMB} (0)$; $\Delta T_{\rm obs} = 1\ \Kelv$ should be regarded as a normalization factor only.  I use the Rayleigh-Jeans limit for the hot blackbox flux, which does not account for the (slight when $T_{\rm obs} \gg T_{\rm CMB} (0)$) fall-off in flux in the CMB Wien tail.  Note that at a fixed luminosity, the surface area of a blackbox or Chilly Way decreases as its temperature increases.\\
{\bf References used:} NVSS -- \citet{Condon98}, where the given flux limit applies for 50\% completeness; AT20G -- \citet{Murphy10}, where the given flux limit in their Table 1; \emph{Planck} -- \citet{Ade16-PCCS2}; SPT-SZ -- \citet{Mocanu13}; AKARI -- \citet{Kawada07}, using the expected $5\sigma$ point source sensitivities; WISE -- \citet{Wright10}, using the $5\sigma$ point source sensitivities.}
\label{table:FiguresOfMerit}
\end{deluxetable*}

I calculated these figures of merit for some radio, microwave, and infrared surveys and have listed them in Table~\ref{table:FiguresOfMerit}.  The quantities differ for cold blackboxes, with a flux given by equation~\ref{eqn:FCold}, and hot blackboxes, with a Rayleigh-Jeans spectrum (equation~\ref{eqn:FExcessRJ}).  Of the all-sky surveys, the \emph{Planck} 217 GHz PCCS2 catalog is the best for cold blackboxes.  Other \emph{Planck} bands, from 100 to 353 GHz, have a respectable sensitivity too.  But the South Pole Telescope Sunyaev-Zeldovich (SPT-SZ) survey has a much greater depth still at the cost of survey area; it is much better at finding smaller, fainter blackboxes.  Although not anywhere near as sensitive as the microwave surveys, the NVSS (NRAO VLA Sky Survey) catalog still has an effective reach of millions of $\Mpc^3$, despite being at 1.4 GHz.  If a blackbox were found, a radio telescope could check whether it has a thermal spectrum out to long wavelengths.  The Australia Telescope 20 GHz (AT20G) survey is about as effective at \emph{Planck}'s 30 GHz channel.  Furthermore, radio interferometers have a much better angular resolution than \emph{Planck}, and are better at measuring the size of a blackbox.  Cold blackboxes emit no detectable infrared flux, and so are missed by the infrared surveys in Table~\ref{table:FiguresOfMerit}.

Infrared surveys are extremely effective at finding hot blackboxes.  When searching for blackboxes with a fixed high temperature and radius, AKARI's far infrared (FIR) survey has a volume reach that is thousands of times better than \emph{Planck}.  The volume reach of WISE is about a billion times times that of \emph{Planck}.  The apparent weakness of \emph{Planck} comes about because at high temperatures, very little of the luminosity comes out in microwaves, since $F_{\nu} \propto \nu^2$.  If we fix the luminosity only and compare the reach of \emph{Planck} for cold blackboxes and the reach of WISE for blackboxes with $T_{\rm obs}$ of a few hundred K, the survey volumes are roughly the same.  This demonstrates that WISE is effective for finding classical Dyson spheres and Chilly Ways.  All of these surveys are all-sky, so we do not have to worry that a blackbox is hiding outside of their survey field.  But these figures assume that the blackboxes are hot, with $x = h \nu / (k_B T_{\rm obs}) \la 1$.  The FIR surveys are useful only if $T_{\rm obs} \ga 20\ \Kelv$ and WISE is useful only if $T_{\rm obs} \ga 50\ \Kelv$.

\subsection{Results from the \emph{Planck} Catalog of Compact Sources}

\begin{figure}
\centerline{\includegraphics[width=9cm]{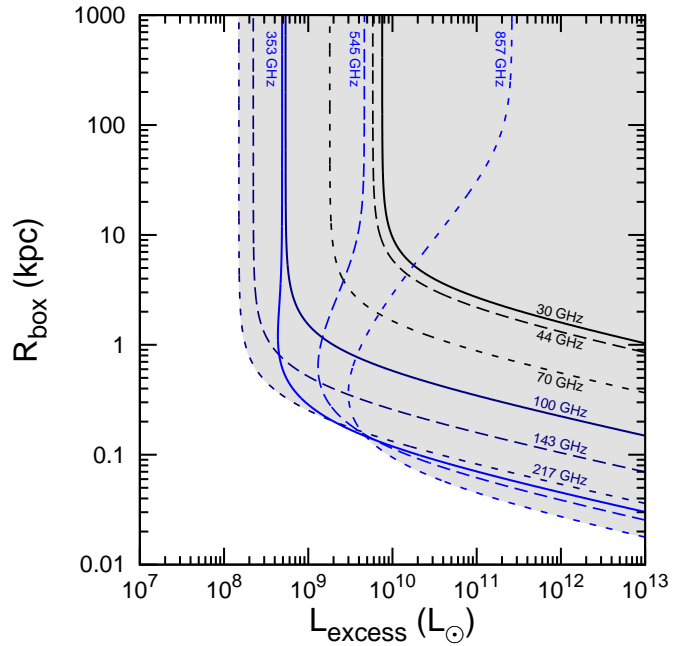}}
\figcaption{A comparison of the bounds that can be set by point source searches in each frequency channel of \emph{Planck}.  Blackboxes with luminosities and radii that fall right of these curves can be detected.  These channels are 30 MHz (black, solid), 44 MHz (black, dashed), and 70 MHz (black, dotted) for the Low Frequency Instrument, and 100 GHz (dark blue, solid), 143 GHz (dark blue, dashed), 217 GHz (dark blue, dotted), 353 GHz (light blue, solid), 545 GHz (light blue, dashed), and 857 GHz (light blue, dotted) for the High Frequency Instrument.  The plotted limits are derived for blackboxes within 100 Mpc comoving distance.\label{fig:PlanckBandComparisons}}
\end{figure}

I concentrated my efforts on looking for blackboxes in the PCCS2 catalog \citep{Ade16-PCCS2}.  Although it is not optimal in all cases, the \emph{Planck} PCCS2 survey performs fairly well over a wide range of blackbox sizes and luminosities.  The effectiveness of each channel is shown in Figure~\ref{fig:PlanckBandComparisons}.  In addition, the spectral slope of blackboxes is confined to a relatively narrow range at low frequency, and can be used to select candidate blackboxes.  Finally, the PCCS2 also has only a few thousand sources at most in each band, so it can be examined quickly.  I restrict my search to the primary PCCS2 catalog and do not include those in the ``excluded'' PCCS2E catalogs covering regions contaminated by Galactic dust.  This greatly reduces the number of sources at the cost of sacrificing some sky area at high frequencies.

I matched sources in each pair of channels in order to check their spectral index.  Each source in the first channel was provisionally matched to the source in the second channel that was nearest to its position.  The match was considered valid if (1) the potential matches were less than 30 arcminutes apart (the approximate size of the 30 GHz \emph{Planck} beam; \citealt{Ade16-PCCS2}) and (2) the match was consistent both ways -- that is, the nearest source in the first channel to the potential match in the second channel was the original source.  Generally, those sources that matched had positions consistent to within an arcminute or two, and it was rare that a source in the second channel was the closest to one in the first channel but not the other way around.  

For the photometry, I used the DETFLUX values listed in the PCCS2, which are the flux levels derived by the PCCS2 detection pipeline \citep{Ade16-PCCS2}.  These were measured under the assumption that the sources are pointlike.  This is basically true for galaxy-sized blackboxes, which cover a solid angle of $\sim 0.3\ (R_{\rm box}/10\ \kpc)^2 (D_A/100\ \Mpc)^{-2}$ square arcminutes, compared to the 100 GHz beam of about 100 square arcminutes \citep{Ade16-PCCS2}.  The DETFLUX values have the smallest uncertainties, minimizing the errors in the spectral slope.  I do also consider the other three photometry methods, but the uncertainties in those measurements are often as big as the fluxes themselves.  Many sources that simply have poor flux determinations pass the cut on $\alpha$, diluting those results.  

I then calculated the spectral slope of the source between the two channels.  The uncertainty in the spectral slope was calculated according to
\begin{equation}
\sigma_{\alpha} = \sqrt{\left(\frac{\sigma_1}{F_1}\right)^2 + \left(\frac{\sigma_2}{F_2}\right)^2} \big/ \ln\left(\frac{\nu_2}{\nu_1}\right),
\end{equation}
where $F_1$ and $F_2$ were the fluxes in the first and second channels, and $\sigma_1$ and $\sigma_2$ were the flux uncertainties in these channels.  The matched sources were selected if $\alpha_{\rm min} \le \alpha + 2 \sigma_{\alpha}$ and $\alpha - 2 \sigma_{\alpha} \le 2.0$, where the minimum spectral slope $\alpha_{\rm min}$ is that of a cold blackbox (equation~\ref{eqn:alphaColdRange}).

\begin{deluxetable}{cccccccc}
\tablecaption{Selection of sources detected by \emph{Planck} with DETFLUX photometry}
\tablehead{\colhead{$\nu_1$} & \colhead{$\nu_2$} & \colhead{$\alpha_{\rm min}$} & \multicolumn{4}{c}{\emph{Planck} compact sources} \\
\colhead{(GHz)} & \colhead{(GHz)} & & \colhead{At $\nu_1$} & \colhead{At $\nu_2$} & \colhead{Matched} & \colhead{$\alpha$ selected}}
\startdata
 30 &  44 & +1.9309 & 1560 &  934 &  797 &   64\\
 30 &  70 & +1.8803 & 1560 & 1296 &  865 &    0\\
 30 & 100 & +1.8107 & 1560 & 1742 &  835 &    0\\
 30 & 143 & +1.6939 & 1560 & 2160 &  908 &    0\\
 30 & 217 & +1.4598 & 1560 & 2135 &  688 &    0\\
 30 & 353 & +0.9752 & 1560 & 1344 &  242 &    1\\
 30 & 545 & +0.2632 & 1560 & 1694 &  120 &   29\\
 30 & 857 & -0.8680 & 1560 & 4891 &  132 &  132\\
\hline
 44 &  70 & +1.8385 &  934 & 1296 &  752 &    3\\
 44 & 100 & +1.7546 &  934 & 1742 &  528 &    0\\
 44 & 143 & +1.6169 &  934 & 2160 &  530 &    0\\
 44 & 217 & +1.3468 &  934 & 2135 &  397 &    0\\
 44 & 353 & +0.7994 &  934 & 1344 &  195 &    2\\
 44 & 545 & +0.0094 &  934 & 1694 &   95 &   33\\
 44 & 857 & -1.2291 &  934 & 4891 &   84 &   84\\
\hline
 70 & 100 & +1.6453 & 1296 & 1742 &  679 &   31\\
 70 & 143 & +1.4729 & 1296 & 2160 &  679 &    0\\
 70 & 217 & +1.1450 & 1296 & 2135 &  508 &    2\\
 70 & 353 & +0.5012 & 1296 & 1344 &  227 &    4\\
 70 & 545 & -0.4045 & 1296 & 1694 &  102 &   88\\
 70 & 857 & -1.7976 & 1296 & 4891 &  103 &  103\\
\hline
100 & 143 & +1.3009 & 1742 & 2160 & 1132 &   39\\
100 & 217 & +0.9147 & 1742 & 2135 &  781 &    9\\
100 & 353 & +0.1777 & 1742 & 1344 &  266 &   51\\
100 & 545 & -0.8356 & 1742 & 1694 &  145 &  142\\
100 & 857 & -2.3693 & 1742 & 4891 &  193 &  191\\
\hline
143 & 217 & +0.5835 & 2160 & 2135 &  963 &  311\\
143 & 353 & -0.2669 & 2160 & 1344 &  288 &  198\\
143 & 545 & -1.4068 & 2160 & 1694 &  173 &  159\\
143 & 857 & -3.1024 & 2160 & 4891 &  230 &  217\\
\hline
217 & 353 & -0.9957 & 2135 & 1344 &  531 &  416\\
217 & 545 & -2.3081 & 2135 & 1694 &  431 &  270\\
217 & 857 & -4.2215 & 2135 & 4891 &  502 &  326\\
\hline
353 & 545 & -3.7784 & 1344 & 1694 &  779 &  506\\
353 & 857 & -5.9911 & 1344 & 4891 &  782 &  381\\
\hline
545 & 857 & -8.1142 & 1694 & 4891 & 1377 &  853
\enddata  
\label{table:PlanckSelectedCounts}
\end{deluxetable}

The number of sources that pass these cuts for each pair of channels is listed in Table~\ref{table:PlanckSelectedCounts}.  At high frequencies, the spectral slope constraints are not very powerful, and hundreds of sources pass.  But at low frequencies, relatively few sources have slopes in the narrow range.  In fact, there are some pairs of channels that select \emph{no} sources.  In other words, there are \emph{no} blackboxes bright enough to be detected by \emph{Planck} at both of these frequencies -- even though the reach of \emph{Planck} includes millions of $\Mpc^3$.

When I consider the other photometry sets in the PCCS2, the power of the cuts is somewhat weakened because of the greater uncertainties in $\alpha$.  The GAUSSFLUX photometry method was designed to handle extended sources by fitting a surface brightness profile to the maps.  I find that no sources are selected for $\nu_1 =$~70 GHz and $\nu_2 =$~143, 217, or 353 GHz.  In contrast, with the APERFLUX photometry, the only pair of channels with no selected sources is $\nu_1 =$~30 GHz and $\nu_2 =$~217 GHz; with the PSFFLUX photometry, the same pair of frequencies leads to 4 selected sources, the fewest of any pair.  The APERFLUX and PSFFLUX photometry would lead to the poorest results if accepted, since the low frequency channels are so insensitive (as in Figure~\ref{fig:PlanckBandComparisons}).

\subsubsection{The nine (refuted) candidates spanning 100 and 217 GHz}
\label{sec:Planck100and217GHzSearch}

The 100 GHz channel is more sensitive than the lower frequency channels with no matched sources.  As listed in Table~\ref{table:PlanckSelectedCounts}, only nine sources pass all of the cuts with $\nu_1 =$~100 GHz and $\nu_2 =$~217 GHz.  Checking these nine is a very quick matter, and one unlikely to produce false positives.  These nine sources are listed in Table~\ref{table:BlackboxCandidates}.

\begin{deluxetable*}{llllccccc}
\tablecaption{Blackbox candidates at 100 and 217 GHz}
\tablehead{\colhead{$l_{100}$} & \colhead{$b_{100}$} & \colhead{$l_{217}$} & \colhead{$b_{217}$} & \colhead{$\Delta\theta$} & \colhead{$F_{100}$} & \colhead{$F_{217}$} & \colhead{$\alpha$} & \colhead{Identification} \\ \colhead{($^{\circ}$)} & \colhead{($^{\circ}$)} & \colhead{($^{\circ}$)} & \colhead{($^{\circ}$)} & \colhead{(arcmin)} & \colhead{(mJy)} & \colhead{(mJy)} & & }
\startdata
 97.4864 & -87.9704 &  97.5697 & -87.9643 &  0.40 & $ 936 \pm 54$ & $3828 \pm 47$ & $1.82 \pm 0.08$ & NGC 253\\
140.4274 & -17.3910 & 140.3795 & -17.4033 &  2.84 & $ 326 \pm 59$ & $ 831 \pm 31$ & $1.21 \pm 0.24$ & NGC 891\\
141.4130 & +40.5670 & 141.4028 & +40.5627 &  0.53 & $1054 \pm 48$ & $2861 \pm 31$ & $1.29 \pm 0.06$ & M82\\
172.1037 & -51.9171 & 172.1030 & -51.9263 &  0.55 & $ 303 \pm 54$ & $ 712 \pm 32$ & $1.10 \pm 0.24$ & NGC 1068\\
204.0026 & -25.1867 & 204.1594 & -25.6601 & 29.65 & $ 459 \pm 67$ & $ 847 \pm 42$ & $0.79 \pm 0.20$ & [CB88] 28\\
218.0349 & +12.5998 & 218.0615 & +12.6003 &  1.56 & $ 302 \pm 65$ & $ 488 \pm 36$ & $0.62 \pm 0.29$ & UW CMi and Galactic dust (?)\\
221.4591 & +45.0640 & 221.4501 & +45.0775 &  0.90 & $ 878 \pm 61$ & $3695 \pm 44$ & $1.86 \pm 0.09$ & CW Leo\\
300.6434 & -43.7305 & 300.6815 & -43.7428 &  1.81 & $ 403 \pm 49$ & $ 839 \pm 32$ & $0.95 \pm 0.16$ & NGC 460\\
318.6650 & +16.9314 & 318.7068 & +17.0341 &  6.62 & $ 345 \pm 71$ & $1138 \pm 33$ & $1.54 \pm 0.27$ & Dobashi 6193
\enddata
\tablecomments{Fluxes are those derived from the DETFLUX method in \citet{Ade16-PCCS2}.}
\label{table:BlackboxCandidates}
\end{deluxetable*}

Three of the sources (NGC 253, NGC 891, and M82) are nearby infrared-bright starburst galaxies \citep{Sanders03}.  NGC 1068 is mostly powered by its Seyfert nucleus, although it also has a prominent starburst region \citep{Telesco84}.  The 100 GHz luminosity of these galaxies is a combination of synchrotron emission from cosmic ray electrons, free-free emission from H II regions, and thermal emission from dust \citep{Williams10}.  The 217 GHz luminosity is mostly the thermal glow of dust heated by starlight in the starburst (and Seyfert) regions \citep{Williams10,Peel11}.  It is possible that these starburst regions are opaque to their own FIR radiation \citep[c. f.,][]{Thompson05}, making them natural blackboxes of a sort, but they are transparent at the millimeter wavelengths considered here.  This is demonstrated by the positive curvature in their microwave spectra, in contrast to the negative curvature expected of a blackbox spectrum.  

Another of the sources, NGC 460, is also associated with intense star formation.  This open star cluster is contained within the N84 H II region in the Small Magellanic Cloud \citep{Heinze56,Davies76,Testor87}.  The fluxes at 217 GHz and above fall along an approximate Rayleigh-Jeans tail, consistent with dust.  At lower frequencies, the spectrum is flat in $F_{\nu}$, possibly indicating free-free emission.  Thus, NGC 460 has a positive curvature in its millimeter spectrum and is not a natural blackbox at these frequencies.  

Two more of the sources are coincident with the molecular clouds [CB88] 28 \citep{Clemens88} and Dobashi 6193 \citep{Dobashi11}.  The positions of the sources at 100 and 217 GHz are separated by 7\farcm~for Dobashi 6193 and almost half a degree for [CB] 28, probably because the clouds are part of large dusty complexes on the sky.  Both dust complexes are obvious sources in AKARI images, and their obscuration is easily seen in DSS2 (Digitized Sky Survey) images.  \emph{Planck}'s photometry is unreliable at frequencies above 217 GHz in these regions, but the spectra of these sources appears to be roughly flat at 100 GHz for [CB] 28 and then rising for both nebulas. 

\begin{figure*}
\centerline{\includegraphics[width=9cm]{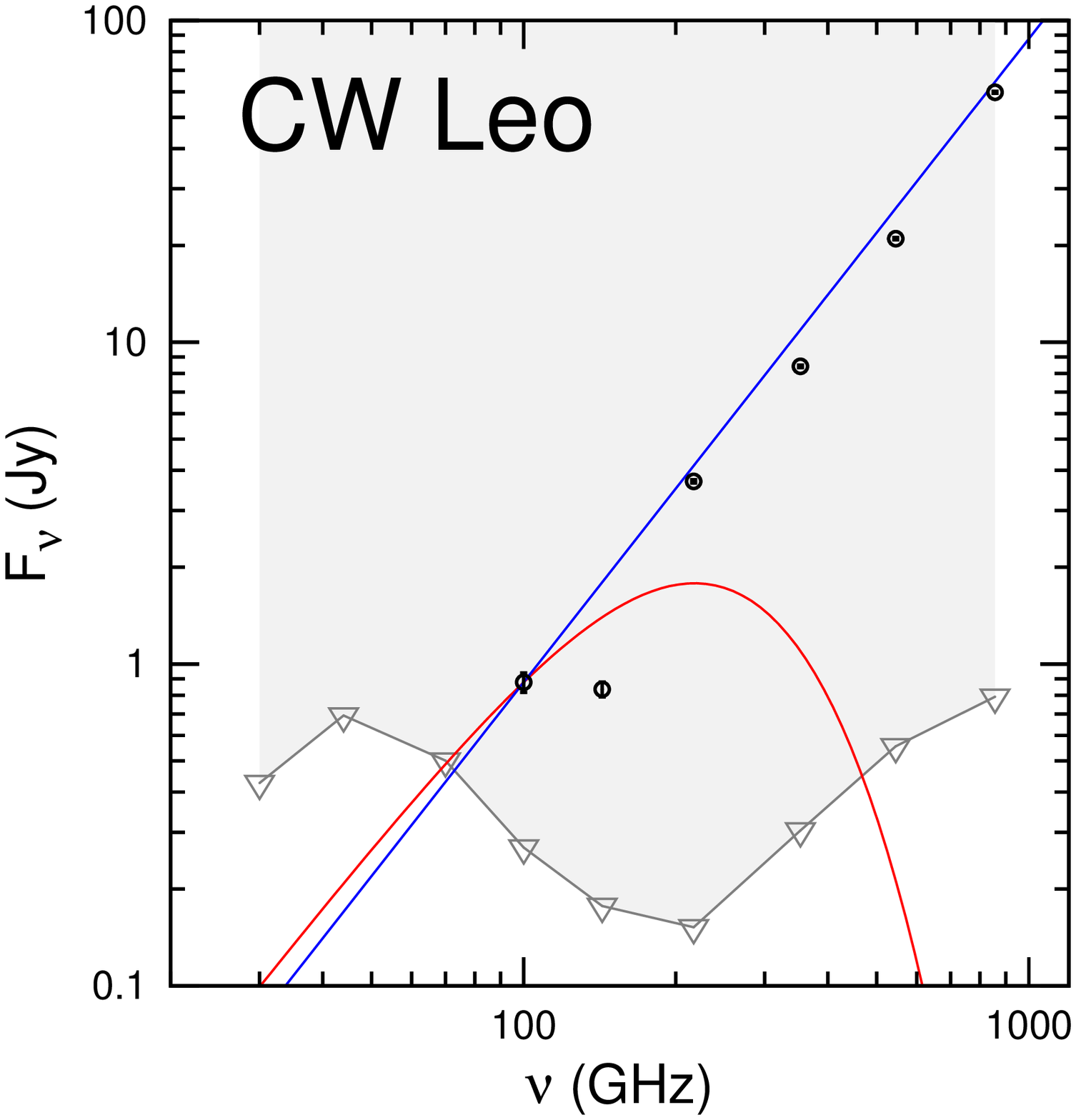}\includegraphics[width=9cm]{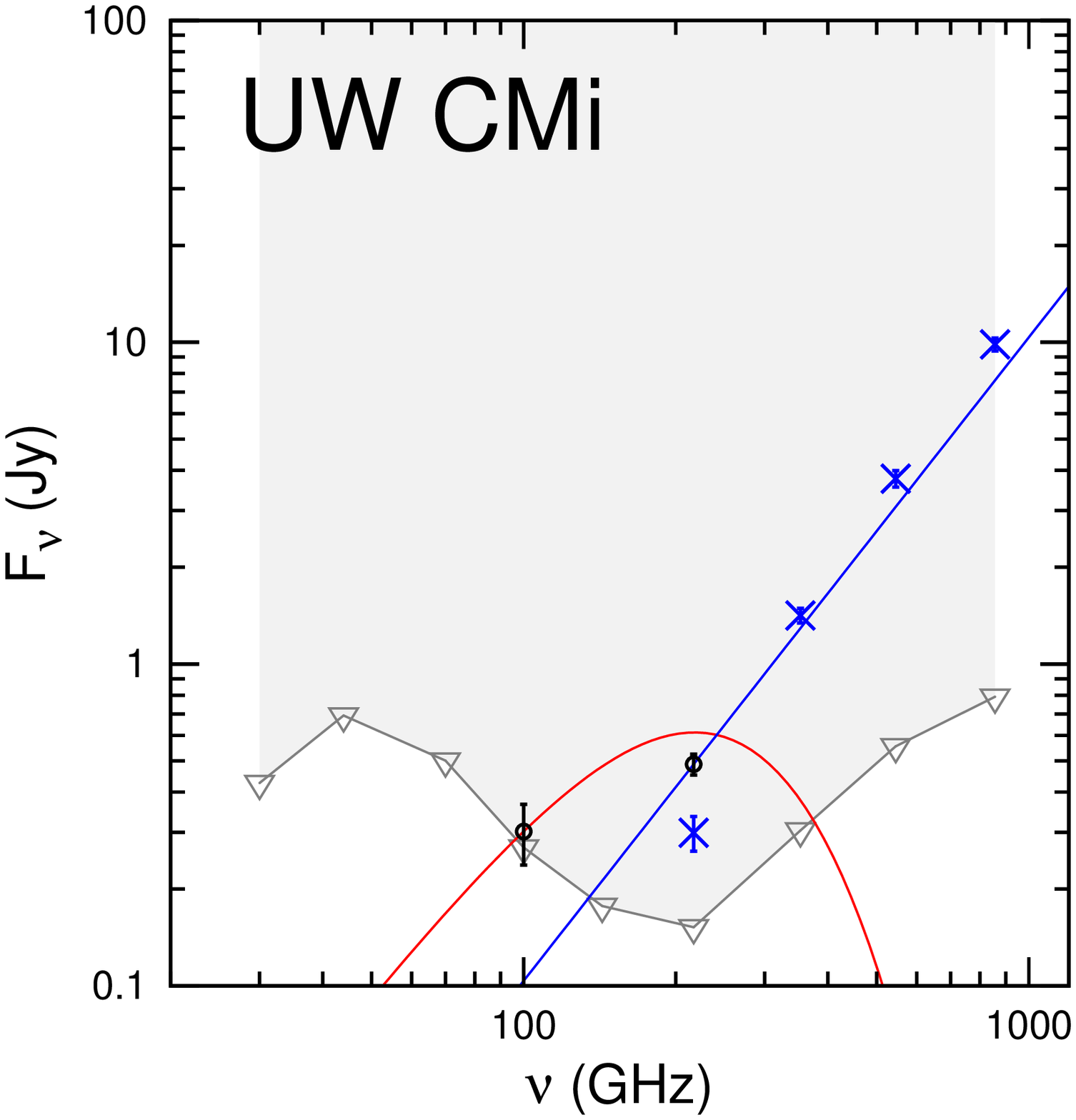}}
\figcaption{The \emph{Planck} PCCS2 spectra of CW Leo (left) and UW CMi (right).  The DETFLUX photometry for each source in the PCCS2 catalog is shown as the black data points.  At high frequencies, the photometry for UW CMi becomes unreliable because of the Galactic dust in the region; only the PCCS2E photometry is available, and it is plotted as the blue data points.  The PCCS2 sensitivity limits are plotted as the grey triangles; fluxes in the light grey shaded region should be detectable.  In both cases, the 217 GHz data points roughly lie on a Rayleigh-Jeans tail (blue lines).  Cold blackbox spectra normalized to the 100 GHz flux are plotted in red.  For UW CMi, a cold blackbox spectrum is inconsistent with the 143 GHz non-detection.\label{fig:CWLeoSpectrum}\label{fig:UWCMiSpectrum}}
\end{figure*}

The eighth source is the carbon star CW Leo (IRC+10216), one of the most brilliant stars in the sky in the MIR \citep{Becklin69}.  Dusty, massive circumstellar shells surround the star, effectively shrouding it in a natural stellar-scale blackbox \citep{Toombs72}.  The \emph{Planck} photometry indicates a Rayleigh-Jeans tail (Figure~\ref{fig:CWLeoSpectrum}).  That is in line with other measurements, spanning the radio to the MIR, although the spectrum may start to diverge from $\alpha = 2$ at millimeter and radio wavelengths as the shells become transparent \citep{Sahai89,Griffin90}.  CW Leo and similar stars are known to be potential false positives in searches for Dyson spheres \citep{Carrigan09}, much like how it turns up here in a blackbox search.

\begin{figure}
\centerline{\includegraphics[width=3cm]{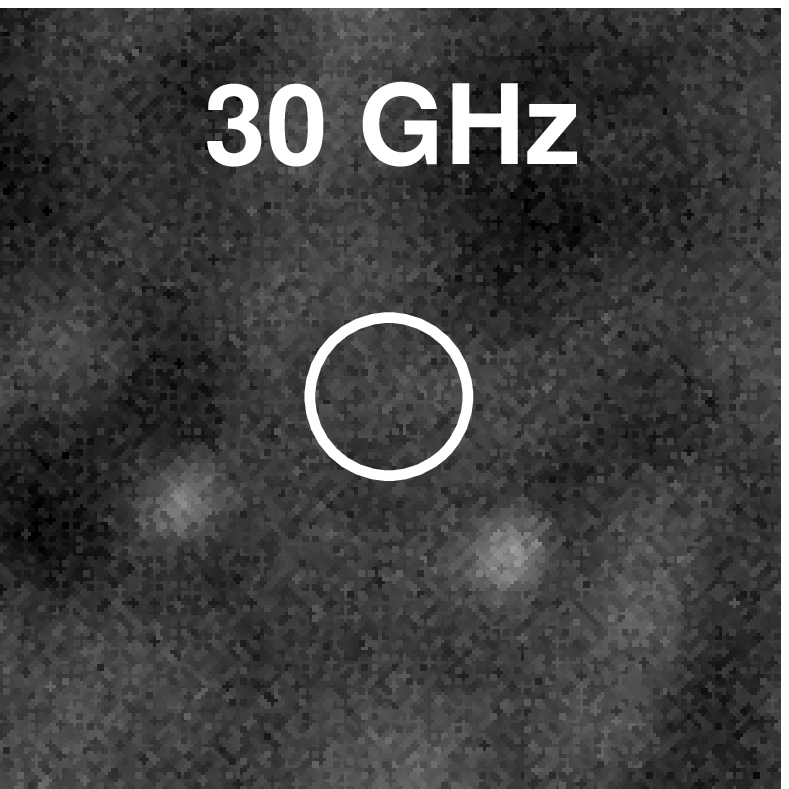}\includegraphics[width=3cm]{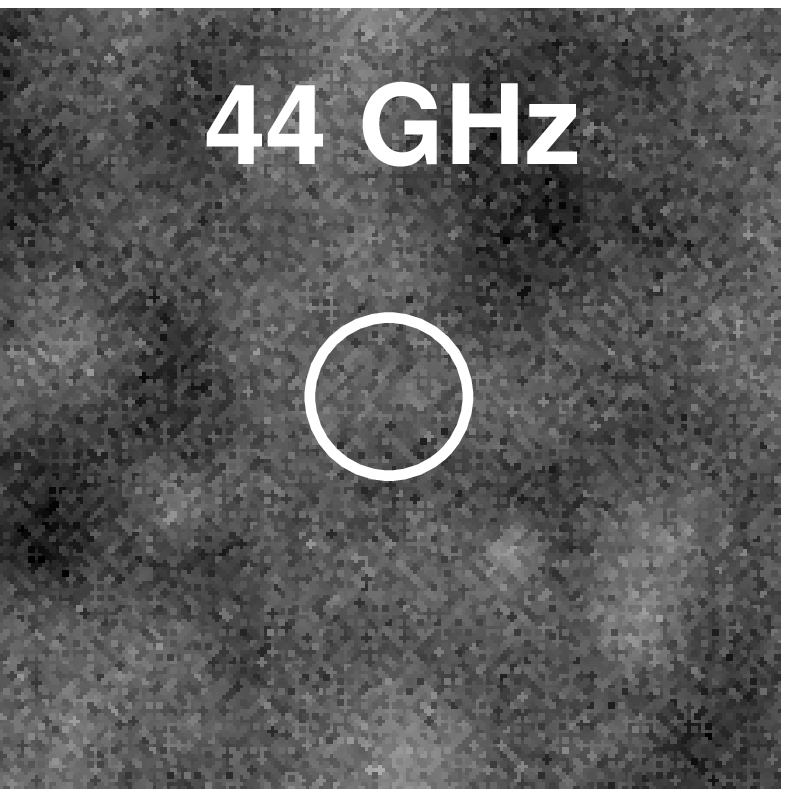}\includegraphics[width=3cm]{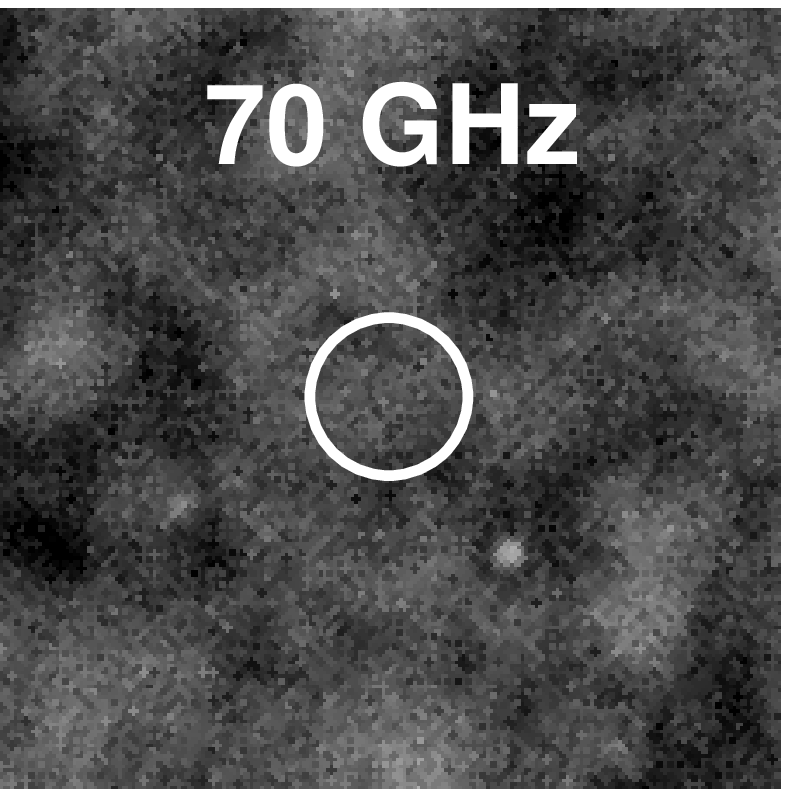}}
\centerline{\includegraphics[width=3cm]{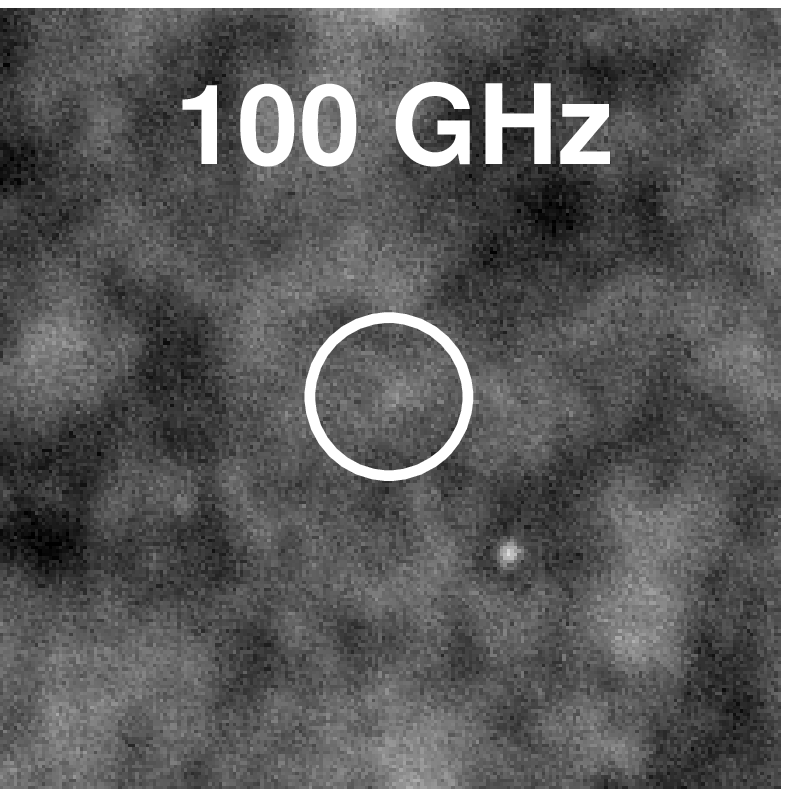}\includegraphics[width=3cm]{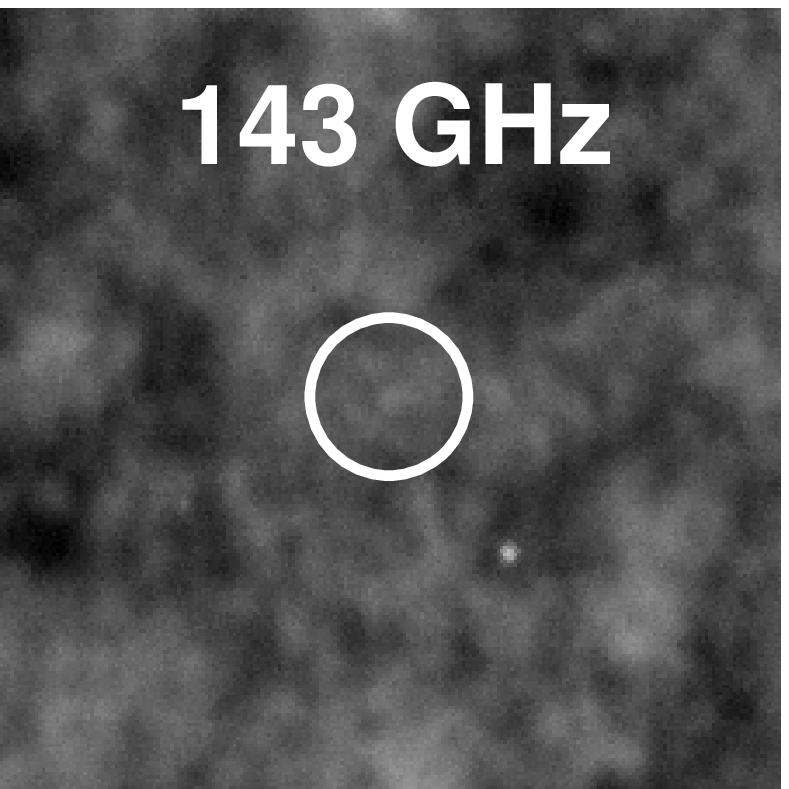}\includegraphics[width=3cm]{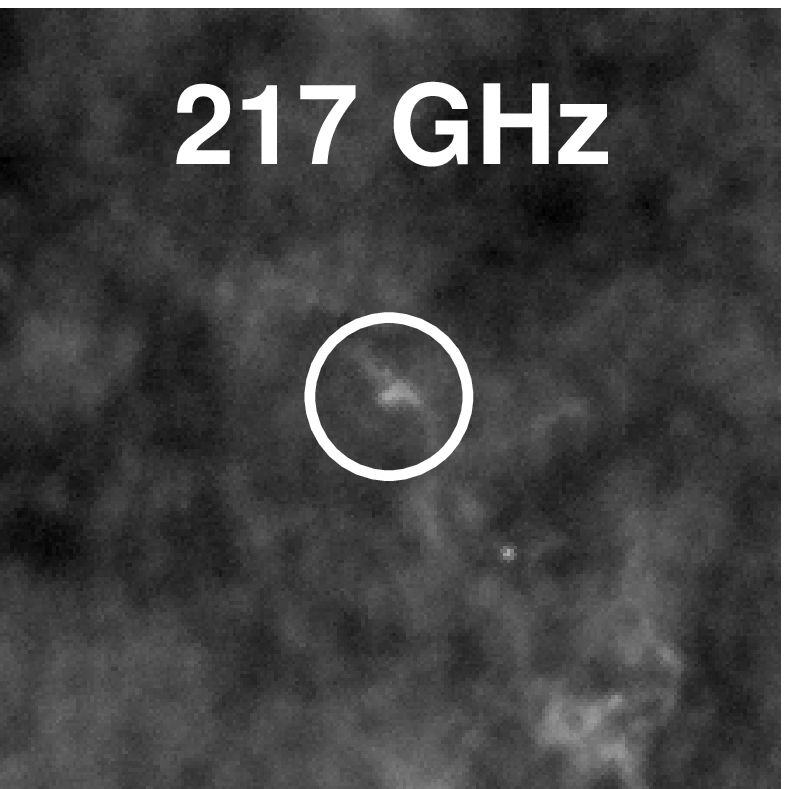}}
\centerline{\includegraphics[width=3cm]{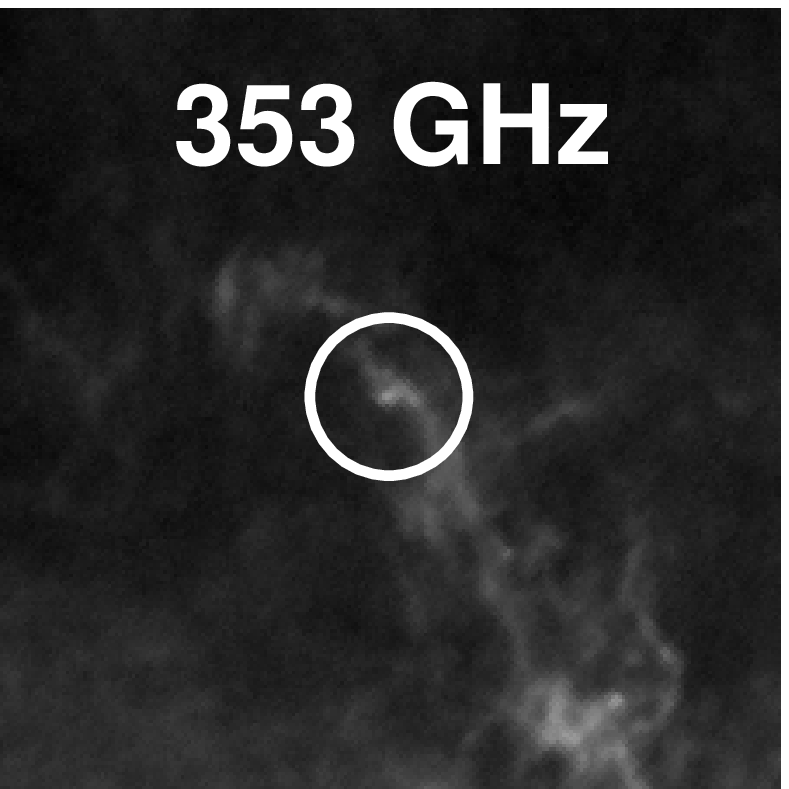}\includegraphics[width=3cm]{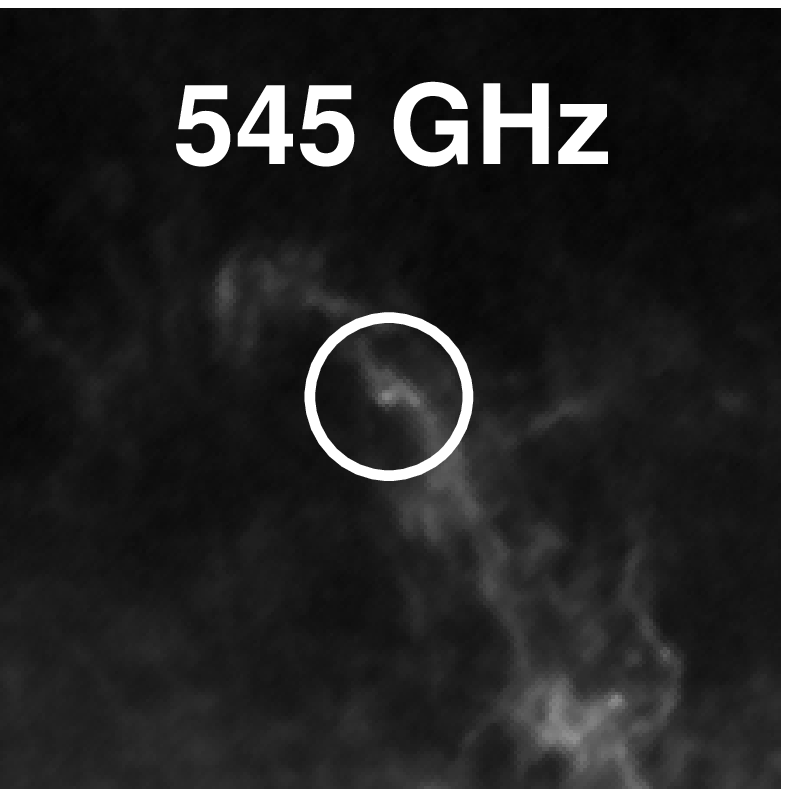}\includegraphics[width=3cm]{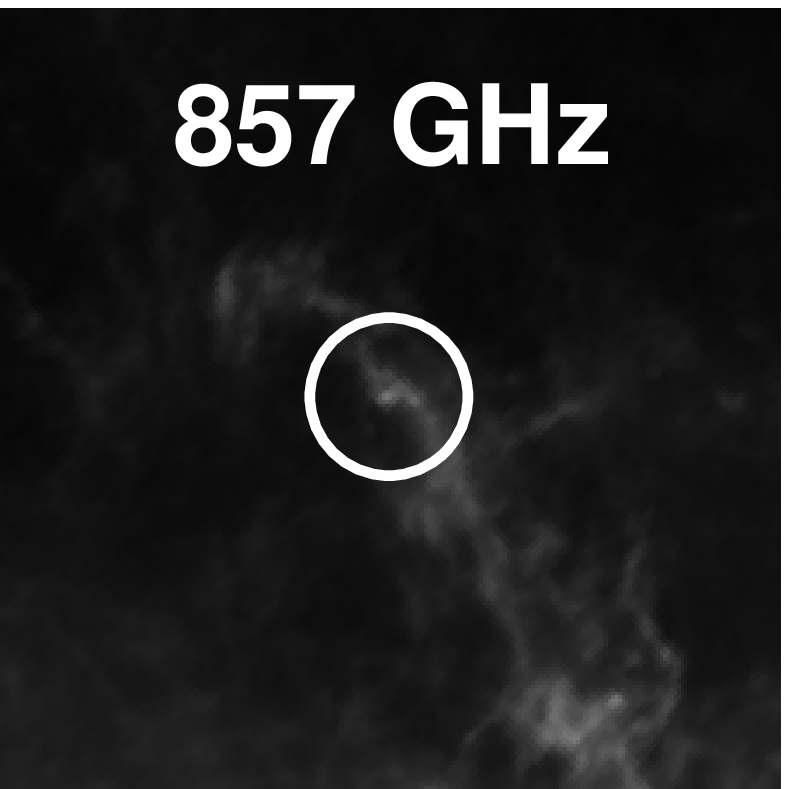}}
\figcaption{UW CMi as it appears in the \emph{Planck} maps. The white circles have radii of 1$^{\circ}$, centered on the 100 GHz source position.\label{fig:UWCMi}}
\end{figure}

The last source is harder to identify, but it is in the vicinity of the Mira variable star UW CMi \citep{Smak65} and a filament of Galactic dust.  The source is faintly detected at 100 GHz, but not in the more sensitive 143 GHz channel (Figure~\ref{fig:UWCMiSpectrum}).  Then, starting at 217 GHz, it has a Rayleigh-Jeans spectrum (blue line in Figure~\ref{fig:UWCMiSpectrum}), if the PCCS2E photometry is included.  On the \emph{Planck} maps, the source appears faint and unresolved at 100 GHz (Figure~\ref{fig:UWCMi}).  At 217 GHz and higher, it is clear that a Galactic dust filament runs through the region.  There is also what appears to be an unresolved point source (Figure~\ref{fig:UWCMi}).  Maybe the point source is just a knot or projection effect of the Galactic dust filament, but Mira variables like UW CMi occasionally are obscured by dusty shells, showing up as infrared and radio sources \citep{Bedijn87,Reid97}.  Something identified as UW CMi (=07451611+0110558, IRAS 07426+0118) is in fact included in the COBE (Cosmic Background Explorer) DIRBE (Diffuse Infrared Background Experiment) point source catalog, although the photometry is only reliable at 4.9$\ \um$ and bluer \citep{Smith04}.

Whatever this remaining source is, it is probably not a blackbox.  If it were, it should be brighter at 143 GHz than at 100 GHz.  This is seen from the red curve in Figure~\ref{fig:UWCMiSpectrum}, which is the cold blackbox flux spectrum normalized to run through the 100 GHz point.  A hot blackbox would be even brighter at 143 GHz than a cold blackbox.  The nature of the 100 GHz detection remains unclear, though.

\subsection{The significance of the null results}

My quick look through the \emph{Planck} source catalog revealed no blackboxes.  There is, of course, plenty of room for improving the methods I used, by investigating every source in the catalog at all frequencies and eliminating those with known counterparts.  The 100 GHz band is by no means the most sensitive of \emph{Planck}'s channels to blackboxes, as seen in Figure~\ref{fig:PlanckBandComparisons}.  But even this cursory search's null result is remarkable.  As seen in Table~\ref{table:FiguresOfMerit}, the 100 GHz band of \emph{Planck} has a reach that includes about a $\Gpc^3$ for Milky Way-containing blackboxes.

\begin{figure}
\centerline{\includegraphics[width=9cm]{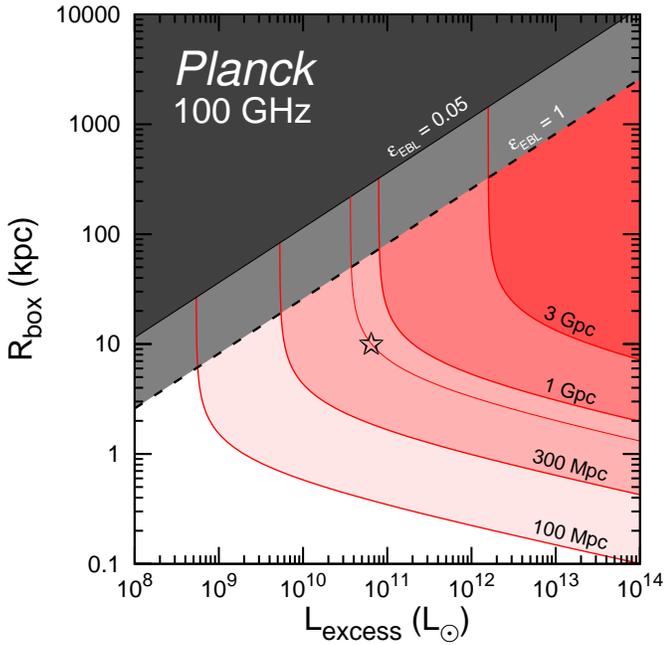}}
\figcaption{No blackboxes are found by examining the few sources with thermal spectral slopes between 100 GHz and 217 GHz in the \emph{Planck} Compact Source Catalog.  The resulting limits on the luminosities and radii of blackboxes, from \emph{Planck}'s 100 GHz source sensitivity, are plotted here.  The exclusion regions are shaded for comoving distances of 100 Mpc, 300 Mpc, 1 Gpc, and 3 Gpc in increasing darkness.  For comparison, a shrouded Milky Way is plotted as the star, with a luminosity given by \citet{Strong10}; the thin red line running through the star is \emph{Planck}'s 100 GHz sensitivity at 720 Mpc comoving distance.  Heating by the EBL sets a minimum luminosity for a blackbox at each radius, depending on $\varepsilon_{\rm EBL}$.  Luminosities smaller than this minimum luminosity are shaded medium grey ($\varepsilon_{\rm EBL} = 1$) and dark grey ($\varepsilon_{\rm EBL} = 0.05$).  I use the present-day EBL energy density from \citet{Finke10} on this plot.
\label{fig:Planck100GHzLimits}}
\end{figure}

A more precise bound depends on the size of the blackbox and its excess luminosity.  The parameter space ruled out by \emph{Planck}'s 100 GHz source catalog is shown in Figure~\ref{fig:Planck100GHzLimits}.  Big and bright blackboxes are ruled out.  The bound at a given distance asymptotes to a constant excess luminosity when the blackbox is big -- it generally does not matter whether it is 1 Mpc or 10 Mpc wide, if the excess luminosity remains constant.  The excess heating that is provided by the EBL sets a minimum luminosity, excluding the possibility of big but dim blackboxes.  Thus, \emph{Planck} is already effective at ruling out cluster-sized blackboxes, for example, even if the entities inside wish to conserve energy.  When the blackbox is small (and therefore hot), the bound has the asymptotic form of $L_{\rm excess} \propto R_{\rm box}^{-6}$ -- there is a weak dependence on the internal luminosity, but \emph{Planck} is not sensitive to blackboxes that are only a parsec wide even if $L_{\rm excess} \approx 10^{14}\ \Lsun$ that are a few hundred Mpc away.  

From Figure~\ref{fig:Planck100GHzLimits}, it is clear that my null result rules out much of the relevant parameter space for blackboxes out to Gpc distances.  Specifically, blackboxes that are brighter than a trillion $\Lsun$ are ruled out to a comoving distance of 1.5 Gpc ($z = 0.37$) as long as they are at least 10 kpc in radius.  Trillion $\Lsun$ luminosities are attained in ultra-luminous infrared galaxies, active galactic nuclei, and galaxy clusters.  Galaxy clusters shrouded in blackboxes are detectable out to much further still, 2.6 Gpc ($z = 0.71$) if $\varepsilon_{\rm EBL} = 0$ and 7.3 Gpc ($z = 3.9$) if $\varepsilon_{\rm EBL} = 0.05$.    But the lookback time at $z = 0.7$ includes almost half of the Universe's history.  Unless some aliens evolved early in the Universe's history, there are no such blackboxes anywhere in the observable Universe.  

The bounds on blackboxes with smaller luminosities more like those of present-day galaxies remain impressive.  When the excess luminosity is 100 billion $\Lsun$ and the radius is bigger than about 10 kpc, blackboxes should be detected by \emph{Planck}'s 100 GHz channel out to 820 Mpc ($z = 0.19$).  The Milky Way is slightly dimmer than this, with a total luminosity of around 65 billion $\Lsun$ \citep{Strong10}, but if it were shrouded in a blackbox, it would have been detected out 720 Mpc.  Even something only as bright as one of the Magellanic Clouds, with a billion $\Lsun$, would have been seen if it was within 140 Mpc.

\begin{figure}
\centerline{\includegraphics[width=9cm]{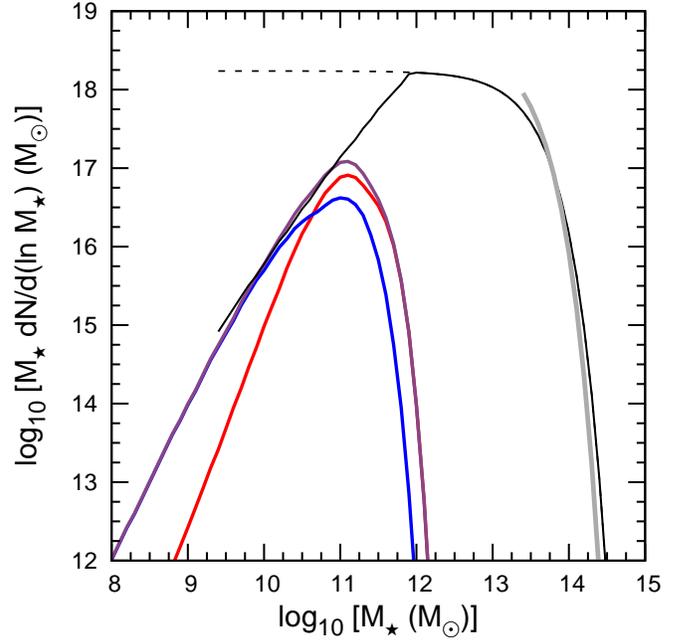}}
\figcaption{The stellar mass in galaxies and clusters that would have been present in the 100 GHz PCCS2 catalog if they were converted into blackboxes, shown as a function of the stellar mass of those galaxies and clusters.  This measures the power of the survey.  The distributions are shown for individual red and blue galaxies as red and blue lines, as according to \citet{Baldry12} and \citet{Muzzin13} and with $R_{\rm box} = 10\ \kpc$, with the total shown as the violet line.  The thick grey line shows the sensitivity to rich clusters with $R_{\rm box} = 3\ \Mpc$, using the luminosity distribution of \citet{Bahcall79} and any $\varepsilon_{\rm EBL}$.  Finally the black lines represent $R_{\rm box} = 3\ \Mpc$ blackboxes around any galaxy, group, or cluster, with $\varepsilon_{\rm EBL} =$~0 (solid) or 0.05 (dotted). \label{fig:ReachVsMStar}}
\end{figure}

I find the number of galaxies within the survey volume by integrating galaxy mass functions: 
\begin{equation}
\label{eqn:NGalaxyReach}
N_{\rm reach} = \int \frac{\Omega_{100-217} D_{\rm C,max}^3}{3} \frac{dn}{dM_{\star}} dM_{\star}.
\end{equation}
I used the mass function number densities $dn/dM_{\star}$ from \citet{Baldry12} for $z < 0.2$ and \citet{Muzzin13} for $z \ge 0.2$.  The maximum comoving distance a blackbox is detectable $D_{\rm C, max}$ depends on the internal galaxy luminosity.  I calculate these by assuming a bolometric stellar mass-to-luminosity ratio of 2 $\Msun/\Lsun$ for red galaxies (from the models of \citealt{Maraston98} and \citealt{Vazquez05} for a 10 Gyr old stellar population, adjusted for IMF) and 0.5 $\Msun/\Lsun$ for blue galaxies (the $B$-band mass-to-light ratio quoted in \citealt{Baldry12}; naively extrapolating the values given in \citealt{Leitherer99} for a continuously star-forming population to 10 Gyr gives $\sim$0.4~$\Msun/\Lsun$).  The blackbox radius is assumed to be 10 kpc. The PCCS2 only covers a part of the sky at high frequencies because of possible contamination from the emission of Galactic dust.  The $\Omega_{100-217}$ factor accounts for this; it is the solid angle covered in both the 100 GHz and 217 GHz catalogs.  Table~\ref{table:FiguresOfMerit} lists how much of the sky is included in each channel.    Since my search in Section~\ref{sec:Planck100and217GHzSearch} includes sources that are in both the 100 GHz and 217 GHz catalogs, the best possible solid angle covered by both is 65.9\% if the 100 GHz exclusion area is contained entirely within the 217 GHz exclusion area, while the worst possible solid angle is $100 - 15.0 - 34.1 = 50.9\%$.  It is clear from the maps in \citet{Ade16-PCCS2} that the optimistic case is more accurate, so I use 65.9\%.

A quantity more directly related to the number of inhabited planets is the stellar mass in these galaxies, since its unlikely that there are as many in small galaxies as in large galaxies:
\begin{equation}
\label{eqn:MStarReach}
M_{\rm reach} = \int \frac{\Omega_{100-217} D_{\rm C,max}^3}{3} M_{\star} \frac{dn}{dM_{\star}} dM_{\star}.
\end{equation}
\citet{Baldry12} uses a Chabrier initial mass function (IMF) for the stellar population, so that's what I use as well.  \citet{Zackrisson16} estimates that there are about 1.5 terrestrial planets per $\Msun$ of stellar mass in the Universe for this IMF, mostly around M dwarfs.  The number of terrestrial planets for F, G, and K class stars is much smaller at 0.064 per $\Msun$.  These rates allow me to estimate the number of terrestrial planets in the survey volume.  Finally, I can derive an effective volume of the survey, $V_{\rm eff} = M_{\rm reach} / \mean{\rho_{\star}}$, from the average cosmic density of stars at $z = 0$, $\mean{\rho_{\star}} = 2.3 \times 10^8\ \Msun\ \Mpc^{-3}$ \citep{Baldry12}.

From equation~\ref{eqn:NGalaxyReach}, I find that there are 3.0 million galaxies that would have been found in my search if they had been converted into blackboxes.  These galaxies contain $1.8 \times 10^{17}\ \Msun$ of stars.  The contributions of galaxies of different masses to this limit (but not accounting for the smaller sky coverage at 217 GHz) are shown in Figure~\ref{fig:ReachVsMStar}.  The search is biased towards galaxies with $\sim 10^{11}\ \Msun$ of stars.  This comes out to $2.6 \times 10^{17}$ terrestrial planets within the surveyed galaxies, with $1.1 \times 10^{16}$ terrestrial planets around F, G, and K stars.  I calculate an effective volume of 0.98 $\Gpc^3$.

There are a few factors that weaken these constraints when setting actual upper bounds, but the basic point still stands.  The PCCS2 is estimated to be 90\% complete.  For the sake of argument, suppose that the completeness $f_{\rm complete}$ in both the 100 GHz and 217 GHz channels together is $0.9^2 = 81\%$.  It is also possible that the null result is a fluke of Poisson statistics.  The upper bound on the abundance of blackboxes is then
\begin{equation}
\label{eqn:PoissonUpperLimit}
{\cal A}_{\rm blackbox} = \frac{\overline{N}}{f_{\rm complete} N_{\rm reach}},
\end{equation}
where $\overline{N}$ is the largest average number of detections compatible with a null result at the desired confidence level.  For a 95\% confidence level, $\overline{N} = 3.0$ \citep{Olive15}.  By combining the results of equation~\ref{eqn:NGalaxyReach} with equation~\ref{eqn:PoissonUpperLimit}, I find that my search sets a 95\% confidence level upper limit of $1.2 \times 10^{-6}$ on the frequency of blackboxes per galaxy, or less than one per 810,000 galaxies.  There are roughly $4.8 \times 10^{16}\ \Msun$ of stars expected in this many galaxies, with about $7.1 \times 10^{16}$ terrestrial planets, $3.0 \times 10^{15}$ of them around F, G, and K stars \citep{Baldry12,Zackrisson16}.  In other words, less than one in seventy quadrillion terrestrial planets has been the home of blackbox-builders.

\begin{figure}
\centerline{\includegraphics[width=9cm]{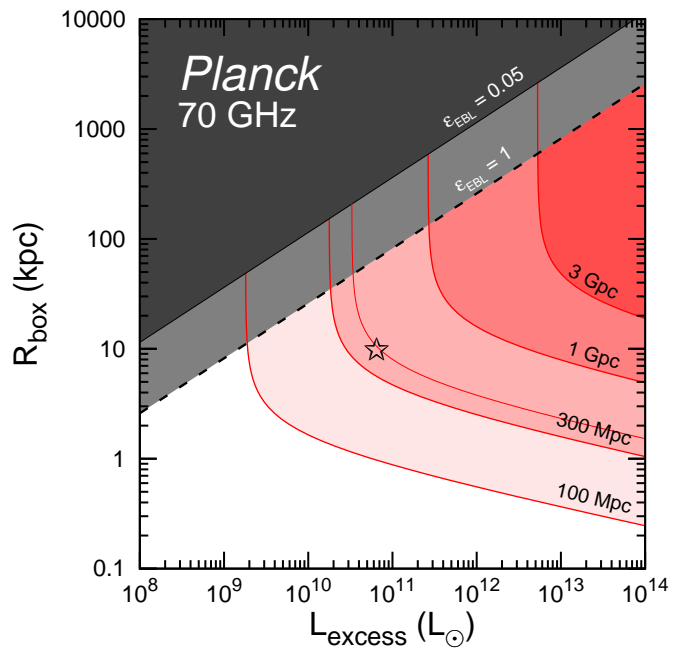}}
\figcaption{Same as Figure~\ref{fig:Planck100GHzLimits}, except now showing the weaker sensitivity from \emph{Planck} at 70 GHz. The thin red line now shows \emph{Planck}'s 70 GHz sensitivity at 400 Mpc comoving distance.\label{fig:Planck70GHzLimits}}
\end{figure}

A more conservative limit is derived using the null detections for the 70 and 143 GHz catalogs using the GAUSSFLUX (and DETFLUX) photometry (Figure~\ref{fig:Planck70GHzLimits}).  These limits require replacing $\Omega_{100-217}$ with $\Omega_{70-143}$, the fraction of the sky covered by the PCCS2 catalog at both 70 and 143 GHz, which is 85\%.  Then blackboxes are ruled out in 610,000 galaxies with $3.2 \times 10^{16}\ \Msun$ of stars and $4.7 \times 10^{16}$ terrestrial planets.  These figures lead to a 95\% confidence level upper limit of one blackbox per 166,000 galaxies that contain $8.5 \times 10^{15}\ \Msun$ of stars and $1.3 \times 10^{16}$ terrestrial planets.  If I am even more conservative and only consider the null detections for the 30 and 217 GHz pair using the APERFLUX photometry, then my limits are about an order of magnitude times weaker than these values.

These results are favorable compared to other searches for extragalactic cosmic engineering.  Most searches before \GHAT~only included about a hundred galaxies, thousands of times fewer than the cold blackbox limit.  \citet{Griffith15} estimated that there were roughly 100,000 resolved galaxies in \GHAT, out of a possible 230,000 sources not close to the Galactic plane.  This translates to a 95\% confidence level upper limit of about 1 Chilly Way per 33,000 galaxies that captures at least 85\% of the host's starlight.  The blackbox limit is twenty (four) times constraining for the DETFLUX (GAUSSFLUX) null results.  From the lack of extragalactic cosmic engineers in the Milky Way, \citet{Olson15-Clustering} constrained the abundance of expanding Type III+ societies to be $\la 1\ \Gpc^{-3}$, depending on how fast they expand and when they evolved.  Those values are in line with the null result presented here.

On a final note, the limits are far more powerful still on galaxy clusters shrouded in blackboxes (with $R_{\rm box} = 3\ \Mpc$).  These can be detected out to $z \ge 0.5$ in any of the low frequency channels of \emph{Planck}.  I estimate the number of galaxy clusters in the survey volume with the luminosity function fits of \citet{Bahcall79}.  Two cases are considered: rich galaxy clusters only with $L_{\rm int} \ge 10^{13}\ \Lsun$ using the fit for Abell clusters, and galaxy clusters and groups with $L_{\rm int} \ge 10^{12}\ \Lsun$ using the fit for single galaxies, groups, and clusters.  As with red galaxies, I assume that the mass-to-light ratio is 2~$\Msun/\Lsun$; although the luminosities are given for V-band, I assume that they are equal to the bolometric luminosity.  The 70 and 143 GHz pair of channels is actually more constraining than 100 and 217 GHz, since they cover more of the sky (cluster blackboxes are still easily visible at 70 GHz, as seen in Figure~\ref{fig:Planck70GHzLimits}).  Only clusters with $z \le 0.5$ are included, but otherwise I ignore evolution effects.

As seen in Figure~\ref{fig:ReachVsMStar}, more stellar mass resides in detectable galaxy clusters than in normal galaxies.  There are roughly 30,000 rich clusters with $7.9 \times 10^{17}\ \Msun$ of stars (95\% confidence level limits: 1 in 8,200 rich clusters; 1 in $2.1 \times 10^{17}\ \Msun$ of stars).  Naively applying the cosmic planet abundance rate of \citet{Zackrisson16}, there are 1.2 quintillion planets in these clusters, of which, 50 quadrillion orbit F, G, and K stars (95\% confidence level limits: 1 in 320 quadrillion terrestrial planets, 14 quadrillion around F, G, and K stars).  

From the \citet{Bahcall79} luminosity function, there are about 700,000 galaxy cluster and groups with $L_{\rm int} \ge 10^{12}\ \Lsun$ out to $z \ge 0.5$, almost all of which could have been detected in this survey.  There is some weak dependence on $\varepsilon_{\rm EBL}$, which affects the limits below 70 GHz; a 3 Mpc wide blackbox would be visible simply by re-radiating 5\% of the incident EBL.  In any case, there are about $3 \times 10^{18}\ \Msun$ of stars with 5 quintillion planets (95\% confidence level limits: 1 in 200,000 clusters and groups, $9 \times 10^{17}\ \Msun$ of stars, and 1.4 quintillion planets).  

About 1 in 5 stars at $z \approx 0$ resides in a galaxy cluster\footnote{I derive this from the cosmic baryon density, derived from \citet{Ade15-Params}, a fraction of 4\% of baryons in ICM according to \citet{Shull12}, and an approximate mass ratio of 5 to 1 of ICM to stars in clusters with a few $10^{14}\ \Msun$ of total mass \citep{Gonzalez13}.  The resulting average cluster stellar density is compared to the total value from \citet{Baldry12}.  The \citet{Bahcall79} luminosity functions imply that $\sim 10\%$ of stars reside in rich clusters with $L \ge 10^{13}\ \Lsun$, but $\sim 60\%$ are in groups and clusters with $L \ge 10^{12}\ \Lsun$.}; that is the appeal of limits on shrouded clusters.  On the other hand, the amount of materials required may be prohibitive according to Section~\ref{sec:MetalDipoles}.  Collecting the necessary materials from the ICM may take much longer than the age of the Universe too (Appendix~\ref{sec:Manufacture}).  Furthermore, it is possible that some constraints prevent intelligence from evolving in cluster environments, in analogy with the suggestion of a ``galactic habitable zone'' \citep{Gonzalez01}.  I therefore do not dwell on them too much, but their habitability and cosmic engineering potential may be worth exploring in more detail.

\subsection{How to sharpen the null result with extant data}

And yet the 100 GHz channel of \emph{Planck} is not the most sensitive to blackboxes by any means (Figure~\ref{fig:PlanckBandComparisons}).

\begin{figure}
\centerline{\includegraphics[width=9cm]{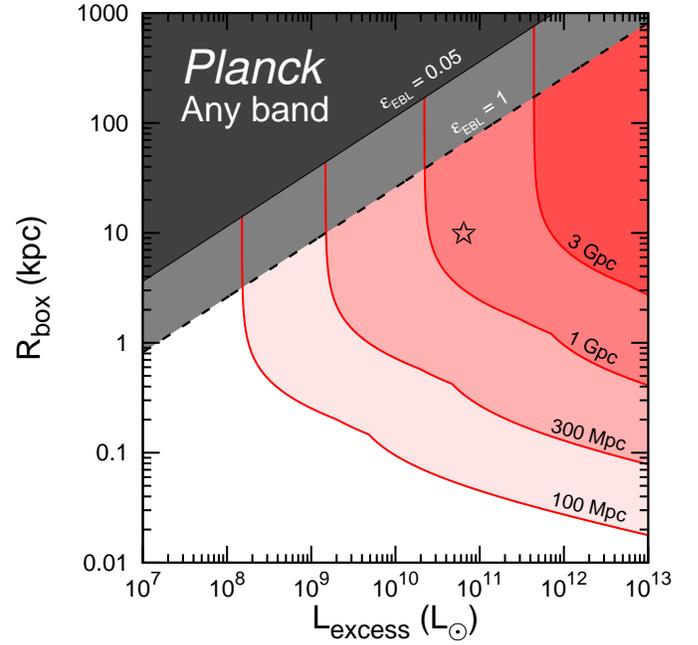}}
\figcaption{Same as Figure~\ref{fig:Planck100GHzLimits}, but now including constraints from all frequencies, assuming there are no blackboxes in the PCCS2 at any frequency.  Note the shifted values along the axes.
\label{fig:PlanckAllBandLimits}}
\end{figure}

Tighter limits are possible if we assume that none of the PCCS2 sources in any channel are blackboxes.  Nothing like a blackbox has ever been reported by \emph{Planck} or anyone else, so they are probably very rare, but to be sure each source should be identified and/or examined.  Figure~\ref{fig:PlanckAllBandLimits} shows the final sensitivity of \emph{Planck} that is possible if there are no blackboxes in the PCCS2 catalogs.  The PCCS2 should include any blackboxes with excess luminosity of at least 500 billion $\Lsun$ and radius greater than 10 kpc out to a comoving distance of 2.8 Gpc ($z = 0.76$).  Shrouded Milky Ways are detected out to 1.4 Gpc ($z = 0.35$).  Even a shrouded Magellanic Cloud-like galaxy with $L_{\rm int} = 10^9\ \Lsun$ is visible out to 250 Mpc.  Blackboxes larger than 300 kpc in radius are visible just from the EBL heating alone.

Using equation~\ref{eqn:NGalaxyReach}, I find the reach of the 217 GHz PCCS2 catalog is about 27 million galaxies including $1.8 \times 10^{18}\ \Msun$ of stars with about $3 \times 10^{18}$ terrestrial planets ($1.2 \times 10^{17}$ of them not orbiting M dwarfs).  Even taking into account the possibility of statistical flukes, the PCCS2 is capable of setting an upper limit of about 1 blackbox per 7 million galaxies and per $5 \times 10^{17}\ \Msun$ at 95\% confidence level.  There are essentially no improvements for shrouded galaxy clusters at $z \le 0.5$, because they would already have been detected at 70 to 143 GHz.

\begin{figure}
\centerline{\includegraphics[width=9cm]{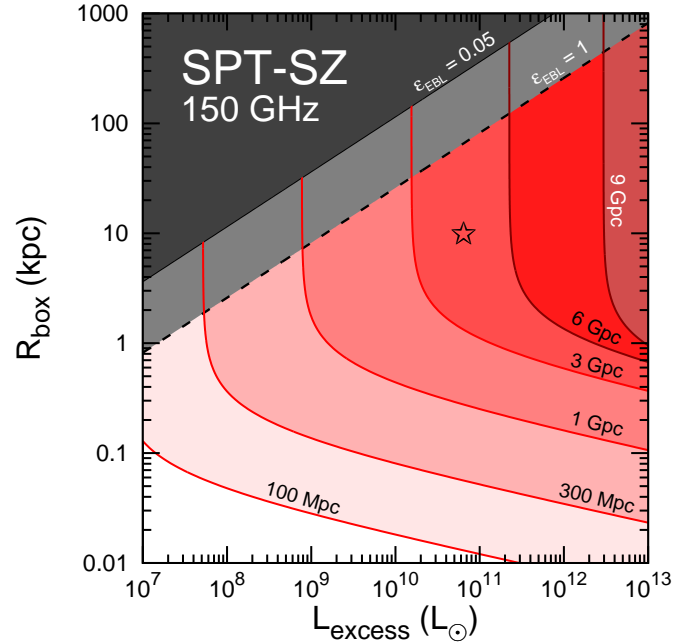}}
\figcaption{Same as Figure~\ref{fig:Planck100GHzLimits}, but for the SPT-SZ survey at 150 GHz, assuming there are no blackboxes in its catalogs.  The deeper shaded regions are for comoving distances of 6 Gpc and 9 Gpc.  Note the shifted values along the axes.
\label{fig:SPTSZ150GHzLimits}}
\end{figure}

Whereas \emph{Planck} covers most of the sky to low sensitivity, the SPT-SZ survey reaches much higher sensitivity over about two percent of the sky (Table~\ref{table:FiguresOfMerit}).  Within that sky region, its reach is truly enormous.  This is seen in Figure~\ref{fig:SPTSZ150GHzLimits}, where the brightest blackboxes with $10^{13}\ \Lsun$ are detected beyond 10 Gpc of comoving distance -- at redshift 13, when the Universe was just 300 million years old.  A trillion $\Lsun$ blackbox is detectable out to 8 Gpc of comoving distance ($z = 4.7$, at a cosmic age of 1.3 Gyr).  The range for shrouded Milky Ways is 4.5 Gpc ($z = 1.5$).  SPT-SZ can see a shrouded Magellanic Cloud with $L_{\rm blackbox} \approx 10^9\ \Lsun$ out to 1 Gpc.  For most of these galaxy types, the limiting factor is no longer the distance but the time it takes for starfaring entities to evolve.  Further improvement on SPT-SZ for median-sized shrouded galaxies is better done with a larger sky area rather than a fainter flux sensitivity.  

If I ignore the cosmological evolution of galaxies and starfarers, I find that the deeper flux limits roughly balances the smaller sky area compared to \emph{Planck}.  The 150 GHz band of the SPT-SZ survey can sample about 22 million galaxies with 0.8 quintillion $\Msun$ and 1.2 quintillion terrestrial planets.  

Together, these two surveys have the capability to set incredible limits on cosmic engineers.  Most of the galaxies in the SPT-SZ survey's reach are further away than those that are within the reach of PCCS2, so if nothing is found in either, that rules out blackboxes in about 50 million galaxies with about 4 quintillion terrestrial planets (150 quadrillion not orbiting M dwarfs).  The combined effective volume is around $10\ \Gpc^3$.  In other words, \emph{Planck} and the SPT can sample about 6\% of the galaxies in the later half of the observable Universe's history.

\section{An Argument for Cosmic Pessimism}
There have been a few attempts to reconcile the Fermi-Hart argument with the possibility of prevalent, long-lived starfarers.  Following \citet{Grinspoon13}, these scenarios might be called ``Cosmic Optimism''.  In this section, I interpret the null results for the blackbox search as evidence for the opposite scenario, Cosmic Pessimism.  That is, there are no aliens in the local Universe, and we (and other societies) have a large chance of destroying ourselves, although starfaring is technologically possible.  As with all interpretations of SETI results, support for Cosmic Pessimism involves a lot of subjective judgments and so should be regarded as suggestive rather than definitive.

\subsection{We are (probably) alone}
\label{sec:Alone}
All useful searches for Type III societies are able to rule out trillions of planets as their potential birth sites.  Waste heat is generally believed to be the most robust sign of massive engineering, as it is a consequence of thermodynamics, and thermodynamics suggests that the optimal temperature for many projects is around 3 K, the coldest possible without exotic physics.  My search in the \emph{Planck} source catalog for galactic-sized structures with 3 K waste heat instead rules out their existence among millions of galaxies.  Furthermore, rich galaxy clusters shrouded in blackboxes simply do not exist within the observable Universe.  

Even if one does not believe that Type III societies are motivated by thermodynamics, all other searches have found nothing.  Perhaps waste heat from Type III societies is warmer, but \GHAT~found no sign of them among about a hundred thousand galaxies in the WISE catalog.  Or perhaps we could abandon the waste heat idea entirely.  But as I pointed out in \citet{Lacki15}, there are no signs of non-thermal emission from Type III societies either.  The quietness of the cosmic radio background, for example, implies that less than $10^{-5}$ of the Universe's luminosity is converted into radio broadcasts.  Similar constraints exist for gamma rays, neutrinos, low frequency gravity waves, and the highest energy cosmic rays.  Nor are there signs of expanding, intergalactic Type III societies, which would have spread to the Milky Way by now if their abundances were much greater than my constraints \citep{Olson15-Clustering}.  Finally, one could look for signs that they are moving stars and interstellar matter \citep{Voros14}, or otherwise altering their galaxies' appearance \citep{Carrigan12,Wright14-SF}.  To my knowledge, there has not been a survey reported in the literature for artificial morphologies among many galaxies.  But thousands of people working for the Galaxy Zoo program have examined the shapes of nearly a million galaxies and classified them.  Users have identified new and unusual objects, like Hanny's Voorwerp and Green Pea galaxies, but nothing artificial has been reported \citep{Lintott11}.  Arguably, the very fact that we are able to find empirical relations that apply to galaxies throughout the Universe hints that they are mostly governed by the same non-artificial processes.  These results all imply that Type III societies are not present in the observable Universe.

If Type III societies do not exist, does that mean that societies like our own are rare?  My own take is a provisional, cautious yes: we may in fact be alone in the observable Universe.\footnote{If we live in an infinite Universe, which at least seems plausible, then aliens almost certainly exist somewhere or somewhen, because we exist, implying the probability of technological societies emerging on any given planet is nonzero (as noted in \citealt{Wesson90}).  When I say that we may be alone, I am referring to the observable Universe only.}

The implications of the null result can be interpreted in the form of an altered Drake equation, which relates the number of societies with the number of planets, the chances of a society arising, and their lifespans:
\begin{equation}
\label{eqn:Type3Drake}
\frac{N_{\rm III}}{(\rho_{\rm SFR} V_{\rm eff} / \mean{m_{\star}}) n_p f_h} = f_l f_i f_c f_{\rm III} {\cal L}_{\rm III}.
\end{equation}
A survey finds $N_{\rm III}$ Type III societies in a volume $V_{\rm eff}$ that includes many galaxies.  The first term that limits this number is the rate that stars are formed in the volume, which can be found from the comoving cosmic star-formation rate $\rho_{\rm SFR}$, and the typical mass of a star, $m_{\star}$, which is about $0.35\ \Msun$ for the Chabrier IMF used by \citet{Baldry12} \citep{Draine11}.  Further limiting factors are the number of planets per star ($n_p$), the fraction of those planets that are habitable ($f_h$), the probability that life appears on one of the habitable planets ($f_l$), the probability that life evolves intelligent species ($f_i$), and the probability that intelligence develops technology comparable to our own and possibly capable of interstellar communication ($f_c$).  I have multiplied another factor $f_{\rm III}$, the fraction of technological species that go on to become Type III societies.  The limiting terms together multiply out to the rate at which Type III societies appear; at any given time, $N_{\rm III}$ is found by multiplying it with ${\cal L}_{\rm III}$, the lifespan of the evidence for Type III societies or the age of the Universe, whichever is shorter.  I have written the equation this way to emphasize that the null observation is being used to constrain the unknown factors on the right, rather than attempting to come up with $N_{\rm III}$ by estimating all of the factors.

Searches for the artifacts of starfarers are powerful because the relevant lifetime is not the lifetime of the society, but the lifetime of the artifacts \citep{Carrigan12}.  For example, many of the surfaces in the Solar System are billions of years old, and they all appear completely pristine.  Some structures, like abandoned cities or strip mines, would be visible even after all of this time.  The lack of these structures says something not just about the number of societies in the Galaxy today, but all societies that have ever lived in the past 4 Gyr.  If $f_l$, $f_i$, and $f_c$ are all around 0.1, this amounts to about a hundred million societies in the Milky Way alone \citep[c.f.,][]{Freeman75}.  One reason Fermi's Paradox is so powerful is that it bypasses ${\cal L}$ in a way that searches for living societies don't (but at the added cost of an unknown probability that societies become starfaring).  

Galaxy-spanning Type III societies have a minimum lifespan of 100 kyr because of causality.  In principle, they could survive for billions of years.  Effective immortality may not be inevitable, but a blackbox may outlive its creators as a functioning ecosystem.  Even without any kind of self-repair, the natural survival time of a dipole cloud with $\tau_{\rm abs} = 1$ is a fraction of a percent of the galaxy's or cluster's dynamical time: a few hundred kyr for a galaxy, and a few hundred Myr for a cluster (Appendix~\ref{sec:Collisions}).  Thus, the null result implies 
\begin{multline}
\label{eqn:DrakeResults}
f_l f_i f_c f_{\rm III} \la 7 \times 10^{-13} N_{\rm III} \left(\frac{V_{\rm eff}}{\Gpc^3}\right) \left(\frac{{\cal L}_{\rm III}}{\Myr}\right)^{-1} \left(\frac{n_p f_h}{0.01}\right)^{-1} \\
\times \left(\frac{\rho_{\rm SFR}}{0.05\ \Msun\ \yr^{-1}\ \Mpc^{-3}}\right)^{-1} \left(\frac{\mean{m_{\star}}}{0.35\ \Msun}\right).
\end{multline}
I have used a cosmic star-formation rate that was characteristic of $z \sim 0.5$, when the Solar System formed \citep{Hopkins06}.  The value of $n_p f_h$ used is motivated by the planet abundances around F, G, and K stars estimated by \citet{Zackrisson16}.

This confirms the need for some kind of ``Great Filter'', some process that sets at least one of the remaining factors in equation~\ref{eqn:DrakeResults} to nearly zero \citep{Hanson98,Bostrom02}.  Without any further observational evidence, we have no reason to conclude that $f_{\rm III}$ must be zero.  By symmetry, we might come up with a Bayesian prior that says it's equally likely for any one of the four factors to be zero while the rest are one, for example.  

This kind of argument depends on the grouping of the parameters; it matters whether $f_{\rm III}$ represents one threshold or many.  There are similar arguments about $f_{\rm before}$.\footnote{\citet{Lineweaver02} interpreted the development of life soon after the Earth's formation to mean that $f_l \approx 1$, but this has been disputed \citep{Brewer08,Spiegel12}.  \citet{Carter83} argued for one strong filter step based on the apparent coincidence between the time it took for us to appear and the Sun's lifetime, but later revised it to five or six as the estimated habitable lifetime of the Earth dropped \citep{Carter08}.  Against the estimates of Carter, \citet{Livio99} disputed that humanity's evolution timescale and the Sun's lifetime are independent, while \citet{Cirkovic09-Reset} argued that short evolutionary timescales are possible because catastrophes can reset any evolutionary development.}  Without good information on the number of possible filters, I suggest, as a very simple parameterization, we group them into $f_{\rm before} = f_l f_i f_c$ and $f_{\rm after} = f_{\rm III}$.  We have no idea what the ratio $f_{\rm after}/f_{\rm before}$ is, but for a symmetrical prior, there's an equal chance that it is greater than 1 rather than less than 1.  When the ratio is 1, $f_{\rm after} = f_{\rm before} \la 8 \times 10^{-7} (n_p f_h/0.01)^{-1/2}$.  This median value implies that the number of societies in our galaxy comparable to our own, $N_{\rm MW} = {\rm SFR}_{\rm MW} / \mean{m_{\star}} \times n_p f_e f_{\rm before} {\cal L}$, is
\begin{multline}
\label{eqn:MWDrake}
N_{\rm MW} \la 16 \left(\frac{\rm SFR}{2\ \Msun\ \yr^{-1}}\right) \left(\frac{\mean{m_{\star}}}{0.35\ \Msun}\right)^{-1} \left(\frac{n_p f_e}{0.01}\right)^{1/2} \\
\times \left(\frac{f_{\rm before}}{8 \times 10^{-7}}\right) \left(\frac{\cal L}{\Gyr}\right).
\end{multline}
Unless we have good reason to suspect $f_{\rm after} \ll f_{\rm before}$, the lack of Type III societies suggests that at most a few technological societies have ever arisen in the Milky Way.  SETI efforts focusing on nearby stars are probably not going to find aliens, especially since they may broadcast recognizable signals for only a few centuries before dying or going on to other things.  The power of these null results and Fermi's Paradox in general is that they actually use the largeness of the Universe as an argument against the prevalence of aliens.

\emph{Reasons for a relatively high $f_{\rm after}$} -- I think there are plausible motives for cosmic engineering, suggesting a relatively large $f_{\rm after}$.  Aside from sheer expansionism, some scientific experiments might require cosmic engineering to be carried out, in particular building particle accelerators to test Planck-scale physics \citep{Lacki15}.  Although very esoteric now, Planck-scale physics could determine the ultimate limits of technology, like whether traversable wormholes can be built \citep[e.g.,][]{Morris88}, which might become an especially pressing issue as the cosmic expansion accelerates and most of the Universe falls out of contact.

The accelerating cosmic expansion provides another motivation for cosmic engineering: the Universe as we know it is dying.  Already galaxies that were once in contact have receded back over the horizon.  If they wish to exchange any messages or visit each other (barring wormholes or other unlikely possibilities), this is the last chance they have \citep{Heyl05,Krauss07}.  Likewise, galaxies are fed by intergalactic gas.  As the expansion accelerates, the galaxies will be starved, with most of the baryonic matter in the Universe simply diluting away \citep{Nagamine03}.  Already, the cosmic star-formation rate per comoving volume has fallen by an order of magnitude from its peak at $z \ga 1$ \citep{Hopkins06}.  In the long run, the amount of resources that they will have for the eternal winter ahead depends on how much they gather now.  Intelligence is out of time.

I described how blackboxes might be ecosystems in Section~\ref{sec:GeniusLoci}, evoking the Gaia hypothesis for Earth's ecosystems, in which biological feedback processes stabilize the Earth's environment on geological timescales.  One of the original motivations for the Gaia hypothesis was the Earth's climatic stability as the Sun's luminosity increased over billions of years \citep{Lovelock74}.  Like Earth, galaxies are dependent on the external cosmos, for gas instead of energy.  Natural galaxies are poor at self-regulation, in that they tend to grow over about a hundred dynamical times before quenching star-formation.  A blackbox might be built as a way of creating a galactic-scale Gaia, to manage the flow of external gas at an optimal rate for the galaxy's present and future.   

The suggestion of blackboxes as the ``optimal'' form of a Type III society may seem to be ascribing identical motives to all societies.  I think this apparent homogeneity is illusory.  A high entropy macrostate actually has the widest variety of microstates.  From the point of view of fundamental physics, two microstates in a high entropy macrostate are as different from each other as they are from a microstate in a low entropy macrostate.  The difference lies in what we can observe and know -- entropy measures our ignorance more than their similarities.

It's well known that optimally compressed information appears indistinguishable from white noise \citep{Pierce60}.  But to the members of the society, the ``noise'' can be resolved into meaning.  Thermal blackbody radiation is simply white noise at low frequencies.  If we knew the story of each photon a blackbox generates, what made them and who, if we could give them names, the thermal emission would tell rich stories that would not at all seem alike.

My own guess is that the motives of aliens would seem largely random to us.  The question is whether this randomness in goals corresponds to thermodynamic randomness.  Landauer's Principle suggests there is a connection.  Although framed in terms of computation, itself usually thought of as a numerical task, computation can be interpreted as dynamics \citep[e.g.,][]{Margolus98}.  Any process with a rhythm is bounded by Landauer's Principle, if it responds to its environment.  There are more possible high-entropy rhythms than low entropy ones.  Even if aliens do not care about maximizing choice or diversity for their own sakes, what they do care about doing is more likely to produce a lot of entropy than a little.

\emph{What are the remaining holes in the argument?} --  The title of this paper is a reference to the provocative \citet{Tipler80}, which argued that aliens would launch self-replicating probes that would soon visit and colonize every star in the Galaxy, and that the lack of such outposts in the Solar System rules out the existence of aliens in the Milky Way.  Although he didn't directly apply the argument to other galaxies, Tipler would later predict that humanity itself would sweep through the cosmos, and that we are actually alone in the entire Universe, even asserting that SETI was useless on this basis \citep[e.g.,][]{Tipler95}.  \citet{Sagan83} rejected this argument, saying that replicator probes would soon devour whole sections of the Universe if they existed, and thus would be banned as an obvious threat to all intelligent life.  In many ways, this paper does follow Tipler's reasoning: it works off the premise that at least some societies do spread across their galaxy (at least) and engineer it, with the lack of visible engineering implying that such societies are rare.

Still, I think both \citet{Tipler80} and \citet{Sagan83} were too quick to write off the existence of cosmic engineers.  For all we knew in 1980, the Universe really could have been filled with blackboxes and Chilly Ways.  The lack of such engineering in the Milky Way itself was sufficient evidence that Type III societies were fairly rare \citep[c.f.,][]{Olson15-Clustering}, but didn't prove their nonexistence.  In the decades that have followed, infrared and microwave surveys have become enormously more sensitive.  But is it still too soon to conclude that we are alone?  Are there possible Type III structures that we would not have detected yet?

\begin{figure}
\centerline{\includegraphics[width=9cm]{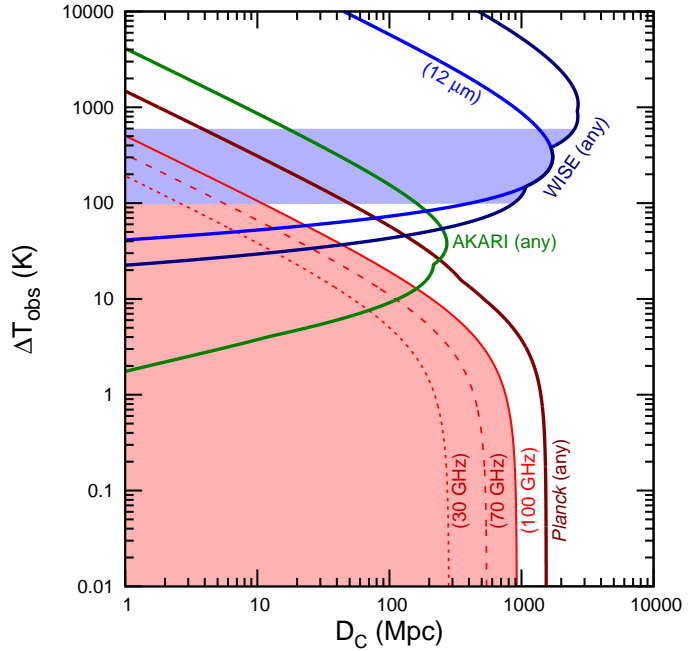}}
\figcaption{Maximum distance a shrouded Milky Way ($L_{\rm excess} = L_{\rm MW}$, $R_{\rm box} = 10\ \kpc$) of various excess temperatures would appear in some band of \emph{Planck} (red), AKARI (green), or WISE (blue).  This paper (based on the 100 GHz survey) excludes the pink region.  \GHAT~looked for Chilly Ways with temperatures of 100 to 600 K, which is only a subset of the parameter space that WISE can exclude (blue shaded region).
\label{fig:TSensitivity}}
\end{figure}

The first possible loophole is emitting temperature.  This paper sets stringent limits on blackboxes with excess temperatures of less than a few K (pink region in Figure~\ref{fig:TSensitivity}).  \GHAT~excludes Chilly Ways with temperatures of 100 to 600 K (within the blue shaded region in Figure~\ref{fig:TSensitivity}, although \GHAT~only examined resolved, relatively nearby galaxies).  That leaves a rather large window around FIR-emitting temperatures of tens of K.  

To some extent, existing surveys should be able to set constraints on FIR-emitting blackboxes.  A careful examination of the sources detected by higher frequency channels of \emph{Planck} should improve sensitivity on the cold side of the range.  WISE has good sensitivity to Chilly Ways as cold as $\sim 50\ \Kelv$.  In between, the best extant survey is that of AKARI's Far-Infrared Surveyor \citep{Kawada07}, which could rule out 30 K Chilly Ways with the luminosity of the Milky Way out to about 200 Mpc (Figure~\ref{fig:TSensitivity}).  

Star-forming galaxies contain dust with temperatures around 20 K \citep{Draine11}.  Distinguishing between this natural dust and artificial dust is a challenge when searching for Chilly Ways of these temperatures.  The apparent gas-to-dust ratio can be used as a signature of artificial activity.  Dipole antennas can have a far greater opacity than the same mass of natural, three-dimensional dust, so a galaxy filled with them will appear to have an anomalously high dust mass, perhaps exceeding the gas mass itself.  

There is also a window above 600 K that has not been explored in SETI.  Although organic beings cannot survive in places that hot, technology might.  Chilly Ways with $600\ \la T_{\rm obs} \la 3000\ \Kelv$ could be built with Dyson spheres fairly close to the surfaces of stars, or a larger cloud around a bright active nucleus.  They could be detected with WISE, which was designed to find similarly hot brown dwarfs.  It is not possible to build a Dyson sphere hotter than the host's star surface, so Type III societies with $T_{\rm obs} \ga 3000\ \Kelv$ cannot be powered by starlight unless there is extensive stellar engineering.  Shells around accretion-powered objects like an active nucleus or X-ray binaries could be hotter, if they can survive the temperature, although it may be hard to tell apart artificial and natural obscuration.

Another possibility not yet examined is that of partial blackboxes with $\tau_{\rm abs} \ll 1$.  My calculations in Section~\ref{sec:DipoleClouds} indicate that the Milky Way has just enough metals in its ISM to reach $\tau_{\rm abs} \approx 1$ with metallic wire dipole antennas, and much more if carbon nanotube antennas can be used.  But there may be unforeseen engineering difficulties, and low metallicity galaxies may lack enough materials to build a full blackbox.  Alternatively, a blackbox under construction may still be mostly transparent.

The flux from a partial blackbox scales roughly as $\tau_{\rm abs}$.  Thus, the density limits on partial blackboxes goes as $\tau_{\rm abs}^{-3/2}$.  Very roughly, the null detections in \emph{Planck}'s 100 GHz channel imply 95\% confidence limits of less than 1 in 30,000 galaxies with a $\tau_{\rm abs} = 0.1$ partial blackbox, and 1 in 800 with a $\tau_{\rm abs} = 0.01$ partial blackbox.  Further improvements can be set mainly by a more careful examination of the \emph{Planck} catalogs and more sensitive microwave surveys.  Beyond a certain point, though, emission from artificial dust is buried by that of natural dust, both the Rayleigh-Jeans tail of small grains and the blackbody emission of very large grains equilibrating to the same temperature as any artificial dust.  

The thermodynamic motivations for blackboxes ignore one very important possibility.  Astronomical black holes are far, far colder than the CMB, and so are ideal cold reservoirs.  Vastly more entropy is generated by sending energy into a black hole than by running it through a blackbox.  The uses of black holes for technically advanced societies as a kind of negative sun has been noted by \citet{Opatrny16}.  If Type III societies are bound only by thermodynamics and Landauer's Principle, they may end up tightly packed around the frigid supermassive black holes in the centers of their galaxies, rather than puffed out in diffuse clouds.

So why bother with blackboxes at all?  Blackboxes are much easier to construct; the basic technology involves relatively simple physics.  In contrast, it is not clear how feasible megastructures girdling supermassive black holes are.  While in principle more entropy is produced by sending energy into a black hole than out through a blackbox, it is not clear how to actually use that entropy increase for computation.  After all, most matter falling into black holes actually becomes very hot.  Shifting this power down to low temperatures would require a large emitting area -- in other words, a blackbox.  The entropy is ultimately released in the Hawking radiation, and could be used for computation then, but this could take more than a googol years, and it is not even clear baryonic matter will still be around then \citep{Adams97}.

In addition, typical supermassive black holes cannot actually sustain Type III, or even Type II societies, as defined by their power consumption.  The total amount of power that can be harvested from the CMB by sending it into a black hole is limited:
\begin{align}
\nonumber L_{\rm CMB \to BH} & \approx 4 \pi R_S^2 \sigma_{SB} T_{\rm CMB} (z)^4 \\
                             & = 9 \times 10^{-9}\ \Lsun\ (1 + z)^4 \left(\frac{M_{\rm BH}}{10^8\ \Msun}\right)^2,
\end{align}
with $R_S = 2 G M_{\rm BH} / c^2$ as the Schwarzschild radius of a black hole of mass $M_{\rm BH}$ \citep[see also][]{Opatrny16}.  Trying to harvest more power requires either hotter material surrounding the black hole or non-thermal emission that would produce enormous waste heat, and is self-defeating.  The thermodynamic advantages of black holes require the aliens to be very patient.  

Finally, black holes are ultimately fueled by the matter around them, which comes from the galaxy at large and ultimately the intergalactic medium.  Although Type III societies may decide on black holes as their ultimate home, they still have to move gas and possibly stars into the centers of their galaxies.  In order to funnel the ISM of their home galaxies inwards, they could very well build a blackbox as an intermediate step, using its artificial dust to interface with the gas.

\subsection{We are (probably) doomed}
\label{sec:Doom}
If we are alone in the observable Universe, then one might conclude that the ``Great Filter'' is safely behind us, and that humanity's future beyond current crises is bright.  It is not.  Even if $f_{\rm after}$ is much greater than $10^{-7}$, it can still be much less than 1.  In fact, the very same dipole cloud technology may also spell our doom.  For consider, what would happen if we placed a blackbox around the Earth itself?

The outcome would depend on $\tau_{\rm abs}$ and which wavelengths it absorbed.  A $\tau_{\rm abs} = 1$ planetary blackbox that absorbed optical light would reduce solar insolation to about 30\% of its current value.  As a result, the surface temperature of the planet would drop to $\sim e^{-1/4} \approx 80\%$ of its current value, a fall of roughly 60 K.  The surface of the planet would freeze within a few months, and the oceans would freeze in about a millennium, as the world artificially transitioned into a Snowball Earth phase.  In the end, most life would go extinct, except that living deep underground or near oceanic vents.

Alternatively, a $\tau_{\rm abs} = 1$ planetary blackbox that absorbed mid-infrared light would cause an enormous greenhouse effect.  The Earth's greenhouse gases, particularly water vapor, already push the surface temperature up about 30 K \citep{Kasting03}.  Since the temperature only increases as $(1 + \tau)^{1/4}$, the effects of the next optical depth are thus reduced somewhat.  Still, a mid-infrared planetary blackbox could raise the Earth's surface temperature by another few tens of K.  By itself, this would make the surface inhospitable to complex life.  It could also trigger a runaway greenhouse effect, in which the oceans boil into water vapor, itself a greenhouse gas.  If this happened, all life on the planet would go extinct, including bacteria.

\begin{deluxetable*}{lcccccccc}
\tablecaption{Minimum masses of planetary and Solar System blackboxes}
\tablehead{\colhead{Material} & \colhead{$\sigma_{\rm bulk}^{-1}$} & \colhead{$\nu_{\sigma}^{\rm bulk}$} & \colhead{$\Sigma_{\rm \Omega-bulk}$} & \colhead{$\Sigma_{\rm atomic}$} & \colhead{$\Sigma_{\rm Drude}$} & \colhead{$\Sigma_{\rm blackbox}$} & \multicolumn{2}{c}{Dipole cloud minimum masses} \\ & & & & & & & \colhead{10000 km} & \colhead{1 AU} \\ & & & & & & & \colhead{$M_{\rm box}$} & \colhead{$M_{\rm box}$} \\ & \colhead{$(\Ohm\ \meter)$} & \colhead{$(\THz)$} & \colhead{$(\kgm2)$} & \colhead{$(\kgm2)$} & \colhead{$(\kgm2)$} & \colhead{$(\kgm2)$} & \colhead{$(\Mton)$} & \colhead{$(10^{20}\ \kg)$}}
\startdata
Lithium   & $9.5 \times 10^{-8}$ & 20  & $4.0 \times 10^{-5}$ & $8.3 \times 10^{-9}$ & $6.0 \times 10^{-6}$ & $4.0 \times 10^{-5}$ & 50  & 0.11\\
Beryllium & $3.7 \times 10^{-8}$ & 42  & $5.6 \times 10^{-5}$ & $3.0 \times 10^{-8}$ & $1.0 \times 10^{-5}$ & $5.6 \times 10^{-5}$ & 70  & 0.16\\
Sodium    & $4.9 \times 10^{-8}$ & 5.6 & $3.8 \times 10^{-5}$ & $1.5 \times 10^{-8}$ & $1.4 \times 10^{-5}$ & $3.8 \times 10^{-5}$ & 50  & 0.11\\
Magnesium & $4.5 \times 10^{-8}$ & 17  & $6.1 \times 10^{-5}$ & $2.7 \times 10^{-8}$ & $1.5 \times 10^{-5}$ & $6.1 \times 10^{-5}$ & 80  & 0.17\\
Aluminum  & $2.7 \times 10^{-8}$ & 22  & $5.8 \times 10^{-5}$ & $4.2 \times 10^{-8}$ & $1.7 \times 10^{-5}$ & $5.8 \times 10^{-5}$ & 70  & 0.16\\
Potassium & $7.5 \times 10^{-8}$ & 4.5 & $5.1 \times 10^{-5}$ & $1.3 \times 10^{-8}$ & $1.7 \times 10^{-5}$ & $5.1 \times 10^{-5}$ & 60  & 0.14\\			
Calcium   & $3.4 \times 10^{-8}$ & 7.3 & $4.3 \times 10^{-5}$ & $2.5 \times 10^{-8}$ & $1.8 \times 10^{-5}$ & $4.3 \times 10^{-5}$ & 50  & 0.12\\
Iron      & $1.0 \times 10^{-7}$ & 76  & $6.3 \times 10^{-4}$ & $1.2 \times 10^{-7}$ & $5.0 \times 10^{-5}$ & $6.3 \times 10^{-4}$ & 800 & 1.8\\
Nickel    & $7.2 \times 10^{-8}$ & 59  & $5.1 \times 10^{-4}$ & $1.4 \times 10^{-7}$ & $5.5 \times 10^{-5}$ & $5.1 \times 10^{-4}$ & 600 & 1.4\\
Copper    & $1.7 \times 10^{-8}$ & 6.4 & $1.2 \times 10^{-4}$ & $1.4 \times 10^{-7}$ & $7.8 \times 10^{-5}$ & $1.2 \times 10^{-4}$ & 150 & 0.3\\
\hline
Carbon nanotubes  & \nodata & \nodata & \nodata & \nodata & \nodata & $1.1 \times 10^{-5}$ & 14 & 0.03
\enddata  
\tablecomments{The resistivities are for temperatures of 293 to 300 K and come from the same references as in Table~\ref{table:ColdMetalResistivities}.  The values listed assume $\tau_{\rm abs} = f_{\ell} = \overline{\eta_{\rm abs}} = 1$, $f_{\rm pol} = 1/2$, $\beta_a = 1/2$, $A_{\rm atomic} = \pi (\nm)^2$, $\lambda_{\rm min} = 5\ \um$, and $\lambda_{\rm max} = 0.5\ \mm$.}
\label{table:PlanetaryBlackboxes}
\end{deluxetable*}

Dipole clouds can act as planetary blackboxes.  The amount of mass required for various metals is listed in Table~\ref{table:PlanetaryBlackboxes}.  At 300 K, the resistivities of metals is much greater than at 3 K.  Thus, the limiting factor for these constructs is bulk Ohmic losses, as given in equation~\ref{eqn:SigmaOhmic}.  Roughly $10^8$ metric tons of spaceborne metallic dipoles would be necessary to shroud the Earth in a $\tau_{\rm abs} = 1$ blackbox.  This is orders of magnitude more than has been launched into orbit so far, so the threat is not immediate.  Assuming the current price of US \$10,000 per kilogram into orbit, the launch costs would be about US \$100 trillion, about two years of the gross world product.  The amount of copper is about roughly a decade's worth of current production on the Earth.\footnote{http://minerals.usgs.gov/minerals/pubs/commodity/copper/mcs-2016-coppe.pdf}  Carbon nanotube antennas, if they have ideal properties, would require just one tenth the mass.  While out of the question now, humanity probably can build a planetary blackbox within a few decades if it is determined to do so, and if the world's economy and launch capabilities continue to grow.

Aside from the material and launch costs, there are a lot of engineering difficulties.  Foremost is the collision problem -- if the dipoles are scattered randomly into random Earth orbits, they would collide with each other in just seconds (Appendix~\ref{sec:Collisions}).  They would either have to be launched into precisely defined orbits so that they don't run into each other, or have some kind of collision avoidance system, which could be jammed to cause the cloud to self-destruct.  In addition, the conductivity of metals generally becomes imaginary at infrared frequencies, preventing them from being used at high frequencies.  Still, infrared and optical frequency antennas are under development; carbon nanotubes might be suited for this purpose.  Finally, the blackbox would prevent space travel beyond itself, unless the blackbox had some holes (as in a toroidal geometry).  While these problems delay it, the basic problem remains -- dipole clouds are planet killers.

In the medium term, optically thin dipole clouds could still wreak plenty of havoc on their own.  Even a $\tau_{\rm abs} \approx 0.01$ cloud is capable of adjusting the Earth's temperature by a K -- the same as all industrial global warming to the present.  Roughly 500 kilotons of metals, or 140 kilotons of carbon, are required, which is competitive with space-based sunshades to cool the Earth by the same amount \citep{Angel06}.  The costs are already becoming feasible, although the total masses still far exceed everything that has been launched so far.  Building such a dipole cloud may become tempting if global warming is not stopped by other means in the next few decades, but the technology could also be used to double global warming, or overshoot and trigger an Ice Age.

Even a dipole cloud with $\tau \approx 10^{-5}$ in visible light may prove disastrous.  If the cloud is wide enough, light scattered from this cloud would reach everywhere on Earth.  The resulting light pollution would be comparable to a constant full Moon.  All ground-based optical astronomy would end for as long as the cloud survived.  Stargazing itself would suffer.  No one would ever see the Milky Way again, for as long as the cloud remained.  There would also be ecological consequences from the light pollution.  

Nor is the threat hypothetical: there already has been a dipole cloud.  The West Ford Project of the 1960s placed a temporary ring of dipoles around the Earth to act as a GHz-frequency ionosphere, allowing radio signals to be bounced around the planet \citep{Morrow61,WestFord61}.  Only a few kilograms of copper were used, but it was enough to change the Earth's radio environment.  The negative impacts on both radio and optical astronomy motivated protests by astronomers and the world at large \citep{Goldberg61,Liller61,Lilley61,Lovell62}.  As a result, radio astronomers mobilized to protect the radio sky from interference \citep{Robinson99}.  In the end, the impact of the West Ford dipole belt was relatively small \citep{Sandage63,Tifft63}, and most of the dipoles crashed to Earth as designed a few years later \citep{Shapiro66}.  Still, if the political situation changes, far worse is easily achieved.

Dipole clouds are just one method of geoengineering, the deliberate alteration of a planet's global properties.  \citet{Cirkovic04} has suggested that geoengineering in general contributes to the destruction of technological societies; they specifically discussed a proposal to drill into the Earth's deep interior.  Geoengineering has been gaining prominence in recent years as a possible method of thwarting climate change, attracting controversy about its efficacy and ethics \citep[see][]{Caldeira13}.  While the focus is likely to remain on mitigating climate change for a while, simply because of the size and cost of geoengineering projects, the potential for much more radical alterations in the coming centuries might be worth examining before they become feasible.

\subsection{One minute from midnight, forever}
\label{sec:1159}

It is something of a truism that humanity is passing through a ``bottleneck'' phase of our history.  For the first time, we have the power to destroy ourselves with nuclear weapons and ecological catastrophes, but if we survive these crises and spread into space, there will always be human settlements that survive, giving us a kind of species immortality.  The potential of dipole clouds and cosmic engineering suggests otherwise, I think -- there is nothing particularly special about our current situation.  Certainly, planetary blackboxes and radical geoengineering can pose new kinds of threats in a few decades, and dipole cloud technology can be scaled to far larger scales.

An optically thick Solar System-sized dipole cloud is a Dyson sphere, for all intents and purposes.  The amount of material to build such a cloud is listed in Table~\ref{table:PlanetaryBlackboxes}.  Unlike normal Dyson sphere concepts, which require the disassembly of entire planets to build, the required masses are of order $\sim 10^{19}\ \kg$ for metal wire dipoles or $\sim 10^{18.5}\ \kg$ for carbon nanotube dipoles.  For comparison, the former mass is comparable to that of the asteroid 16 Psyche, or about 5\% of 4 Vesta \citep{Baer11}.  The values listed assume that the sphere is opaque only at MIR wavelengths, but extending the range covered down to $\lambda_{\rm min} = 250\ \nm$ to capture the Sun's optical light merely doubles the mass required.  The \citet{Dyson60} sphere around the Sun requires the disassembly of Jupiter, a feat needing a millennium's worth of Solar luminosity, but the unbinding of 16 Psyche requires less than a millisecond's worth of Solar luminosity, and a few seconds' more worth of Solar energy to alter the material's heliocentric orbital energy.  It does not take some inconceivably advanced society bending the laws of physics to build a Dyson sphere -- it's the kind of thing we might build in a few centuries using fairly conservative technology.

Dyson spheres bring with them enormous potential and enormous challenges.  The availability of so much power suggests that building beacons to signal extraterrestrial societies is actually very easy, even if only a minority of a society chooses to do it.  The environmental cost is the enormous waste heat produced by the Dyson sphere, which acts as a greenhouse for the entire inner Solar System.  Just as a planetary blackbox could easily devastate the Earth, all of the inner planets could be devastated by a Dyson swarm (something like this scenario was described in \citealt{Hoyle57}).  The thermal effects of a Dyson sphere refutes the idea that the Solar System is too big to pollute.

There is no reason to stop at 1 AU radii.  Very large Dyson swarms have the advantage that they are cold, so the resistivities of the dipoles go down, and a smaller $\Sigma_{\rm blackbox}$ is needed.  Since the resistivities generally scale as $\sigma_{\rm bulk}^{-1} \propto T$ or faster \citep{Chi79-Alkali,Chi79-Alkaline}, $\Sigma_{\rm \Ohm-bulk} \propto R_{\rm box}^2 T^{-1} \propto R_{\rm box}^{3/2}$.  Dyson swarms with radii of 100 AU or larger could be built using the resources of the Kuiper Belt alone.  

Spreading to other stars is not a guarantee of immortality either.  Wherever a starfaring society goes, dipole clouds and their potential for environmental damage may follow.  Furthermore, Dyson swarms could be used to channel much of a sun's output from one stellar system to another (a concept proposed by the science fiction critic James Nicoll).  This could be used cooperatively, for assisting cosmic engineering projects, but obviously it can be used destructively.  The only way to ensure survival of the society against war would be to disperse from planets and star systems entirely and spread into the interstellar medium itself.  In other words, societies may be compelled to develop interstellar and then galactic-scale dipole clouds.  As in \citet{Carrigan12} and \citet{Wright14-SF}, a starfaring society's development may appear as a growing cloud of darkness, but the interior would be diffuse and chaotic, not the result of an organized program based around stars.

Like an ecosystem, a Type III blackbox is a tightly coupled system.  It subsists on the flow of energy and gas through the galaxy.  What happens in one part of a galaxy could affect the rest.  For example, I show in Appendix~\ref{sec:AvailablePower} that in a highly optically thick blackbox, the amount of available power is basically constant at all depths, provided the power is always dissipated.  But it is possible for an inner shell of the blackbox to hoard the available power, leaving none for the outer shells -- much like how a community upstream on a river can divert the water and leave it dry for communities downstream.  This suggests that resources would have to be managed carefully even in a Type III blackbox.  

In this sense, even a Type III society may face similar problems as our own, with what we might call war and pollution, but more generally positive feedback.  This could be interpreted as an argument against cosmic engineering, but the issues remain the same whether societies remain Type I, or go on to Type II or Type III, because dipole clouds and blackboxes can be built on any of these stages.  Whether one remains small or grows big, the potential for devastation always is there.

Planetary or stellar or galactic doom is not inevitable.  We have managed to avoid destroying ourselves.  But nor have our crises gone away, and they could persist even for Type II and Type III societies.  The Doomsday Clock may never reach midnight, but it could always remain just a minute away, no matter what form a society takes.  Life may always be teetering on the brink of annihilation.

We may not be facing a special bottleneck, but instead simply be facing the challenges that all mature societies have to deal with, now and forever.

\emph{Implications for SETI} -- The ease of building dipole clouds suggests an underexplored avenue for SETI, the search for short-lived but extremely powerful societies.  A Dyson sphere can send radio signals bright enough to be detected at extragalactic distances.  This might be done for the express purpose of communication, or it might be done in the hopes of some kind of remote sensing if the society lasts long enough \citep[e.g.,][]{Scheffer14}.  Even if aliens are very rare, they might still be found if we search distant galaxies.

While ${\cal L}$ might be very short, the ruins of these societies might survive much longer.  For a dipole cloud, the survival time is only a few months for a 1 AU Dyson sphere, but thousands of years for a very diffuse cloud thousands of AU wide.  Even an inactive cloud would emit thermal emission.  If the cloud is located within the Milky Way and is opaque at frequencies below a GHz, the cloud will appear as a negative flux source, since the temperature of the cloud is much smaller than the brightness temperature of the Galactic synchrotron background.

As noted before, galaxy-sized dipole clouds can be expected to last for around a Myr before self-destructing, and cluster-sized dipole clouds would persist for around 100 Myr.  After that time, if there is no automation or self-repair, collisions would grind the dipoles down into small dust grains, which may be hard to distinguish from natural dust grains.  Some dipoles might naturally be ejected in galactic winds, or they might be deliberately propelled out of their home galaxies.  These outlying dipoles would cause the outer reaches of the galaxy to have an abnormally high opacity.  The total effect would be subtle since the optical depth may be very small, but high angular resolution studies of the CMB might still set constraints on these diffuse, persistent halos.

\subsection{A final note on SETI}
Despite these conclusions, I nonetheless believe that SETI programs are still useful.  After decades of debate, there remains no consensus on the values of $f_{\rm before}$ and $f_{\rm after}$.  Aside from waiting to see what the future development of humanity is like, SETI provides the only way to grapple with this issue.  Even if there are fairly convincing arguments that $f_{\rm after}$ is big and $f_{\rm before}$ is small, the nature of science demands that they be tested whenever it's reasonable.  Indeed, there are plenty of searches for unlikely and unexpected phenomena in astronomy and physics; why not planetbound alien societies? 

SETI also tends to examine underexplored regions of parameter space, as noted by \citet{Harwit81}.  For example, it provides a motivation for searching for very brief pulses of radio emission \citep{Siemion10,Messerschmitt14} and visible light \citep{Howard04,Borra12}, photon orbital angular momentum \citep{Harwit03}, and YeV neutrinos \citep{Lacki15}. From a purely commensal point of view, SETI brings the possibility of finding unexpected natural phenomena.

Finally, I think there is a misconception that the null results mean that SETI has failed.  I would argue that the null results are actually a spectacular success.  As far as we knew in 1960, when modern interstellar SETI began \citep{Cocconi59,Drake61}, the Universe really could have been filled with aliens.  Practically every star system could have had a radio beacon; there could have been millions of Dyson spheres in the Milky Way; every other galaxy could have been concealed in a blackbox.  Now we know that all of these phenomena are rare.  The limits on the abundances of Type II and Type III societies have now improved by a factor of a million, for example (although we must be careful about false negatives, as noted above).  SETI has delivered decisive and improving results on a profound question, even if those results are not the ones originally hoped for.  What remains is to close the window and examine the remaining possibilities, and improve the limits by further orders of magnitude in the coming decades. 

\acknowledgements
I used catalogs and maps derived from \emph{Planck}, an ESA science mission with instruments and contributions directly funded by ESA Member States, NASA, and Canada.  During my examination of the nine candidate blackboxes in the \emph{Planck} compact source catalog, I used Aladin, Simbad, and the NASA/IPAC Extragalactic Database (NED).  The Aladin sky atlas was developed at CDS, Strasbourg Observatory, France.  The SIMBAD database is operated at CDS, Strasbourg, France.  NED is operated by the Jet Propulsion Laboratory, California Institute of Technology, under contract with the National Aeronautics and Space Administration.  In addition, I wish to acknowledge the use of NASA's Astrophysics Data System and arXiv.

%==========================
%==========================

\appendix
\section*{A Peek Inside the Blackbox}
The advantage of looking for blackboxes is that their properties do not depend much on what's going on inside of them.  As long as the inhabitants are bound by thermodynamics, and as long as the galaxy is optically thick, it will emit blackbody radiation, no matter how strange the technology behind it.  

But blackboxes, and cosmic engineering in general, might be criticized as completely infeasible; they are not a robust sign of extraterrestrial intelligence if there's no way Type III societies can exist.  I also made some fanciful claims in Section~\ref{sec:GeniusLoci} for why aliens may wish to build one, such as gaining the ability to move the ISM around, without justifying whether those uses are feasible.  I address some of the capabilities and challenges of blackboxes in this Appendix, under the assumption that they are dipole clouds.  This is not to say that actual blackboxes would be dipole clouds -- just that dipole clouds use physics that is well understood and known materials, bypassing objections that vast, unproven technology gains are required to make them.  As such, this Appendix is meant as a plausibility argument rather than a definitive prediction of the form of Type III societies.

\section{Skin effects and the length-width ratio of dipole motes}
\label{sec:SkinEffects}
When I calculated the resistance of a dipole antenna in Section~\ref{sec:DipoleClouds}, I assumed that it was proportional to the cross-sectional area of the antenna.  But this estimate is a naive lower limit when dealing with high frequencies, and thermal radiation is very high frequency indeed by radio standards.  Instead, depending on the electric and magnetic properties of the conductor, skin effects may confine the induced electrical currents to regions near the surface of the conductor.  If the conductor is simply a single filled rod, this manifests as an apparent resistance, as the inner regions of the disk are essentially useless for carrying currents \citep{Carr12}.

The skin depth of a material to radio waves of frequency $\nu$ is
\begin{equation}
r_{\delta} = \sqrt{\frac{\sigma^{-1}}{\pi \nu \mu_r \mu_0}},
\end{equation}
where $\mu_0$ is the magnetic permeability of free space and $\mu_r$ is the relative permeability of a material \citep{Jackson99}.  For non-ferromagnetic materials, $\mu_r$ is usually very close to 1, but ferromagnetic materials like iron or nickel may have $\mu_r$ of thousands or bigger, depending on their purity.  For millimeter wavelength radiation, the skin depth is 
\begin{equation}
r_{\delta}= 0.92\ \nm\ \mu_r^{-1/2} \left(\frac{\sigma^{-1}}{10^{-12}\ \Omega\,\meter}\right)^{1/2} \left(\frac{\lambda}{\mm}\right)^{1/2}.
\end{equation}
A good quality iron conductor with $\mu_r \approx 10^4$ has a skin depth of about 100 nm at these frequencies.

The skin effect must be taken into account when designing the dipole cross-section if the radius of a filled disk with a given area is smaller than the skin depth:
\begin{equation}
\pi f_{\rm skin} r_{\delta}^2 > A_{\rm dipole}.
\end{equation}
The $f_{\rm skin}$ factor accounts for the fact that a fraction of the current still flows below one skin-depth.  In Section~\ref{sec:DipoleClouds}, I assumed that the cross-sectional area was at least $A_{\Omega} = \sigma^{-1} \ell / \RRad$.  Then, the skin effect matters if $f_{\rm skin} \RRad \nu / (\mu_r \mu_0) > \ell$, or 
\begin{equation}
\left(\frac{\ell}{\lambda}\right) < 0.19 f_{\rm skin} \mu_r^{-1} \left(\frac{\RRad}{\RDipole}\right).
\end{equation}
Regardless of the conductor's resistivity or the resonant frequency, the skin effect can be basically ignored for a half-wave dipole with $\ell / \lambda \approx 0.5$ if the conductor is non-ferromagnetic.  However, it clearly cannot be ignored for a ferromagnetic material like iron.  

There is no reason that the antenna conductor must have a filled cross-section.  High frequency dipoles often are manufactured with ring cross-sections, effectively only including the current-bearing skin of the antenna \citep{Carr12}.  A ring cross-section may also be used to create a relatively wide dipole, which has a larger bandwidth near resonance \citep{Laport52,Carr12}.  Even wider antennas can be effectively achieved because the conducting ring itself does not have to be solid, but can consist of disconnected elements.  For example, a few conducting rods can be arranged in a sparse ring; when wired together correctly, they form a cage dipole antenna \citep{Carr12}.  With a little engineering, the skin effect does not prevent the efficient use of ferromagnetic conductors like iron in a blackbox.

\section{The collision problem and the fate of dipole clouds}
\label{sec:Collisions}
Like most conceptions of Dyson spheres, a dipole cloud involves an astronomical number of swarm elements swarming in a limited space.  And like Dyson spheres \citep{Carrigan09}, actually maintaining a blackbox runs into a problem of traffic control -- how does one keep all of those elements from crashing into one another?

Simply allowing the motes to crash into one another won't work, not if the elements are half-wavelength dipoles.  When two thin dipoles collide, they almost always cross one another like an X shape.  If the radius of each dipole is $r$, the contact area where the dipoles touch is of order $4 r^2$.  Assuming the relative velocity of the elements is $v$, the collision lasts for a duration of $r / v$.  A significant fraction $f_{\cal S}$ of the momentum of each dipole ($m v \approx \pi \rho \ell r^2 v$) is transferred to the other during this collision.  This setup leads to a stress on the contact area of ${\cal S} \approx f_{\cal S} \times (\pi \rho \ell r^2 v) \times (4 r^2)^{-1} \times (r / v)^{-1} \approx (\pi/4) f_{\cal S} \rho v^2 (\ell / r)$:
\begin{equation}
{\cal S} \approx 8 \times 10^6\ \MPa\ f_{\cal S} \left(\frac{\rho}{\rhoUnits}\right) \left(\frac{v}{\kms}\right)^2 \left(\frac{\ell/r}{10^4}\right).
\end{equation}
For comparison, most materials fail under stresses greater than a few hundred MPa.

The characteristic collision timescale for a blackbox is in fact the orbital time of an element, because if a blackbox has an optical depth of $\sim 1$, any path through the host galaxy typically passes through $\sim 1$ element.  Although dipole antennas have needle-like shapes, with geometrical cross-sections much smaller than their radio cross-sections, two elements are likely to meet at an angle with one another when they collide, so their effective cross sections are still of order $\ell^2$.

In more detail, the average time a mote of length $\ell$ goes between collisions is
\begin{equation}
t_{\rm coll} (\ell) = \left[\int \frac{dN_{\rm motes}}{d\ell^{\prime}} \mean{A_{\rm coll} (\ell, \ell^{\prime})} v d\ell^{\prime}\right]^{-1}, 
\end{equation}
where $V_{\rm box}$ is the volume of the region filled by motes, $dN_{\rm motes}/d\ell^{\prime}$ is the mote length distribution function, and $\mean{A_{\rm coll} (\ell, \ell^{\prime})}$ is the typical collision cross section for two motes of length $\ell$ and $\ell^{\prime}$ averaged over all possible orientations.  The mean cross section for two infinitely thin wires is $(\pi/8) \ell \ell^{\prime}$, after taking into account projection effects, so I use $\mean{A_{\rm coll} (\ell, \ell^{\prime})} = f_{\rm coll} (\pi / 8) \ell \ell^{\prime}$.  Then, using equation~\ref{eqn:dNdlnl} for $dN_{\rm motes}/d\ell^{\prime}$,
\begin{equation}
t_{\rm coll} (\ell) = \left[\frac{\pi^2 \tau_{\rm abs} f_{\rm coll} A_{\rm box} v \ell}{8 f_{\rm pol} \eta_{\rm abs} \beta_a f_{\ell}^2 V_{\rm box}} \left(\frac{1}{\ell_{\rm min}} - \frac{1}{\ell_{\rm max}}\right)\right]^{-1}.
\end{equation}

For a spherical blackbox with $A_{\rm box} = 4 \pi R_{\rm box}^2$ and $V_{\rm box} = (4/3) \pi R_{\rm box}^3$, and in the usual case that $\ell_{\rm max} \gg \ell_{\rm min}$, 
\begin{align}
\nonumber t_{\rm coll} (\ell) & \approx \frac{2}{3 \pi^2} \frac{R_{\rm box}}{v} \frac{\ell_{\rm min}}{\ell} \left(\frac{f_{\rm pol}}{1/2}\right) \left(\frac{\beta_a}{1/2}\right) \frac{f_{\ell}^2 \eta_{\rm abs}}{\tau_{\rm abs} f_{\rm coll}}\\
                              & \approx 330\ \kyr\ \left(\frac{R_{\rm box}}{10\ \kpc}\right) \left(\frac{v}{200\ \kms}\right)^{-1} \left(\frac{\ell}{10\ \ell_{\rm min}}\right)^{-1} \left(\frac{f_{\rm pol}}{1/2}\right) \left(\frac{\beta_a}{1/2}\right) \frac{f_{\ell}^2 \eta_{\rm abs}}{\tau_{\rm abs} f_{\rm coll}}.
\end{align}
The value of $\ell$ corresponds to the middle of the absorption band, where most of the thermal emission comes out, for the values of the parameters that are typically used in this paper.

It is clear that the collision problem is far more severe for planetary-scale blackboxes ($t_{\rm coll} \approx 10\ \sec$) and Dyson swarms ($t_{\rm coll} \approx 1\ \dayUnit$) than for a galaxy-scale blackbox ($t_{\rm coll} \approx 300\ \kyr$), if the motes are released into random orbits.  Nonetheless, the collision time is still very short compared to the millions or billions of years that a Type III society may last.  There are a few obvious solutions.  The first is giving the dipoles some kind of navigation system to steer around each other.  Getting one mote to detect another and exchange navigational information with it in time to avoid collision could have practical difficulties that I discuss later.  A second solution is to give the antennas some kind of self-repair capability.  A third is to launch the motes in organized patterns with cold velocity dispersions, so that the relative velocities of nearby motes is very small compared to their total orbital velocities.  For example, a velocity dispersion of $10\ \kms$, more typical of the Galactic interstellar medium, prolongs the mote survival time to $70\ \Myr$.  As \citet{Carrigan09} notes, gravitational instabilities may arise that disrupt these patterns, and the stellar population of a galaxy will stir the population of motes and heating their relative velocities.   

The lifetime of the blackbox is important when setting limits on the abundance of blackbox builders (equation~\ref{eqn:Type3Drake}).  Even if the motes start with zero relative velocity to each other, once any guidance is turned off, they will be dragged by the turbulent gas flow of the ISM.  Generally, turbulence is stirred up on an outer scale, and the energy cascades down to smaller scales in one flow crossing time on the outer scale.  Supersonic turbulence would probably dissipate the blackbox very quickly for three reasons: (1) it is characterized by shocks, with velocity discontinuities on small scales, resulting in relatively large velocities between the motes; (2) it concentrates most of the mass into transient, small clumps, where the higher mote density will result in much faster destruction; and (3) even if the motes survive in the clumps, the clumpiness of the ISM will leave most of the galaxy unobscured \citep[e.g.,][]{Ostriker01}.  This is the situation for turbulence in cold and warm gas in galaxies \citep{MacLow04,Hill08}.  So the maximum unguided blackbox lifetime is probably of order $t_{\rm turb} \approx 30\ \Myr$ for an outer scale of $300\ \pc$ and a velocity dispersion of $10\ \kms$.

Subsonic turbulence is not as clumpy, and it is characterized by a Kolmogorov spectrum, with small velocity dispersions on small length scales.  This is the case for the ICM of galaxy clusters.  The flow crossing time may not limit the blackbox lifetime, but for completeness, it is of order $300\ \Myr$ for an outer scale of $100\ \kpc$ and a velocity dispersion of $300\ \kms$ \citep{Subramanian06}.

The collisional destruction of the motes does not proceed at the same rate for all of them.  Large motes are destroyed first.  If the motes break into only a few pieces when they shatter, then the debris from large motes should not accelerate the destruction of the small motes much.  For metallic dipoles at very cold ISM temperatures, the dipole wire radius is set by nanoscale effects on the Drude conductivity, so that the mass goes as $m \propto \ell^{5/3}$.  Then, since $m\ dN_{\rm motes}/d\ell \propto \ell^{-1/3}$, the small motes outnumber the fragments of large motes, as long as the size of the debris is bigger than $\ell_{\rm min}$.  For carbon nanotube antennas, this conclusion is only strengthened since $m \propto \ell$ and $m\ dN_{\rm motes}/d\ell \propto \ell^{-1}$.

\section{Thermodynamic efficiency deep in the blackbox}
\label{sec:AvailablePower}
When I motivated blackboxes as a form that Type III might take in Section~\ref{sec:GeniusLoci}, I implied that individual smart dust elements could harness the ambient radiation to do useful work on the ISM.  I also suggested that a Type III society might build a blackbox with an enormous optical depth, to create a greenhouse environment with relatively high temperatures.  But while a marginally opaque blackbox can harness the huge temperature difference between starlight and the CMB, the temperature gradients deep in an opaque atmosphere are small.  Unless there's something very wrong with our knowledge of physics (Section~\ref{sec:2LoT}), only a small fraction of the ambient radiation energy density can be harnessed.  Furthermore, the temperature of a cold blackbox asymptotes to $T_{\rm CMB}$; the energy density of reprocessed energy may be just a tiny fraction of the total energy density.  Does the Second Law of Thermodynamics render blackboxes useless at high optical depths?  As it turns out, no.

When there is only a small temperature difference $\Delta T$ to exploit in an environment with an ambient temperature $T$, the maximum Carnot efficiency of a heat engine is
\begin{equation}
\eta_{\rm Carnot} \approx \frac{\Delta T}{T}.
\end{equation}
At any given place deep in a blackbox atmosphere, an element receives radiation from cooler material that is up to $\sim 1$ optical depth above it and hotter material that is up to $\sim 1$ optical depth below it.  This provides the temperature difference that can be used for work.

I consider the case where heat transport in the atmosphere is through radiative diffusion.  An element is located at an optical depth $\tau^{\prime}$ (including absorption and scattering) from the blackbox's surface, at a distance $R^{\prime}$ from the blackbox's center.  The luminosity generated inside the blackbox within $R^{\prime}$ is $L^{\prime}$.  This situation has the famous grey atmosphere solution if (1) the opacity is the same at all frequencies, (2) $dL^{\prime}/dR^{\prime} = 0$, (3) the mean free path of photons is much smaller than $R^{\prime}$, and (4) there is no ambient radiation field outside of the blackbox.  Then the energy density here is $u^{\prime} = 3 L^{\prime} /(4 R^{\prime 2} c) \times [\tau^{\prime} + q(\tau^{\prime})]$, where $q(\tau^{\prime})$ is a slowly varying function \citep{Chandrasekhar60}.  At large optical depths, $q(\tau^{\prime}) \approx 2/3$.  For a non-grey atmosphere, the Rosseland optical depth $\tau_R$ can be used for $\tau^{\prime}$ \citep{Mihalas99}.  Of course, an actual blackbox is embedded in the CMB and the EBL.  This adds a constant $u_{\rm ext}$ to the energy density everywhere in the blackbox.  So the ambient radiation energy density around the element is
\begin{equation}
\label{eqn:GreyEnergyDensity}
u^{\prime} \approx \frac{3 L^{\prime}}{4 R^{\prime 2} c} (\tau^{\prime} + 2/3) + u_{\rm ext}.
\end{equation}
And if the radiation field is in a thermodynamic equilibrium, 
\begin{equation}
T^{\prime 4} \approx \frac{3 L^{\prime}}{16 \pi R^{\prime 2} \sigma_{\rm SB}} (\tau^{\prime} + 2/3) + \frac{c u_{\rm ext}}{4 \sigma_{\rm SB}}.
\end{equation}

The temperature differential can be estimated by finding $dT^{\prime}/d\tau^{\prime}$:
\begin{equation}
4 T^{\prime 3} \frac{dT^{\prime}}{d\tau^{\prime}} = \frac{3 L^{\prime}}{16 \pi R^{\prime 2} \sigma_{\rm SB}} \left[1 - \frac{\tau^{\prime} + 2/3}{\chi R^{\prime}} \left(\frac{d\ln L^{\prime}}{d\ln R^{\prime}} - 2\right)\right],
\end{equation}
where $\chi$ is the extinction coefficient.  If the mean free path of photons $\chi^{-1}$ is much smaller than $R^{\prime}$ and $dR^{\prime}/d\ln L^{\prime}$, as assumed when deriving the grey atmosphere solution, then $4 T^{\prime 3} dT^{\prime}/d\tau^{\prime} = 3 L^{\prime} / (16 \pi R^{\prime 2} \sigma_{\rm SB})$.  The Carnot efficiency of the element is then
\begin{equation}
\eta_{\rm Carnot} \approx (1/4) [(\tau^{\prime} + 2/3) + (1/3) u_{\rm ext} (4 \pi R^{\prime 2} c / L^{\prime})]^{-1} \times d\tau^{\prime},
\end{equation}
where $d\tau^{\prime} \approx 1$.

But while the thermodynamic efficiency is indeed small, the flux of useful energy through the atmosphere is
\begin{equation}
\label{sec:FHarvest}
F_{\rm harvest} \approx \eta_{\rm Carnot} c u^{\prime} / 4 \approx \frac{3 d\tau^{\prime}}{16} \frac{L^{\prime}}{4 \pi R^{\prime 2}}.
\end{equation}
Deep in the atmosphere, the ambient energy density is greater by a factor $\tau^{\prime}$ than at the surface, canceling out the smaller thermodynamic efficiency.  Likewise, if there is a lot of background radiation, a mote cannot harvest that background, but it can still harvest a much smaller flux from the blackbox's own luminosity.  The use of the grey atmosphere solution assumes that the flux through the atmosphere is constant, not converted into a non-thermal form nor escaping the blackbox directly.  This is true if the work performed is ultimately dissipated as waste heat within the blackbox.  If a layer of the blackbox converted all of $F_{\rm harvest}$ into neutrinos that simply escaped, for example, shallower layers above it have no useful flux to do more work. 

The upshot is that a Type III society can, in principle, always extract more work from its starlight by shrouding a blackbox in another layer of material, as long as all the work is converted back into waste heat.  The total amount of work extracted is proportional to the number of motes and the total optical depth $\tau_{\rm total}$, since all of the motes within the atmosphere are exposed to the same amount of useful flux.  This is in accord with the reasoning behind the Matrioshka brain (Section~\ref{sec:GeniusLoci}).  It works even if the excess temperature is much smaller than $T_{\rm CMB}$.

A mote with a single dipole antenna only harnesses a fraction of $F_{\rm harvest}$ because (1) it catches light of a single polarization, (2) the antenna is not perfectly efficient, and (3) it may have a narrow bandwidth.  The total amount of power derived can be written in terms of an effective aperture $A_{\rm work}$:
\begin{equation}
\label{eqn:EHarvestBest}
\EDotHarvest = F_{\rm harvest} f_{\rm pol} A_{\rm work} \approx \frac{3 L^{\prime} f_{\rm pol}}{256 \pi^2 R^{\prime 2}} A_{\rm work}.
\end{equation}
$A_{\rm work}$ depends on the spectral response of the mote, the effectiveness it converts thermodynamically useful power into work, and the total background of thermal radiation available:
\begin{equation}
A_{\rm work} = \int_0^{\infty} \eta_{\rm work}(\lambda) \frac{\lambda^2 B_{\lambda} (T^{\prime}, \lambda)}{B(T^{\prime})} d\lambda.
\end{equation}
The $\eta_{\rm work}$ factor is the antenna's efficiency at converting the thermodynamically useful flux in light with wavelength $\lambda$ into work.  Note that $\eta_{\rm work}$ is the efficiency at converting the total flux of light divided by $\eta_{\rm Carnot}$, so it's possible for $\eta_{\rm work}$ to equal 1.

Throughout most of this paper, I've been considering narrow bandwidth antennas.  Continuing that line of thought, I model the mote efficiency as 
\begin{equation}
\label{eqn:etaWork}
\eta_{\rm work} (\lambda, \ell) = \left\{ \begin{array}{ll} 
                                  \etaWork & (\lambda_a e^{-\beta_a/2} \le \lambda \le \lambda_a e^{\beta_a/2}) \\
																  0                    & (\lambda < \lambda_a e^{-\beta_a/2}~{\rm or}~\lambda > \lambda_a e^{\beta_a/2})
																  \end{array} \right.
\end{equation}
in analogy with $\etaAbs$ (equation~\ref{eqn:etaAbs}).  Note that $\etaWork$ is not necessarily equal to $\etaAbs$, even if $\eta_{\rm Carnot} = 1$; $\etaAbs$ includes all power that is dissipated as heat, while $\etaWork$ only includes the extraction of work.

For these narrowband antennas, the aperture $A_{\rm work}$ has a maximum value when $x_a \equiv h c / (\lambda_a k_B T^{\prime})$ equals 1.5936.  At this value, the aperture attains a value of  
\begin{equation}
A_{\rm work}^{\rm max} = 0.03174 \beta_a \etaWork \left(\frac{h c}{k_B T^{\prime}}\right)^2.
\end{equation}
Not all motes will be resonant at the optimal wavelength in a blackbox, so I find it convenient to define a scaling variable $f_{\rm work} \equiv A_{\rm work} / A_{\rm work}^{\rm max}$ that accounts for the smaller aperture.  Thus, for all narrowband antennas,
\begin{equation}
\label{eqn:EHarvestBestNarrow}
\EDotHarvest = 1.7 \times 10^{-13}\ \Watt\ \etaWork \beta_a f_{\rm work} \left(\frac{L^{\prime}}{10^{11}\ \Lsun}\right) \left(\frac{R^{\prime}}{10\ \kpc}\right)^{-2} \left(\frac{T^{\prime}}{3\ \Kelv}\right)^{-2}.
\end{equation}

For convenience, I will give values for two broadband cases here.  If an antenna has a constant $\eta_{\rm work} (\lambda) = \etaWork$ for a broad frequency range around $k_B T^{\prime} / h$, then because the effective aperture of an antenna scales as $\lambda^2$, it is more effective at absorbing the Rayleigh-Jeans tail of the thermal radiation.  It then has an aperture of
\begin{equation}
A_{\rm work} = \frac{5 \etaWork}{2 \pi^2} \left(\frac{h c}{k_B T^{\prime}}\right),
\end{equation}
and the substitution $f_{\rm work} = 7.98 \beta_a^{-1}$ can be used in equations.  The amount of power that can be harvested by these antennas is
\begin{equation}
\EDotHarvest^{\rm broad} = 1.4 \times 10^{-12}\ \Watt\ \etaWork \left(\frac{L^{\prime}}{10^{11}\ \Lsun}\right) \left(\frac{R^{\prime}}{10\ \kpc}\right)^{-2} \left(\frac{T^{\prime}}{3\ \Kelv}\right)^{-2}.
\end{equation}

The other case demonstrated here is a highly resistive short dipole antenna of length $\ell$, much shorter than $h c / (k_B T^{\prime})$.  It has an internal resistance of $\RResist$ which is much greater than the radiation resistance $\RRad = (\pi/6) \ZImp_0 (\ell/\lambda)^2$ \citep{Jackson99,Condon10} and the antenna reactance $\XReact$ over a broad frequency range around $k_B T^{\prime} / h$.  Thus, $\eta_{\rm work} = \etaWork \RRad/\RResist = (\pi/6) (\ZImp_0/\RResist) (\ell/\lambda)^2$.  These antennas have an aperture of
\begin{equation}
A_{\rm work} = \frac{\pi \etaWork \ZImp_0}{6 \RResist}
\end{equation}
and can harvest
\begin{equation}
\EDotHarvest^{\rm resistive} = 6.4 \times 10^{-13}\ \Watt\ \etaWork \left(\frac{L^{\prime}}{10^{11}\ \Lsun}\right) \left(\frac{R^{\prime}}{10\ \kpc}\right)^{-2} \left(\frac{T^{\prime}}{3\ \Kelv}\right)^{-2} \left(\frac{\RResist}{\RDipole}\right)^{-1}
\end{equation}  
of power.

\section{Literal cloud computing with dipole clouds}
\label{sec:Computation}
Like our own Internet, a blackbox may take the form of many computers that are connected to one another.  Let's compare the power of the individual computers that make up these networks, as well as the rate that information is exchanged between them.

According to Landauer's Principle, irreversible computation is directly related to entropy.  The fundamental source of dissipation is the erasure of bits, each erasure costing $(\ln 2) k_B T^{\prime}$ of energy.  So, the maximum number possible of these erasures that a mote can perform is $\dot{\cal I} = \dot{E}_{\rm harvest} / [(\ln 2) k_B T^{\prime}]$:
\begin{equation}
\dot{\cal I} = 6.1\ \Gbit\ \sec^{-1} \etaWork \beta_a f_{\rm work} \left(\frac{L^{\prime}}{10^{11}\ \Lsun}\right) \left(\frac{R^{\prime}}{10\ \kpc}\right)^{-2} \left(\frac{T^{\prime}}{3\ \Kelv}\right)^{-3}.
\end{equation}
The computational power of a single mote within a blackbox is actually comparable to our present-day computers.  There just happen to be a lot more of them in a galaxy, about $10^{50} \tau_{\rm abs}$, so the total computational power of a galaxy-scale blackbox is $\sim 10^{60} \tau_{\rm abs}\ \bit\ \sec^{-1}$.

Like our own Internet, there does not have to be a direct connection between any two given elements; rather, information can be passed along from one mote to the next until it reaches the destination.  But information transfer between two relatively nearby motes faces some big challenges.  The typical distance between a mote with resonant wavelength $\lambda_a$ and its nearest neighbor with an overlapping resonance is
\begin{equation}
\dBar \approx \left[\frac{\beta_a}{V_{\rm box}} \frac{dN_{\rm motes}}{d\ln \ell}\Big|_{\ell = \lambda_a/(2 f_{\ell})}\right]^{-1/3} = \left(\frac{f_{\rm pol} \etaAbs \lambda_a^2 V_{\rm box}}{4 \pi \tau_{\rm abs} A_{\rm box}}\right)^{1/3} = 16\ \km \left(\frac{2 f_{\rm pol} \etaAbs}{\tau_{\rm abs}}\right)^{1/3} \left(\frac{R_{\rm box}}{10\ \kpc}\right)^{1/3} \left(\frac{\lambda_a}{\mm}\right)^{2/3}
\end{equation}
in a spherical blackbox.  If any two motes can communicate, regardless of resonant wavelength, then the typical neighbor distance is just slightly smaller:
\begin{equation}
\overline{d}_{\rm all} = \left[\frac{2 \pi \tau_{\rm abs} A_{\rm box}}{f_{\rm pol} \etaAbs \beta_a V_{\rm box}} \left(\frac{1}{\lambda_{\rm min}^2} - \frac{1}{\lambda_{\rm max}^2}\right)\right]^{-1/3} = 13\ \km\ \left(\frac{2 f_{\rm pol} \etaAbs}{\tau_{\rm abs}}\right)^{1/3} \left(\frac{R_{\rm box}}{10\ \kpc}\right)^{1/3} \left(\frac{\lambda_{\rm min}}{0.5\ \mm}\right)^{2/3}
\end{equation}
for $\lambda_{\rm max} \gg \lambda_{\rm min}$.  This distance is much bigger than the motes themselves, and unless the transmission is highly beamed, most of the energy is wasted.

Let's calculate the power transfer between two elements.  Suppose the transmitting element broadcasts a signal with a power spectrum of $P_{\rm send} (\nu)$ over a frequency range $\nu_1$ to $\nu_2$.  The duty cycle of the sender's transmissions is $f_{\rm duty}$, so the time-averaged power spectrum is $\mean{P_{\rm send} (\nu)} = f_{\rm duty} P_{\rm send} (\nu)$.  Furthermore, the transmitter beams the signal into a solid angle $\Omega_{\rm send}$ that includes the receiving.

The receiving mote, located a distance $d$ from the sending element, has an effective aperture of $A_{\rm rec}$, a solid angle field of view $\Omega_{\rm rec}$, and can deliver the signal power at frequency $\nu$ to the detector with an efficiency $\eta_{\rm rec} (\nu)$.  The transmitting mote must be in the receiving mote's field of view, and the receiving mote has to be listening while the signal is being sent. The detected power at the receiver mote is 
\begin{equation}
\label{sec:ReceivedPower}
P_{\rm rec} (\nu) = \frac{P_{\rm send} (\nu) A_{\rm rec} \eta_{\rm rec}(\nu)}{\Omega_{\rm send} d^2}.
\end{equation}

Although I have been describing the motes as the same antennas that absorb light in the blackbox, the communication might be through a different class of smart dust entirely or by a different part of the motes than the antennas.  The communication might be sent by optical laser, for example.  But if we are considering broadcasts through individual antennas, then equation~\ref{sec:ReceivedPower} reduces to the Friis transmission equation:
\begin{equation}
\label{sec:Friis}
P_{\rm rec} (\nu) = P_{\rm send} (\nu) \eta_{\rm rec} (\nu) \Gain_{\rm send} \Gain_{\rm rec} \left(\frac{\lambda}{4 \pi d}\right)^2,
\end{equation}
where $\Gain_{\rm send} = 4\pi/\Omega_{\rm send}$ is the gain of the transmitter and $\Gain_{\rm rec} = 4\pi/\Omega_{\rm rec}$ is the gain of the receiver \citep{Heald12,Zavrel16}.  When considering the motes responsible for absorption, the time-averaged power sent by the transmitter is at most $\dot{E}_{\rm harvest}$ in equation~\ref{eqn:EHarvestBest}, when integrated over all frequencies.  For convenience, I define the spectrum shape as
\begin{equation}
\PBarSpect \equiv \frac{\PSendSpect}{\EDotHarvest / f_{\rm duty}}.
\end{equation} 
The received power during a transmission is only
\begin{equation}
\PRecSpect = \frac{3 (12 \pi^2)^{2/3}}{4096 \pi^4} \frac{L^{\prime} \lambda_a^{2/3} A_{\rm work}}{R_{\rm box}^{8/3}} \frac{\Gain_{\rm send} \Gain_{\rm rec}}{f_{\rm duty}} \left(\frac{\sqrt{f_{\rm pol}} \tau_{\rm abs}}{\etaAbs} \frac{\dBar}{d}\right)^{2/3} \left(\frac{\lambda R_{\rm box}}{\lambda_a R^{\prime}}\right)^2 \PBarSpect \eta_{\rm rec}(\nu),
\end{equation}
which comes out to
\begin{multline}
\PRecSpect = 9.2 \times 10^{-30}\ \Watt\ \left(\frac{L^{\prime}}{10^{11}\ \Lsun}\right) \left(\frac{\lambda_a}{\mm}\right)^{2/3} \left(\frac{R_{\rm box}}{10\ \kpc}\right)^{-8/3} \left(\frac{T^{\prime}}{3\ \Kelv}\right)^{-2} \\
            \times \frac{\Gain_{\rm send} \Gain_{\rm rec} f_{\rm work} \etaWork \beta_a}{f_{\rm duty}} \left(\frac{\sqrt{2 f_{\rm pol}} \tau_{\rm abs}}{\etaAbs} \frac{\dBar}{d}\right)^{2/3} \left(\frac{\lambda R_{\rm box}}{\lambda_a R^{\prime}}\right)^2 \PBarSpect \eta_{\rm rec}(\nu).
\end{multline}
If the signal is transmitted around at typical thermal background frequency of $30\ \GHz$, this comes out to only 10 photons per year of active transmissions for the fiducial values.  The time-averaged received power is smaller by a factor of $f_{\rm duty}$.

That's not all.  The interior of the blackbox is filled with thermal radiation and interference from all the other notes, so the receiving mote is in a very noisy environment.  Given a Gaussian noise power spectrum of $P_{\rm noise} (\nu)$, the rate of information transmission is limited by the channel capacity,
\begin{equation}
\label{eqn:ChannelCapacity}
{\cal C} = \int_{\nu_1}^{\nu_2} \log_2 \left[1 + \frac{P_{\rm rec} (\nu)}{P_{\rm noise} (\nu)}\right] d\nu,
\end{equation}
from \citet{Shannon49}.  Here, ${\cal C}$ is the channel capacity during the transmission; the time-averaged channel capacity $\mean{\cal C}$ is again smaller by a factor $f_t$.  According to \citet{Shannon49}, the optimum signal spectrum has a shape such that $P_{\rm rec} (\nu) + P_{\rm noise} (\nu)$ adds to a constant at the frequencies where $P_{\rm noise} (\nu)$ is smallest.  The worst case is white noise, with a flat noise spectrum and a flat signal spectrum.  From an energy point of view, large signal-to-noise ratios are wasteful, since ${\cal C}$ rises logarithmically with $P_{\rm rec} (\nu)$ if $P_{\rm rec} (\nu) / P_{\rm noise} (\nu) \ga 1$.  

The detector is likely to be in thermal equilibrium with the surrounding radiation.  Since I am supposing that it is hooked up to an antenna, it could take the form of an electronic device in a circuit, in which case the thermal noise is the Johnson noise \citep{Nyquist28,Pierce60},
\begin{equation}
P_{\rm thermal} (\nu) = \frac{h \nu}{e^{h\nu / (k_B T^{\prime})} - 1}.
\end{equation}
As long as the electronic detector is in thermal equilibrium with the outside radiation, this formula holds because of the Zeroth Law of Thermodynamics, regardless of the details of the antenna or detector.  

If the transmission is entirely at low frequencies ($\nu \la k_B T^{\prime} / h$), then the Johnson noise is simply thermal noise with $P_{\rm thermal} (\nu) = k_B T^{\prime}$.  If Johnson noise is the primary source of noise and the signal spectrum is $P_{\rm rec} (\nu) = \PRecBol$, ${\cal C} = (\Delta \nu) \log_2 [1 + \PRecBol/(k_B T^{\prime})]$, where $\Delta \nu$ is the bandwidth.  The channel capacity reduces to ${\cal C} = \PRecBol \Delta \nu / [(\ln 2) k_B T^{\prime}]$ when the signal-to-noise ratio is small.  Thus, thermal noise limits the amount of information conveyed per unit energy to 1 bit per $(\ln 2) k_B T^{\prime}$ of energy \citep{Pierce60}, and the channel capacity to
\begin{equation}
\label{eqn:ThermalChannelCapacity}
{\cal C} \le \frac{1}{(\ln 2) k_B T^{\prime}} \int_0^{\infty} P_{\rm rec} (\nu) d\nu = \frac{3 (12 \pi^2)^{2/3}}{4096 \pi^4 \ln 2} \frac{L^{\prime} \lambda_a^{2/3} A_{\rm work}}{k_B T^{\prime} R_{\rm box}^{8/3}} \frac{\Gain_{\rm send} \Gain_{\rm rec} \etaRec}{f_{\rm duty}} \left(\frac{\sqrt{f_{\rm pol}} \tau_{\rm abs}}{\etaAbs} \frac{\dBar}{d}\right)^{2/3} \left(\frac{R_{\rm box}}{R^{\prime}}\right)^2 \PBarSpect \eta_{\rm rec}(\nu),
\end{equation}
with a new mean receiver efficiency defined as
\begin{equation}
\etaRec = \int_{\nu_1}^{\nu_2} \left(\frac{\lambda}{\lambda_a}\right)^2 \PBarSpect \eta_{\rm rec}(\nu) d\nu.
\end{equation}
When averaged over time, this is indeed small for isotropically-emitting motes:
\begin{equation}
\mean{{\cal C}} \le 3.2 \times 10^{-7}\ \bit\ \sec^{-1}\ \left(\frac{L^{\prime}}{10^{11}\ \Lsun}\right) \left(\frac{\lambda_a}{\mm}\right)^{2/3} \left(\frac{R_{\rm box}}{10\ \kpc}\right)^{-8/3} \left(\frac{T^{\prime}}{3\ \Kelv}\right)^{-3} \frac{\Gain_{\rm send} \Gain_{\rm rec} f_{\rm work} \etaWork \beta_a \etaRec}{f_{\rm duty}} \left(\frac{\sqrt{2 f_{\rm pol}} \tau_{\rm abs}}{\etaAbs} \frac{\dBar}{d}\right)^{2/3} \left(\frac{R_{\rm box}}{R^{\prime}}\right)^2.
\end{equation}

Going to higher frequencies to avoid thermal noise may seem like a good idea, but it does not work out.  As it turns out, the channel capacity is limited to 1 bit per $(\ln 2) k_B T^{\prime}$ of energy at all frequencies \citep{Pierce60}.  In principle, this means that high frequencies are exactly as good as low frequencies in terms of information transfer per unit energy, with a $100\ \THz$ near-infrared photon carrying up to 1600 bits in a 3 K environment.  But that kind of information transfer is impractical, since in order to code $N$ bits, the encoder/decoder must be able to distinguish between $2^N$ states.  In the case of the 1600 bit photon, this comes out to $10^{482}$ states.  Instead, it might be practical to get $N_{\rm bit} \approx 10 - 20$ bits per photon by using the timing of the photon or spectroscopy.  Still assuming that the transmitter and receivers are antennas, then the channel capacity is limited to 
\begin{equation}
\label{eqn:QuantumChannelCapacity}
{\cal C} \le N_{\rm bit} \int_0^{\infty} \frac{P_{\rm rec} (\nu)}{h \nu} d\nu = \frac{3 (12 \pi^2)^{2/3}}{4096 \pi^4} \frac{L^{\prime} \lambda_a^{5/3} A_{\rm work}}{h c R_{\rm box}^{8/3}} \frac{N_{\rm bit} \Gain_{\rm send} \Gain_{\rm rec} \eta_{\rm rec}^{\star}}{f_{\rm duty}} \left(\frac{\sqrt{f_{\rm pol}} \tau_{\rm abs}}{\etaAbs} \frac{\dBar}{d}\right)^{2/3} \left(\frac{R_{\rm box}}{R^{\prime}}\right)^2 \PBarSpect \eta_{\rm rec}(\nu),
\end{equation}
this time, the receiver efficiency is defined as
\begin{equation}
\eta_{\rm rec}^{\star} = \int_{\nu_1}^{\nu_2} \left(\frac{\lambda}{\lambda_a}\right)^2 \PBarSpect \eta_{\rm rec}(\nu) (\nu_a/\nu) d\nu.
\end{equation}
So with these assumptions, the data transmission rate for a high frequency antenna system is actually poor:
\begin{equation}
\mean{{\cal C}} \le 4.7 \times 10^{-8}\ \bit\ \sec^{-1}\ \left(\frac{L^{\prime}}{10^{11}\ \Lsun}\right) \left(\frac{\lambda_a}{\mm}\right)^{5/3} \left(\frac{R_{\rm box}}{10\ \kpc}\right)^{-8/3} \left(\frac{T^{\prime}}{3\ \Kelv}\right)^{-2} \frac{N_{\rm bit} \Gain_{\rm send} \Gain_{\rm rec} f_{\rm work} \etaWork \beta_a \eta_{\rm rec}^{\star}}{f_{\rm duty}} \left(\frac{\sqrt{2 f_{\rm pol}} \tau_{\rm abs}}{\etaAbs} \frac{\dBar}{d}\right)^{2/3} \left(\frac{R_{\rm box}}{R^{\prime}}\right)^2.
\end{equation}
As the frequency increases towards the optical regime, the data rate plummets to the point that it takes millions of years just to communicate one bit.

Of course, optical communication is practical in our own technology because neither the detectors nor the receivers are isotropically emitting antennas.  Actual optical detectors have areas far greater than a square wavelength.  Furthermore, because optical wavelengths are much shorter than radio wavelengths, relatively small telescopes can magnify small portions of the sky, boosting the gains of both the transmitter and the receivers.

Aside from thermal and quantum noise, there is also confusion noise from all of the other motes transmitting to each other along the line of sight.  Assuming these motes are oriented randomly, and that the emitting beams are wide enough that they don't all miss the original receiving mote, the time-averaged confusion noise is
\begin{equation}
\label{eqn:ConfusionNoise}
P_{\rm confusion} (\nu) = n_{\rm motes} d_{\rm max} \mean{P_{\rm send}^{\rm all} (\nu)} \eta_{\rm rec}(\nu) \left(\frac{\lambda}{4 \pi}\right)^2.
\end{equation}
The maximum distance along the sightline $d_{\rm max}$ is either the distance to the edge of the blackbox, or the distance until a photon is absorbed, whichever is smaller.  Strictly speaking, each of the motes on the sightline may transmit signals with a distinct spectrum, all of which might be different than the original transmitting mote, so $\mean{P_{\rm send}^{\rm all} (\nu)}$ averages over both time and motes.  If all of the motes are using wideband signals near the Wien peak of the thermal background, then the confusion noise is less than the thermal noise simply from energy conservation.  But the confusion noise is always present, and it is important if (1) the motes' transmissions are all confined to a bandwidth $\Delta \nu \ll k_B T^{\prime} / h$ or (2) the transmissions are at high frequency where there is no thermal noise.

From the form of equation~\ref{eqn:ConfusionNoise}, it is clear that it is enormous compared to the neighboring mote's signal:
\begin{align}
\nonumber \frac{\PRecSpect}{P_{\rm confusion}(\nu)} & = \frac{\Gain_{\rm send} \Gain_{\rm rec}}{f_{\rm duty}} \left(\frac{f_{\rm pol} \etaAbs \lambda_a^2}{12\pi \tau_{\rm abs} R_{\rm box}^2}\right)^{1/3} \left(\frac{R_{\rm box}}{d_{\rm max}}\right) \left(\frac{d}{\dBar}\right)^{-2}\\
           & = 5.2 \times 10^{-17}\ \left(\frac{R_{\rm box}}{10\ \kpc}\right)^{-2/3} \left(\frac{\lambda_a}{\mm}\right)^{2/3} \frac{\Gain_{\rm send} \Gain_{\rm rec}}{f_{\rm duty}} \left(\frac{2 f_{\rm pol} \etaAbs}{\tau_{\rm abs}}\right)^{1/3} \left(\frac{R_{\rm box}}{d_{\rm max}}\right) \left(\frac{d}{\dBar}\right)^{-2}.
\end{align}
It follows that the channel capacity is tiny, ${\cal C} \le \Delta\nu \times P_{\rm send} (\nu) / [(\ln 2) P_{\rm confusion} (\nu)]$, which is
\begin{equation}
\label{eqn:ConfusionCapacity}
{\cal C} \le 2.2 \times 10^{-5}\ \bit\ \sec^{-1}\ \left(\frac{R_{\rm box}}{10\ \kpc}\right)^{-2/3} \left(\frac{\lambda_a}{\mm}\right)^{-1/3} \frac{\Gain_{\rm send} \Gain_{\rm rec} \beta_a}{f_{\rm duty}} \left(\frac{2 f_{\rm pol} \etaAbs}{\tau_{\rm abs}}\right)^{1/3} \left(\frac{R_{\rm box}}{d_{\rm max}}\right) \left(\frac{d}{\dBar}\right)^{-2},
\end{equation}
assuming the frequency bandwidth is $\Delta\nu = \beta_a c/\lambda_a$.

The confusion noise is a fundamental limitation to the rate of information transfer.  It is unaffected by the surrounding temperature.  If one tries to assign each mote its own frequency channel, then the incredibly narrow bandwidth required cancels the improvement in signal-to-noise.  If one tries to stagger the mote transmissions so that only a few are broadcasting at a given time, the time-average channel capacity is reduced in proportion to $f_{\rm duty}$.  More generally, one could try to assign each mote a ``signature'' signal or waveform that a receiver can pick out with a matched filter.  But the signature must have a duration of at least $\sim P_{\rm rec}/P_{\rm confusion} \times (1/\nu)$ to be reliably detected against the noise background, so equation~\ref{eqn:ConfusionCapacity} is enforced.  

Still, the situation is not hopeless for would-be builders of a galactic Internet.  First, the motes could be designed with a high gain, beaming the transmissions.  One way to do this would be to group motes into phased arrays, which could be scaled to arbitrarily large numbers.  Dense phased arrays (with spacings $\la \lambda$) have great freedom in beamforming, with the ability to steer the beam in almost any direction \citep{Tegmark09,Tegmark10}, so neither the transmitter nor the receiver needs to be aimed at the other (although the transmitting array needs information on where the receiver is to steer the beam in its direction).  

Another solution is to arrange the motes in clusters, so that $d/\bar{d} \ll 1$ for motes within the clusters.  In order to send messages between clusters, some kind of relay is necessary.  This relay may simply be all of the motes in the cluster working together.  In this case, when one relay communicates with a neighboring relay, and both consist of $N_{\rm clust}$ motes, the signal-to-noise at the receiver is boosted as $N_{\rm clust}^{1/3}$, even if there are no improvements in gain by beamforming.  Furthermore, there is no reason that the clusters themselves couldn't be grouped into higher order clusters, and so on, until the network has a fractal topology.  Clustering does have the possible disadvantage of reducing the effective opacity from the motes; this can be avoided if the clusters are (at most) two dimensional or are sparse.

If none of the power transmitted in messages misses the receiver, Landauer's Principle sets the maximum channel capacity.  The best possible ${\cal C}$ is just $\dot{\cal I}$, which is of order $10\ \Gbit\ \sec^{-1}$.  Again, this is not much faster than our own network speeds, but the blackbox does it on astronomical scales.  

On a final note, my calculations are based on the assumption that the channel between two motes is open forever.  The optimal coding of information in the presence of noise is based around grouping large chunks of information; otherwise equation~\ref{eqn:ChannelCapacity} is not obtained \citep{Pierce60}.  The successful receipt of a message, of course, requires at least one photon to be detected, but this is already considered in equation~\ref{eqn:QuantumChannelCapacity}, and at low frequencies ($\la k_B T^{\prime} / h$), energy is the more stringent limit since more than one photon is needed to convey one bit in that case \citep{Pierce60}.

\section{Motes as thrusters}
\label{sec:Thrusters}
I also suggested in Section~\ref{sec:GeniusLoci} that motes might be used to direct the flow of matter in the ISM.  The basic principle is the same as the natural driving of winds by radiation pressure on dust grains in stars and winds, but I want to show that it works in a blackbox-like environment too.

If the motes are to push the gas around, they must be able to transfer their momentum to the gas on time scales short compared to the  ISM's dynamical timescale.  The simplest mechanism for this is drag, where the gas particles directly collide with the mote itself.  Consider a mote with a geometrical cross section $A_{\rm geom}$ moving at a speed $v$ relative to gas with mass density $\rho_{\rm ISM}$.  The drag force is the drag pressure $(1/2) \rho_{\rm ISM} v^2$ times the mote's cross section, and the deceleration time is
\begin{equation}
t_{\rm drag} \approx \frac{2 m}{\rho_{\rm ISM} v A_{\rm geom}}.
\end{equation}
If the mote is a solid cylinder, its mass is $m = \pi \rho r^2 \ell = \pi \rho r^2 \lambda_a / (2 f_{\ell})$ and its cross section is $A_{\rm geom} = 2 f_{\rm geom} \ell r$, where $f_{\rm geom}$ is a factor relating to the orientation of the mote.  The drag time is then
\begin{equation}
t_{\rm drag} = \frac{\pi \rho r}{f_{\rm geom} \rho_{\rm ISM} v} = 6.0\ \Myr\ f_{\rm geom}^{-1} \left(\frac{\rho}{\rhoUnits}\right) \left(\frac{r}{100\ \nm}\right) \left(\frac{\rho_{\rm ISM}}{\amu\ \cm^{-3}}\right)^{-1} \left(\frac{v}{\kms}\right)^{-1}.
\end{equation}

One way for a mote to accelerate is to act as a radiative thruster, broadcasting its momentum away in photons.  Only moderate beaming is required for a net force.  The (time-averaged) maximum possible thrust is then ${\cal F} = \dot{E}_{\rm harvest} / c$.  Typical accelerations are
\begin{align}
\nonumber a_{\rm radiative} & \le \frac{\dot{E}_{\rm harvest}}{m c} = \frac{3 L^{\prime} A_{\rm work} f_{\rm pol} f_{\ell}}{128 \pi^3 \rho R^{\prime 2} r^2 \lambda_a c}\\
                            & \le 3.7 \times 10^{-8}\ \meter\ \sec^{-2}\ (2 f_{\rm pol}) f_{\rm work} f_{\ell} \beta_a \etaWork \left(\frac{L^{\prime}}{10^{11}\ \Lsun}\right) \left(\frac{\rho}{\rhoUnits}\right)^{-1} \left(\frac{r}{100\ \nm}\right)^{-2} \left(\frac{\lambda_a}{\mm}\right)^{-1} \left(\frac{R^{\prime}}{10\ \kpc}\right)^{-2} \left(\frac{T^{\prime}}{3\ \Kelv}\right)^{-2}.
\end{align}
In general, this is too slow to avoid colliding with motes unless the warning time is far greater than $\ell/v$ (Section~\ref{sec:Collisions}).

Note that the maximum thrust and the drag time together define a maximum relative velocity of the grains to the gas, $v_{\rm max} = a_{\rm radiative} t_{\rm drag}$.  For a cylindrical antenna,
\begin{align}
\label{eqn:vMaxMote}
\nonumber v_{\rm max} & = \sqrt{\frac{3 L^{\prime} A_{\rm work} f_{\rm pol} f_{\ell}}{128 \pi^2 f_{\rm geom} \rho_{\rm ISM} c \lambda_a r R^{\prime 2}}} \\
                      & = 84\ \kms\ \sqrt{(2 f_{\rm pol}) f_{\rm work} f_{\ell} \beta_a \etaWork / f_{\rm geom}} \left(\frac{L^{\prime}}{10^{11}\ \Lsun}\right)^{1/2} \left(\frac{\rho_{\rm ISM}}{\amu\ \cm^{-3}}\right)^{-1/2} \left(\frac{r}{100\ \nm}\right)^{-1/2} \left(\frac{\lambda_a}{\mm}\right)^{-1/2} \left(\frac{R^{\prime}}{10\ \kpc}\right)^{-1} \left(\frac{T^{\prime}}{3\ \Kelv}\right)^{-1}.
\end{align}
This comes out to about 1 Mpc per 10 Gyr for an ISM density of about $1\ \amu\ \cm^{-3}$.  So, it should be possible to launch motes directly into a galactic halo or even into the IGM without actually sending manufacturing facilities there, especially if they are injected into the hot, rarefied ISM  (note that $a_{\rm radiative}$ is much greater than the pull of gravity, which is usually around $10^{-12}$ to $10^{-11}\ \meter\ \sec^{-2}$ in a disk galaxy).

We might expect the ISM mass to be much greater than the total mass in motes.  Consider a small, optically thin parcel of gas deep within the blackbox.  (The parcel is optically thin so that it does not absorb its own radiative thrust).  Let $dM_{\rm gas}$ be the amount of gas within the parcel, and $dM_{\rm motes}$ be the mass in motes within the parcel.  The maximum thrust the parcel can achieve is $a_{\rm gas} =  a_{\rm radiative} dM_{\rm motes} / (dM_{\rm gas} + dM_{\rm motes}) \approx a_{\rm radiative} dM_{\rm motes} / dM_{\rm gas}$.  Since the motes are, to some extent, locked into the gas, the density in motes is probably roughly proportional to the gas density.  Thus, we can relate $dM_{\rm motes}$ to the total mote mass in the blackbox, $M_{\rm motes}$ (equation~\ref{eqn:DipoleCloudMass}), by a simple factor $\Delta_{\rm motes}$.  Then,
\begin{align}
\label{eqn:aGasOpticallyThin}
\nonumber a_{\rm gas} & \approx a_{\rm radiative} \frac{\Delta_{\rm motes} M_{\rm motes}}{M_{\rm gas}} \le \frac{3 L^{\prime} A_{\rm work} f_{\rm pol} f_{\ell} \Delta_{\rm motes}}{128 \pi^3 \rho R^{\prime 2} r^2 \lambda_a c} \frac{A_{\rm box} \Sigma_{\rm blackbox} \tau_{\rm abs}}{M_{\rm gas}}\\
\nonumber             & \le 2.2 \times 10^{-11}\ \meter\ \sec^{-2}\ (2 f_{\rm pol}) f_{\rm work} f_{\ell} \beta_a \etaWork \Delta_{\rm motes} \tau_{\rm abs}\\
                      & \times \left(\frac{L^{\prime}}{10^{11}\ \Lsun}\right) \left(\frac{\rho}{\rhoUnits}\right)^{-1} \left(\frac{r}{100\ \nm}\right)^{-2} \left(\frac{\lambda_a}{\mm}\right)^{-1} \left(\frac{T^{\prime}}{3\ \Kelv}\right)^{-2} \left(\frac{\Sigma_{\rm blackbox}}{10^{-5}\ \kg\ \meter^{-2}}\right) \left(\frac{M_{\rm gas}}{10^{10}\ \Msun}\right)^{-1} \left(\frac{R^{\prime}}{R_{\rm box}}\right)^{-2}.
\end{align}
This is enough to levitate gas out of the midplane of a disk galaxy if $\tau_{\rm abs} = 1$.  Note that for a small parcel, the maximum thrust is proportional to $\tau_{\rm abs}$.

It's more challenging to move a large, optically thick parcel.  If the parcel is optically thick to its own thrust radiation, then only motes near the surface of the parcel contribute to the thrust of the parcel as a whole.  If the parcel is optically thin to the thrust radiation, the thermodynamically useful energy flux is removed from the parcel within one optical depth of its surface.  Thus, if the optical depth of the parcel to thermal radiation is $\tau_{\rm parcel} \ga 1$, $a_{\rm gas}$ is reduced by a factor $\tau_{\rm parcel}$ compared to equation~\ref{eqn:aGasOpticallyThin}.  When considering the ISM as a whole, the maximum thrust is limited by the total momentum of the blackbox's radiation, so $a_{\rm gas} \approx L_{\rm excess}/(M_{\rm gas} c)$.  

Photons are a poor propellant in an energy-limited environment, since they move at $c$.  If the motes ejected a material propellant at a mass rate $\dot{M}$, they could impart a speed of $u = \sqrt{2 \dot{E}_{\rm harvest} / \dot{M}}$ into it, for a total force of ${\cal F} = \sqrt{2 \dot{E}_{\rm harvest} \dot{M}}$.  Although motes are tiny and unlikely to carry a spare load of propellant within them, they might conceivably harvest the matter around them, analogous to the ramscoop concept of interstellar travel.

Suppose a mote uses all gas that falls onto some capture area $A_{\rm capture}$ as propellant.  It also continues to act as an antenna, heating the propellant with the thermodynamically useful flux of radiation.  Gas enters into $A_{\rm capture}$ with a mean speed of $v_{\rm capture}$.  If the mote is moving with a Mach number greater than 1, then $v_{\rm capture}$ is roughly the speed of the mote, but even if it is at rest with respect to the gas, atoms will still enter the capture area at roughly the speed of sound $c_S$ (which is roughly $0.1\ \kms$ at 3 K).  Let $A_{\rm capture} = f_{\rm capture} 2 \ell r = f_{\rm capture} \lambda_a r / f_{\ell}$.  Then the maximum thrust possible is
\begin{align}
\nonumber a_{\rm ram} & = \frac{1}{\pi^2 \rho} \sqrt{\frac{3 L^{\prime} f_{\rm pol} A_{\rm work} \rho_{\rm ISM} v_{\rm capture} f_{\rm capture} f_{\ell}}{32 R^{\prime 2} \lambda_a r^3}}\\
\nonumber             & = 4.8 \times 10^{-7}\ \meter\ \sec^{-2}\ \left(\frac{\rho}{\rhoUnits}\right)^{-1} \\
                      & \times \sqrt{f_{\rm work} \etaWork \beta_a (2 f_{\rm pol}) f_{\rm capture} f_{\ell} \left(\frac{L^{\prime}}{10^{11}\ \Lsun}\right) \left(\frac{R^{\prime}}{10\ \kpc}\right)^{-2} \left(\frac{\lambda_a}{\mm}\right)^{-1} \left(\frac{r}{100\ \nm}\right)^{-3} \left(\frac{v_{\rm capture}}{\kms}\right) \left(\frac{\rho_{\rm ISM}}{\amu\ \cm^{-3}}\right)}.
\end{align}

The ratio of thrusts from matter and photons is ${\cal F}_{\rm ram}/{\cal F}_{\rm radiative} = \sqrt{2 \dot{M} c^2 / \dot{E}}$:
\begin{align}
\nonumber \frac{{\cal F}_{\rm ram}}{{\cal F}_{\rm radiative}} & = \sqrt{\frac{256 \pi^2 R^{\prime 2} c^2 \rho_{\rm ISM} \lambda_a r v f_{\rm capture}}{3 L^{\prime} A_{\rm work} f_{\rm pol} f_{\ell}}}\\
          & = 9.3 \sqrt{\frac{f_{\rm capture}}{2 f_{\rm work} \etaWork \beta_a f_{\rm pol} f_{\ell}} \left(\frac{L^{\prime}}{10^{11}\ \Lsun}\right)^{-1} \left(\frac{R^{\prime}}{10\ \kpc}\right)^2 \left(\frac{T^{\prime}}{3\ \Kelv}\right)^2 \left(\frac{\rho_{\rm ISM}}{\amu\ \cm^{-3}}\right) \left(\frac{\lambda_a}{\mm}\right) \left(\frac{r}{100\ \nm}\right) \left(\frac{v_{\rm capture}}{\kms}\right)}
\end{align}
If the mote is moving slowly in rarefied ISM and only captures gas that actually hits it, then radiative emission provides a greater possible thrust.  If the mote can capture gas across an area $\sim \ell^2$, perhaps using magnetic fields, then gaseous thrust is around a thousand times more potent.  Using captured gas propellant also is advantageous in low luminosity galaxies.

\section{The growth of motes}
\label{sec:Manufacture}
Planetary systems with small bodies conveniently provide the materials to build a planetary- or stellar-scale blackbox.  A galactic-scale blackbox, though, requires so much materials that a large fraction of the ISM must be mined. Furthermore, the ISM is incredibly diffuse, occupying cubic kiloparsecs.  At first glance, it seems that something the size of a galaxy is needed to catch the metals necessary to build the blackbox in the first place.  

The volume swept up by a blackbox's mote population approaches the total galactic volume over cosmological timescales, though.  That is why the collision problem is so tough for blackboxes, although the collision cross section is much larger than a dipole antenna's geometrical cross section (Appendix~\ref{sec:Collisions}).  Thus, if the motes can replicate themselves using metals they sweep up from the ISM, their population could grow exponentially until the ISM is fully mined.  Alternatively, self-replicating nets might be launched; these would harvest metals in order to grow catchments for ISM metals, devoting a fraction of the consumed ISM for mote manufacture.

The rate that material is harvested is estimated along the same lines as the capture of ISM gas for use in thrusters (Appendix~\ref{sec:Thrusters}).  For a general population of $N_{\rm repl}$ replicators, each capturing ISM that lands on an area $A_{\rm capture}$ as it moves at a speed $v_{\rm capture}$, the rate of growth in the total mass of the replicators is
\begin{equation}
\dot{M}_{\rm repl} = N_{\rm repl} \rho_{\rm ISM} A_{\rm capture} v_{\rm capture} Z,
\end{equation}
where $Z$ is the abundance of the elements that make up the motes.  A general replicating machine might look like a very delicate sheet with surface density $\Sigma_{\rm repl}$ and a mass of $m_{\rm repl} = \Sigma_{\rm repl} A_{\rm capture}$.  The time for population growth is $t_{\rm repl} = m_{\rm repl} N_{\rm repl} / \dot{M}_{\rm repl}$:
\begin{equation}
t_{\rm repl} = \frac{\Sigma_{\rm repl}}{\rho_{\rm ISM} v_{\rm capture} Z}.
\end{equation}

Consider a cylindrical mote that harvests materials that impede onto its geometrical cross section $A_{\rm capture} = 2 f_{\rm capture} \ell r$ (see Section~\ref{sec:Thrusters} for details).  Then, its effective surface density is $\Sigma_{\rm repl} = (\pi/2) \rho r / f_{\rm capture}$.  If it has the capability of building more motes with the harvested materials, the time for the mote population to grow by a factor of $e$ is
\begin{equation}
t_{\rm repl}^{\rm mote} = \frac{\pi \rho r}{2 f_{\rm capture} \rho_{\rm ISM} v_{\rm capture} Z} = 3.0\ \Gyr\ f_{\rm capture}^{-1}\ \left(\frac{\rho}{\rhoUnits}\right) \left(\frac{r}{100\ \nm}\right) \left(\frac{\rho_{\rm ISM}}{\amu~\cm^{-3}}\right)^{-1} \left(\frac{v_{\rm capture}}{\kms}\right)^{-1} \left(\frac{Z}{10^{-3}}\right).
\end{equation} 
Although this is shorter than the age of the Universe, the replication process probably would take several $e$-folds to complete.  The harvest of the ISM could be completed within a few Gyr if the motes achieve $v \approx 100\ \kms$ with the use of radiative thrusting (equation~\ref{eqn:vMaxMote}), even starting from a small population.  Harvesting the ICM of galaxy clusters, which is about a thousand times more rarefied than the ISM, is a much slower process, however.  

The harvesting can also be sped up if the replicators take advantage of the clumping of the ISM.  The supersonic turbulence in the cold, molecular ISM concentrates most of the H$_2$ mass into clouds that are overdense by a factor of $\sim 100$ \citep{Ostriker01,MacLow04}.  Nor do the motes necessarily have to seek out these clouds; they could simply ride along with the gas, following most of it into the clumps.  While only a minority of the Galactic ISM's mass is cold and molecular, gas mixes between the phases on timescales of $\la$ 200 Myr \citep{deAvillez02}.  This strategy will not work in the ICM, however, where the turbulence is subsonic and density fluctuations are relatively small \citep{Subramanian06}. 

Another possibility is to use a separate class of replicators to actually harvest the ISM more efficiently, and only reprocess that material into motes later.  A more efficient geometry for harvesting the ISM would be a very thin sheet only a few nanometers thick, perhaps made of graphene or carbon nanotubes, with $\Sigma_{\rm repl} \approx 1\ \ugm2$.  Then, these sheets would have a population growth time of
\begin{equation}
t_{\rm repl} = 19\ \Myr\ \left(\frac{\rho}{\rhoUnits}\right) \left(\frac{\Sigma_{\rm repl}}{\ugm2}\right) \left(\frac{\rho_{\rm ISM}}{\amu~\cm^{-3}}\right)^{-1} \left(\frac{v_{\rm capture}}{\kms}\right)^{-1} \left(\frac{Z}{10^{-3}}\right).
\end{equation}
If combined with a faster $v_{\rm capture}$, or if they spend most of their time in molecular clouds, an initially small population could collect much of the ISM's metals within the Fermi-Hart timescale of a few Myr.  Once again, though, population growth in a cluster's diffuse ICM is probably too slow to have been achieved yet. 

Further savings in $\Sigma_{\rm repl}$ are possible if the sheets are porous; although this would make it harder to refine gas phase metals in the ISM, the replicator could still capture solid interstellar dust grains and slowly digest them.  Of course, individual replicators could not grow exponentially in size for long, because of the time it would take to circulate materials towards their edges.  Instead, these sheets could be made to fragment once they grew to a certain size, keeping the circulation times reasonable.  A variation of this budding process might release motes into the ISM.  Or, these replicators might be like autotrophs in our own ecosystem, with the motes (or mote-producing machines) being heterotrophs that eat the sheets.

\end{document}